\documentclass[aps,pra,reprint,superscriptaddress,twocolumn]{revtex4-2}
\usepackage[dvipdfmx]{graphicx}
\usepackage{graphicx}
\usepackage{color}
\usepackage[dvipsnames]{xcolor}
\definecolor{DarkGreen}{rgb}{0.0, 0.5, 0.0}
\usepackage[colorlinks,citecolor=DarkGreen,linkcolor=blue,linktocpage]{hyperref}

\usepackage[T1]{fontenc}
\usepackage{textcomp}
\usepackage{bm}
\usepackage{amsmath,amssymb,amsthm}
\usepackage{comment}
\usepackage{grffile}
\usepackage{cancel}
\usepackage{algpseudocode}

\newcommand{\cket}[1]{\left|#1\right\rangle}
\newcommand{\bra}[1]{\left\langle#1\right|}

\begin{document}


\title{Proposal for realizing Heisenberg-type quantum-spin models in Rydberg-atom quantum simulators}

\author{Masaya Kunimi}
\email{kunimi@rs.tus.ac.jp}
\affiliation{Department of Physics, Tokyo University of Science, 1-3 Kagurazaka, Tokyo 162-8601,  Japan}
\author{Takafumi Tomita}
\email{tomita@ims.ac.jp}
\affiliation{Department of Photo-Molecular Science, Institute for Molecular Science,
National Institutes of Natural Sciences, 38 Nishigo-Naka, Myodaiji, Okazaki, Aichi 444-8585, Japan}
\affiliation{SOKENDAI (The Graduate University for Advanced Studies), 38 Nishigo-Naka, Myodaiji, Okazaki, Aichi 444-8585, Japan}


\date{\today}

\begin{abstract}
We investigate the magnetic-field dependence of the interaction between two Rydberg atoms, $|nS_{1/2}, m_J\rangle$ and $|(n+1)S_{1/2}, m_J\rangle$. In this setting, the effective spin-1/2 Hamiltonian takes the form of an {\it XXZ} model. We show that the anisotropy parameter of the {\it XXZ} model can be tuned by applying a magnetic field and, in particular, that it changes drastically near the F\"orster resonance points. Based on this result, we propose experimental realizations of spin-1/2 and spin-1 Heisenberg-type quantum spin models in Rydberg atom quantum simulators, without relying on Floquet engineering. Our results provide guidance for future experiments of Rydberg atom quantum simulators and offer insight into quantum many-body phenomena emerging in the Heisenberg model.
\end{abstract}
\maketitle

{\it Introduction.} The quantum Heisenberg model is one of the most important models for interacting spin systems and plays a crucial role in various fields of physics. In condensed matter physics, the Heisenberg model serves as a fundamental framework for understanding quantum magnetism~\cite{Auerbach1998interacting}, high-temperature superconductivity~\cite{Lee2006doping}, and exotic phases such as quantum spin liquid~\cite{Balents2010}. Moreover, the Heisenberg model has been extensively studied in the context of mathematical physics~\cite{Takahashi1999thermodynamics,Sutherland2004beautiful}, statistical physics~\cite{tasaki2020physics}, and quantum information theory~\cite{Eisert2010}. 

Although the quantum Heisenberg model has been extensively explored, it is difficult to solve it analytically except for the special cases such as the one-dimensional $S=1/2$ chain, which is solvable via the Bethe ansatz~\cite{Bethe1931}. As an alternative to analytical and numerical approaches, quantum simulation is a powerful method for studying the Heisenberg model. In recent years, it has been experimentally realized on various platforms. For example, in the large-$U$ limit, the Fermi-Hubbard model \cite{Mazurenko2017,Brown2017,Xu2023,Shao2024} and the Bose-Hubbard model \cite{Fukuhara2013,Fukuhara2013microscopic,Hild2014,Jepsen2020,Jepsen2021,Sun2021,Jepsen2022,Wei2022} reduces to the Heisenberg model as an effective Hamiltonian. In Rydberg atom quantum simulators~\cite{Browaeys2020}, the Heisenberg model can be engineered using the Floquet technique~\cite{Geier2021,Scholl2022}. Trapped-ion systems~\cite{Morong2023,Kranzl2023experimental,Kranzl2023}, polar molecules~\cite{christakis2023probing,miller2024two,carroll2025observation} and superconducting qubit platforms~\cite{Nguyen2024,Chowdhury2024,Rosenberg2024} have also been used to simulate the Heisenberg model. Moreover, $S=1$ Heisenberg models have been realized in optical lattice systems~\cite{Chung2021,Luo2024}.

In this Letter, we focus on the Rydberg atom quantum simulators. In this platform, various quantum spin models have been experimentally realized, including the Ising model~\cite{Labuhn2016,Zeiher2017,Bernien2017,Leseleuc2018,Lienhard2018,Guardado2018,Keesling2019,Ebadi2021,Semeghini2021,Bluvstein2021,Scholl2021,Hollerith2022,Zhao2023,Bharti2023,Franz2024,Bharti2024,Zhao2025,manovitz2025quantum,datla2025statistical}, the {\it XY} model~\cite{Ravets2014,Ravets2015,Orioli2018,Leseleuc2019,Chew2022,Franz2022a,Chen2023,Franz2024,Bornet2023,Bornet2024,Emperauger2025,Emperauger2025benchmarking,chen2025spectroscopy}, and the {\it XXZ} model~\cite{Signoles2021,Geier2021,Scholl2022,Franz2022a,Steinert2023,Franz2024,Emperauger2025benchmarking}, and others~\cite{Lienhard2020,Kanungo2022,Qiao2025,evered2025probing}. As mentioned above, the Heisenberg model has also been realized using Floquet engineering~\cite{Geier2021,Scholl2022}. More complicated models related to the Heisenberg model have been theoretically proposed \cite{nishad2023quantum,kuji2025proposal,tian2025engineering}. However, this approach has several experimental limitations, such as the requirement for precise control of time-dependent external fields and the constraints on the accessible timescales due to decoherence due to the effects of the finite pulse width. Therefore, it is desirable to realize the Heisenberg model without relying on Floquet techniques. 

One possible route toward this goal is to start from the {\it XXZ} Hamiltonian and tune the anisotropy parameter $\delta$ to unity. The {\it XXZ} model can be implemented in Rydberg atom systems by choosing two Rydberg states with the same parity and treating them as an effective $S=1/2$ system, such as $|nS_{1/2},m_J\rangle$ and $|n'S_{1/2},m_J\rangle$, where $n, n'\gg1$ are principal quantum numbers and $m_J$ is the magnetic quantum number. There are several strategies to tune $\delta$. One is to vary the choice of the principal quantum number. Unfortunately, as shown in Whitlock {\it et al}.~\cite{Whitlock2017} [see Fig.~4(b) in the reference], the Heisenberg point $\delta=1$ cannot be achieved in ${}^{87}$Rb atoms in the absence of external fields. Another approach is to apply the static electric and/or magnetic fields to the Rydberg atoms, which allows continuous tuning of $\delta$. However, to our knowledge, there has been no systematic calculation of the achievable range of $\delta$ under applied external fields.

In this Letter, we calculate the interaction between the Rydberg atom pair $|nS_{1/2},m_J\rangle$ and $|(n+1)S_{1/2},m_J\rangle$ in the presence of a static uniform magnetic field, using second-order perturbation theory. We identify the parameter regime of the Heisenberg point, which appears near the F\"orster resonance. As an application of these results, we propose a method for the experimental realization of a tunable $J_1$-$J_2$ Heisenberg chain, which includes the Majumdar-Ghosh model~\cite{majumdar1969next,majumdar1969next2} as a special point. Furthermore, we also propose a scheme for realizing the $S=1$ Heisenberg model using Rydberg atoms.

\begin{figure}[t]
\centering
\includegraphics[width=5cm,clip]{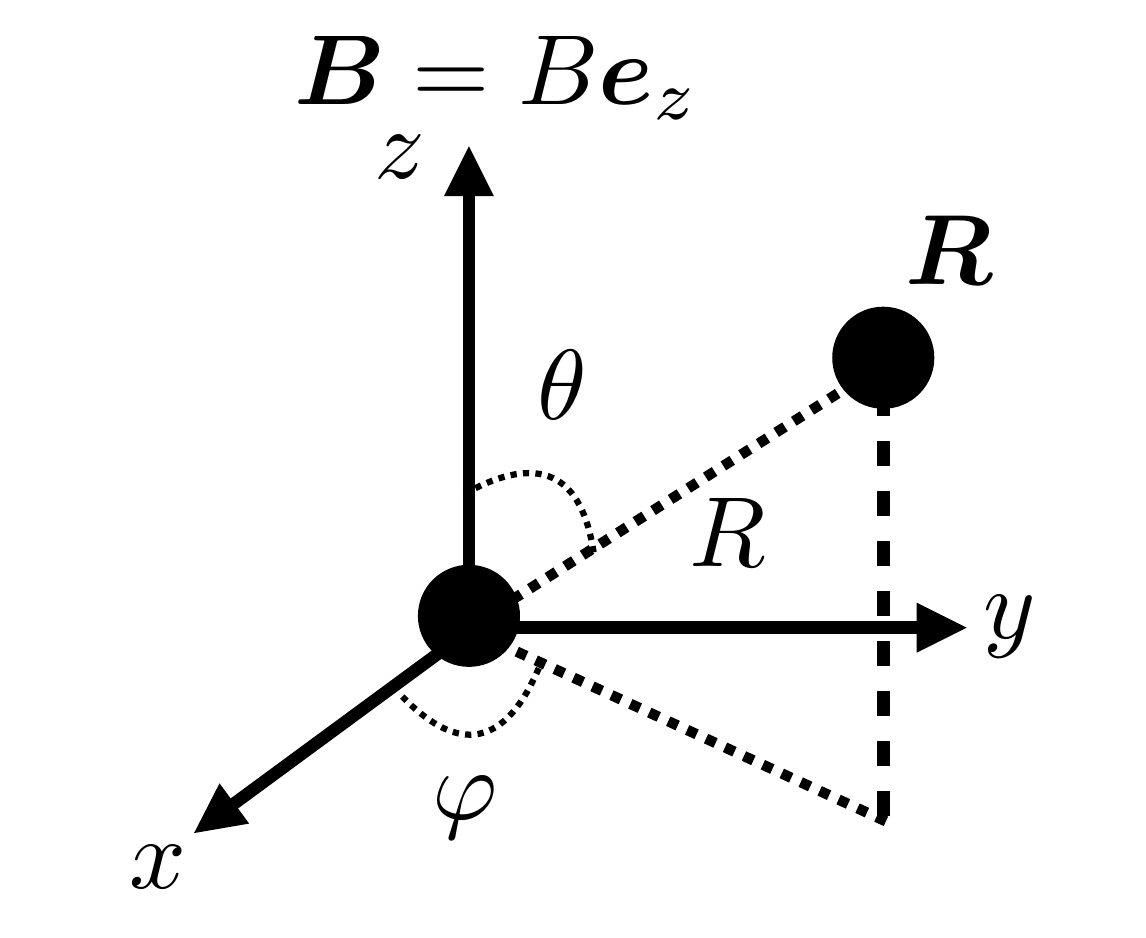}
\caption{Definition of the positions of two Rydberg atoms. One atom is placed at the origin, and the other is placed at the position $\bm{R} \equiv R(\sin\theta \cos\varphi, \sin\theta \sin\varphi, \cos\theta)$, where $R$ is the distance between the atoms, and $\theta$ and $\varphi$ are the polar and azimuthal angles, respectively.
}
\label{fig:position_of_Rydberg_atoms}
\vspace{-0.75em}
\end{figure}

{\it Methods.} We consider two Rydberg atoms in the presence of a static uniform magnetic field $\bm{B}$. In this Letter, the magnetic field is always applied along the quantization axis, i.e., positive $z$ direction: $\bm{B}\equiv B\bm{e}_z\; (B\ge 0)$, where $\bm{e}_z$ is the unit vector (see Fig.~\ref{fig:position_of_Rydberg_atoms}). We assume that the dipole-dipole interaction acts between the two Rydberg atoms. This assumption can be justified when the spatial wave functions of the Rydberg orbits do not overlap~\cite{Sibalic2017,Weber2017}. The interaction Hamiltonian is given by~\cite{Ravets2015}
\begin{align}
\hat{V}_{\rm dd}&\equiv \hat{V}_1+\hat{V}_2+\hat{V}_3,\label{eq:definition_of_dipole-dipole_interaction}\\
\hat{V}_1&\equiv \frac{1-3\cos^2\theta}{4\pi\epsilon_0R^3}\left[\frac{1}{2}(\hat{d}_1^+\hat{d}_2^-+\hat{d}_1^-\hat{d}_2^+)+\hat{d}_1^0\hat{d}_2^0\right],\label{eq:definition_of_V1_dipole_dipole}\\
\hat{V}_2&\equiv \frac{(3/\sqrt{2})\sin\theta\cos\theta}{4\pi\epsilon_0R^3}[e^{-i\varphi}(\hat{d}_1^+\hat{d}_2^0+\hat{d}_1^0\hat{d}_2^+)\notag \\
&\hspace{8.0em}-e^{+i\varphi}(\hat{d}_1^-\hat{d}_2^0+\hat{d}_1^0\hat{d}_2^-)],\label{eq:definition_of_V2_dipole_dipole}\\
\hat{V}_3&\equiv -\frac{(3/2)\sin^2\theta}{4\pi\epsilon_0R^3}\left(e^{-2i\varphi}\hat{d}_1^+\hat{d}_2^++e^{+2i\varphi}\hat{d}_1^-\hat{d}_2^-\right),\label{eq:definition_of_V3_dipole_dipole}
\end{align}
where $\epsilon_0$ is the electric constant, $\hat{d}_i^{\mu}(\mu=x,y,z, i=1,2)$ is the dipole operator of the $i$th atom, $\hat{d}_i^{\pm}\equiv \mp(\hat{d}_i^x\pm i\hat{d}_i^y)/\sqrt{2}$, and $\hat{d}_i^0\equiv \hat{d}_i^z$.

To obtain the effective Hamiltonian, we first calculate the single-atom wavefunction in a uniform magnetic field including diamagnetic terms using the {\it pairinteraction} software~\cite{Weber2017}. In the main text, we consider a relatively weak magnetic field regime, $B\le 200~{\mathrm G}$. In this regime, the single-atom wavefunction in the presence of the magnetic field has strong overlap with the Rydberg wavefunction in the absence of the magnetic field. Therefore, we can identify the magnetic-field-dressed eigenstates by their overlap with the field-free wave function in the absence of the magnetic field $|nL_J,m_J\rangle$, where $L$ denotes the angular momentum of the Rydberg atom. We denote by $|\widetilde{nS_{1/2},m_J}\rangle$ the dressed eigenstate under the magnetic field that has largest overlap with $|nS_{1/2},m_J\rangle$. Since the magnetic field is parallel to the $z$ axis and preserves the space-inversion symmetry, the dressed state is the superposition of bare states with the same parity and the same $m_J$. For example, we obtain the dressed states of the Rb atom for $B=200~{\rm G}$ as follows:
\begin{align}
&|\widetilde{65S_{1/2},-1/2}\rangle\notag \\
&= -0.9999\cket{65S_{1/2},-1/2}+0.008\cket{63D_{5/2},-1/2}\notag \\
&+0.006\cket{63D_{3/2},-1/2}-0.0019\cket{66S_{1/2},-1/2}+\cdots,\label{eq:dressed_S_state_numerical_results}\\
&|\widetilde{65P_{3/2},1/2}\rangle\notag \\
&= -0.9798\cket{65P_{3/2},1/2}+0.1998\cket{65P_{1/2},1/2}\notag \\
&-0.002\cket{63F_{7/2},1/2}-0.001\cket{66P_{3/2},1/2}+\cdots.\label{eq:dressed_P_state_numerical_results}
\end{align}

In the main text, we focus on a pair of dressed states, denoted as $|\widetilde{nS_{1/2},m_J}\rangle$ and $|\widetilde{(n+1)S_{1/2},m_J}\rangle$. We assign these states as spin-1/2 basis states: $\cket{\uparrow}\equiv |\widetilde{(n+1)S_{1/2},m_J}\rangle$ and $\cket{\downarrow}\equiv |\widetilde{nS_{1/2},m_J}\rangle$. We define the target subspace spanned by the following four two-atom states: $\cket{\uparrow\uparrow}, \cket{\uparrow\downarrow}, \cket{\downarrow\uparrow}$, and $\cket{\downarrow\downarrow}$. Using the standard second-order perturbation theory, the effective Hamiltonian in this subspace is given by \cite{Bijnen_PhD_thesis,Whitlock2017,Kunimi2024,Wadenpfuhl2025unraveling}
\begin{align}
\hat{H}_{\rm eff}&=J_{\uparrow\uparrow}\cket{\uparrow\uparrow}\bra{\uparrow\uparrow}+J_{\downarrow\downarrow}\cket{\downarrow\downarrow}\bra{\downarrow\downarrow}\notag \\
&+J_{\uparrow\downarrow}(\cket{\uparrow\downarrow}\bra{\uparrow\downarrow}+\cket{\downarrow\uparrow}\bra{\downarrow\uparrow})\notag \\
&+\frac{J}{2}(\cket{\uparrow\downarrow}\bra{\downarrow\uparrow}+\cket{\downarrow\uparrow}\bra{\uparrow\downarrow}),\label{eq:expression_effective_hamiltonian}\\
J_{\alpha\beta}&\equiv -\sum_{\bm{n}}\frac{|\bra{\bm{n}}\hat{V}_{\rm dd}\cket{\alpha\beta}|^2}{\Delta E_{\rm F}(\bm{n},\alpha,\beta)}\equiv \frac{h C_6^{\alpha\beta}}{R^6},\;\alpha,\beta=\uparrow,\downarrow,\label{eq:definition_of_vdW_interaction}\\
\frac{J}{2}&\equiv -\sum_{\bm{n}}\frac{\bra{\uparrow\downarrow}\hat{V}_{\rm dd}\cket{\bm{n}}\bra{\bm{n}}\hat{V}_{\rm dd}\cket{\downarrow\uparrow}}{\Delta E_{\rm F}(\bm{n},\uparrow,\downarrow)}\equiv \frac{h C_6}{R^6},\label{eq:definition_of_vdW_exchange_interaction}
\end{align}
where $J_{\alpha\beta}$ is the strength of the van der Waals interaction when the initial and final states are in the same spin state, and $J$ characterizes the strength of the exchange interaction ($\uparrow\downarrow\leftrightarrow\downarrow\uparrow$). We also define $C_6^{\alpha\beta}$ and $C_6$ coefficients for these processes. The symbol  $\bm{n}\equiv (n_1, L_1, J_1, m_{J_1}, n_2,L_2, J_2, m_{J_2})$ denotes the quantum numbers of the intermediate pair states, and the F\"orster defect is defined by $\Delta E_{\rm F}(\bm{n},\alpha,\beta)\equiv E_{\bm{n}}-E_{\alpha\beta}\equiv E_{n_1,L_1,J_1,m_{J_1}}+E_{n_2,L_2,J_2,m_{J_2}}-E_{\alpha}-E_{\beta}$, where $E_{n_i,L_i,J_i,m_{J_i}}$ is the energy of the intermediate state and $E_{\alpha}$ is the energy of the $|\widetilde{nS_{1/2},m_J}\rangle$ or $|\widetilde{(n+1)S_{1/2},m_J}\rangle$ state. The spin-1/2 operators are defined as follows: $\hat{S}_j^{+}\equiv \cket{\uparrow_j}\bra{\downarrow_j}$, $\hat{S}_j^-\equiv (\hat{S}_j^+)^{\dagger}$, $\hat{S}_j^x\equiv (\hat{S}_j^++\hat{S}_j^-)/2$, $\hat{S}_j^y\equiv (\hat{S}_j^+-\hat{S}_j^-)/(2i)$, and $\hat{S}_j^z\equiv (\cket{\uparrow_j}\bra{\uparrow_j}-\cket{\downarrow_j}\bra{\downarrow_j})/2$. In terms of these spin operators, the effective Hamiltonian~(\ref{eq:expression_effective_hamiltonian}) takes the form of an {\it XXZ}-type Hamiltonian:
\begin{align}
\hat{H}_{\rm eff}&=\left(E_{\uparrow}-E_{\downarrow}+\frac{J_{\uparrow\uparrow}-J_{\downarrow\downarrow}}{2}\right)\left(\hat{S}_1^z+\hat{S}_2^z\right)\notag \\
&+J(\hat{S}_1^x\hat{S}_2^x+\hat{S}_1^y\hat{S}_2^y+\delta\hat{S}_1^z\hat{S}_2^z)+\frac{1}{4}(J_{\uparrow\uparrow}+J_{\downarrow\downarrow}+2J_{\uparrow\downarrow}),\label{eq:effective_Hamltonian_in_spin_language}\\
\delta&\equiv \frac{J_{\uparrow\uparrow}+J_{\downarrow\downarrow}-2J_{\uparrow\downarrow}}{J}=\frac{C_6^{\uparrow\uparrow}+C_6^{\downarrow\downarrow}-2C_6^{\uparrow\downarrow}}{2C_6}.\label{eq:definition_of_anisotropy_parameter}
\end{align}

Here, we discuss the validity of the perturbation theory. For the perturbative approach to be valid, the ratio between the matrix element and the F\"orster defect must be small. A previous work~\cite{Bijnen_PhD_thesis} proposed a characteristic radius, $R_{\rm c}$, defined as
\begin{align}
R_{\rm c}^3&\equiv \max_{\alpha,\beta}(R_{\rm c}^{\alpha,\beta})^3\equiv \max_{\bm{n},\alpha,\beta}\left|\frac{\bra{\bm{n}}\hat{V}_{\rm dd}\cket{\alpha\beta}}{\Delta E_{\rm F}(\bm{n},\alpha,\beta)}\right|R^3.\label{eq:definition_of_Rc}
\end{align} 
The condition for the validity of the perturbation theory is then given by $R^3 \gg R_{\rm c}^3$ \footnote{We can estimate the next-to-leading-order term, which is given by fourth-order perturbation due to the selection rule. The ratio between the fourth- and second-order terms scales as $(R_{\rm c}/R)^6$. In this Letter, we adopt $R=2R_{\rm c}~[(R_{\rm c}/R)^3=1/8]$. Then, the contribution of the fourth-order term is roughly $1\%$ of the second-order term.}.

We introduce artificial cutoff parameters to perform the numerical calculations, since the number of the Rydberg states is countably infinite. In the single-atom calculations, we consider the following Rydberg states, $|\widetilde{nL_J,m_J}\rangle$ with $[n-\Delta n,n+1+\Delta n]$ and $L=S, P, D,F$, and $G$. For the summations in the interaction strength defined in Eqs.~(\ref{eq:definition_of_vdW_interaction}) and (\ref{eq:definition_of_vdW_exchange_interaction}), we include only the intermediate states that satisfy the condition $-\Delta E+E_{\downarrow\downarrow}\le E_{\bm{n}}\le E_{\uparrow\uparrow}+\Delta E$, where $\Delta E\equiv E_{\uparrow\uparrow}-E_{\downarrow\downarrow}$ denotes the energy difference between the pair states $\cket{\uparrow\uparrow}$ and $\cket{\downarrow\downarrow}$. In the following, we present the results for $\Delta n=2$. We have confirmed that the accuracy of our calculations is approximately at the two-digit level for the anisotropy parameter $\delta$, when we choose $\Delta E=2(E_{\uparrow\uparrow}-E_{\downarrow\downarrow})$, $\Delta n=3$, and $L_{\rm max}=H$, respectively.

{\it Results.} In the main text, we mainly focus on the ${}^{87}$Rb atom and set $\theta = \pi/2$. See the Supplemental Material for results on other atomic species~\footnote{See Supplemental Material for details on the results for other atomic species, a summary of Heisenberg points, details of the atom positions, and the perturbation theory for spin-1 systems.}. In this case, the interaction has no $\varphi$ dependence, i.e., it is isotropic in the $xy$ plane. This is because $\hat{V}_2$ vanishes for $\theta=\pi/2$ and the first and second terms of $\hat{V}_3$ always appear simultaneously in the summation of Eqs.~(\ref{eq:definition_of_vdW_interaction}) and (\ref{eq:definition_of_vdW_exchange_interaction}) due to the selection rules. 

\begin{figure}[t]
\centering
\includegraphics[width=8.6cm,clip]{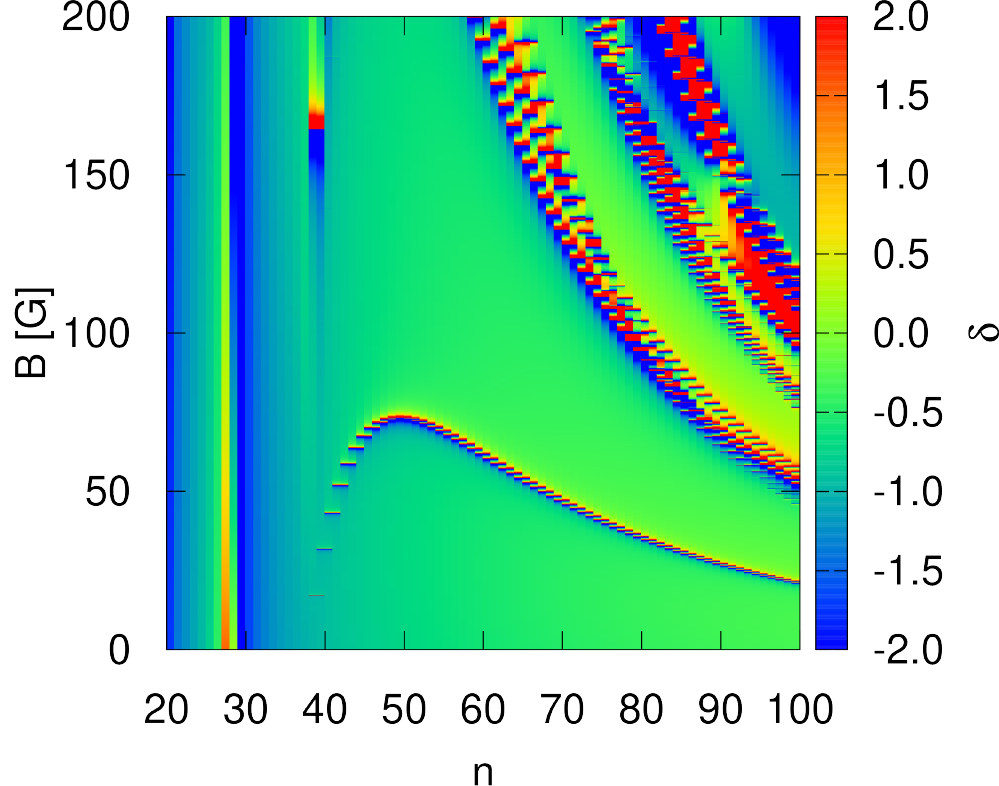}
\caption{Magnetic field and principal quantum number dependence of the anisotropy parameter of the pair $|nS_{1/2},m_J=-1/2\rangle$ and $|(n+1)S_{1/2},m_J=-1/2\rangle$ for $\theta=\pi/2$.
}
\label{fig:delta_b_and_n_dep}
\vspace{-0.75em}
\end{figure}

\begin{figure*}[t]
\centering
\includegraphics[width=13.5cm,clip]{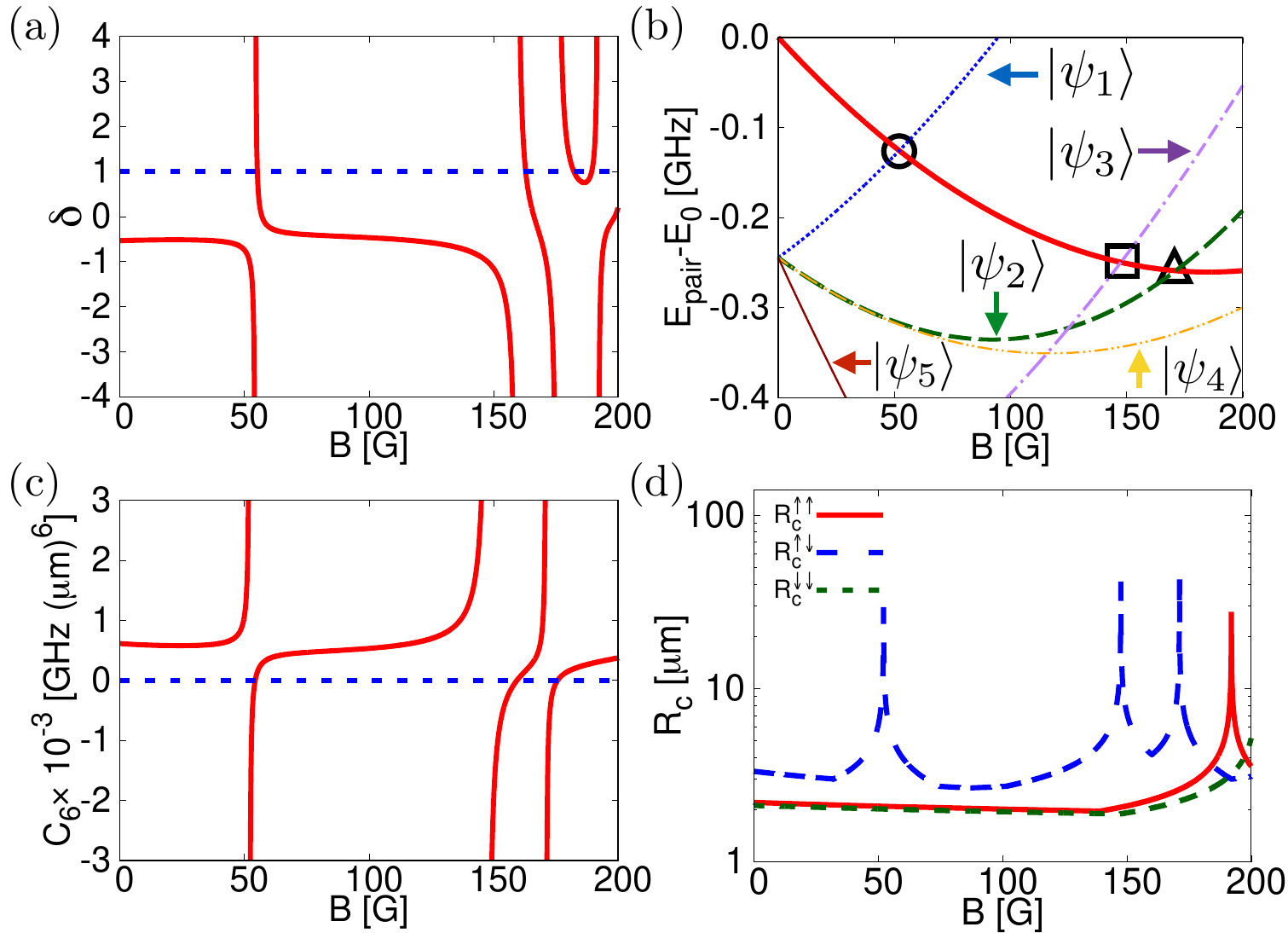}
\caption{Magnetic field dependence of various quantities of ${}^{87}$Rb atoms for $n = 65$ and $m_J = -1/2$. (a) $\delta$ vs $B$. The blue dotted line represents $\delta=1$. (b) Pair energy vs $B$. We plot the pair energy relative to $E_0 \simeq -h \times 1691.81~{\rm GHz}$, which is the pair energy of $\cket{65S_{1/2},-1/2}\cket{66S_{1/2},-1/2}$ at $B=0$. The red solid line represents $E_{\uparrow\downarrow}$. The blue dotted, green dashed, purple dash-dotted, orange dash-dot-dotted, and thin solid red lines represent the pair energies of $\cket{\psi_1}\equiv |\widetilde{65P_{3/2},1/2}\rangle |\widetilde{65P_{3/2},1/2}\rangle$, $\cket{\psi_2}\equiv |\widetilde{65P_{3/2},-1/2}\rangle|\widetilde{65P_{3/2},-1/2}\rangle$, $\cket{\psi_3}\equiv |\widetilde{65P_{3/2},1/2}\rangle|\widetilde{65P_{1/2},1/2}\rangle$, $\cket{\psi_4}\equiv|\widetilde{65P_{3/2},-3/2}\rangle|\widetilde{65P_{3/2},1/2}\rangle$, and $\cket{\psi_5}\equiv |\widetilde{65P_{3/2},-3/2}\rangle|\widetilde{65P_{3/2},-3/2}\rangle$, respectively. The black circles, squares, and triangles indicate the F\"orster resonance points. (c) $C_6$ vs $B$. The blue dotted line represents $C_6 = 0$. (d) $R_{\rm c}$ vs $B$.
}
\label{fig:delta_E_C6_Rc}
\vspace{-0.75em}
\end{figure*}

Figure~\ref{fig:delta_b_and_n_dep} shows the dependence of the anisotropy parameter $\delta$ on the principal quantum number and the magnetic field for ${}^{87}$Rb atoms, for the pair $\cket{65S_{1/2}, -1/2}\cket{66S_{1/2}, -1/2}$. We observe a resonancelike behavior in $\delta$. To understand this behavior, we show the magnetic-field dependence of $\delta$ for $n = 65$ in Fig.~\ref{fig:delta_E_C6_Rc}(a). The anisotropy parameter diverges at certain magnetic-field strengths. Although this divergence is closely related to the F\"{o}rster resonance, it is not at the F\"orster resonance point. To demonstrate this, we plot the energies of the pair of dressed states as a function of the magnetic field in Fig.~\ref{fig:delta_E_C6_Rc}(b). We observe that the divergence point of $\delta$ does not coincide with the F\"orster resonance point. For example, $\delta$ diverges at $B \simeq 54.4~{\rm G}$, whereas the first F\"orster resonance point [marked by the black circle shown in Fig.~\ref{fig:delta_E_C6_Rc}(b)] occurs at $B \simeq 52.1~{\rm G}$. This discrepancy arises from the definition of $\delta$: the divergence of $\delta$ corresponds to the condition $J = 0$ (i.e., $C_6 = 0$). Indeed, the zeros of $C_6$ coincide with the divergence points of $\delta$, as shown in Fig.~\ref{fig:delta_E_C6_Rc}(c). Although there are three F\"orster resonance points as shown in Fig.~\ref{fig:delta_E_C6_Rc}(b), we can see four-divergence behavior in $R_{\rm c}$ shown in Fig.~\ref{fig:delta_E_C6_Rc}(d). This is due to the F\"orster resonance of the other pair $|66S_{1/2},-1/2\rangle |66S_{1/2},-1/2\rangle$.

In any case, the anisotropy parameter $\delta$ exhibits a rapid variation in the vicinity of the F\"orster resonance point. Utilizing this result, we can find the Heisenberg point ($\delta=1$) near the F\"orster resonance point. As shown in Fig.~\ref{fig:delta_E_C6_Rc}(a), there are four Heisenberg points for $n=65$ and $m_J=-1/2$. For example, we obtain $B=182.3~{\rm G}$, $\delta\simeq 0.997$, $R_{\rm c}\simeq 3.8~\mu{\rm m}$, and $J_1\simeq h\times 1.92~{\rm MHz}$ at $R=2R_{\rm c}$, which are realistic experimental parameters in the current experimental situations. We also evaluate the derivative $d\delta/d B$, which represents the sensitivity of $\delta$ to fluctuations in the magnetic field. In this case, we obtain $d\delta/dB\simeq -1.3\times 10^{-4}~{\rm mG}^{-1}$. If we want to achieve $\delta$ with 1\% accuracy, the fluctuations of the magnetic field should be suppressed below 100~mG. This stability can be achieved in the current experimental techniques. References~\cite{Fuchs2008binding,Inada2008collisional,Chang2020collisional,Xie2025Feshbach} have reported a magnetic-field stability of $10~\mathrm{mG}$ \footnote{According to Refs.~\cite{merkel2019magnetic,borkowski2023active}, the ppm-level accuracy of the magnetic field has been achieved for fields in the range of $100$-$1000~{\rm G}$. This means that the fluctuation of the magnetic field is about $0.01~{\rm mG}$.}. Therefore, we can experimentally realize the condition $\delta\simeq 1$ with sufficient stability. In the Supplemental Material, we summarize the parameters of the Heisenberg point for other atomic species and pairs~\footnotemark[1].

\begin{figure}[t]
\centering
\includegraphics[width=8.6cm,clip]{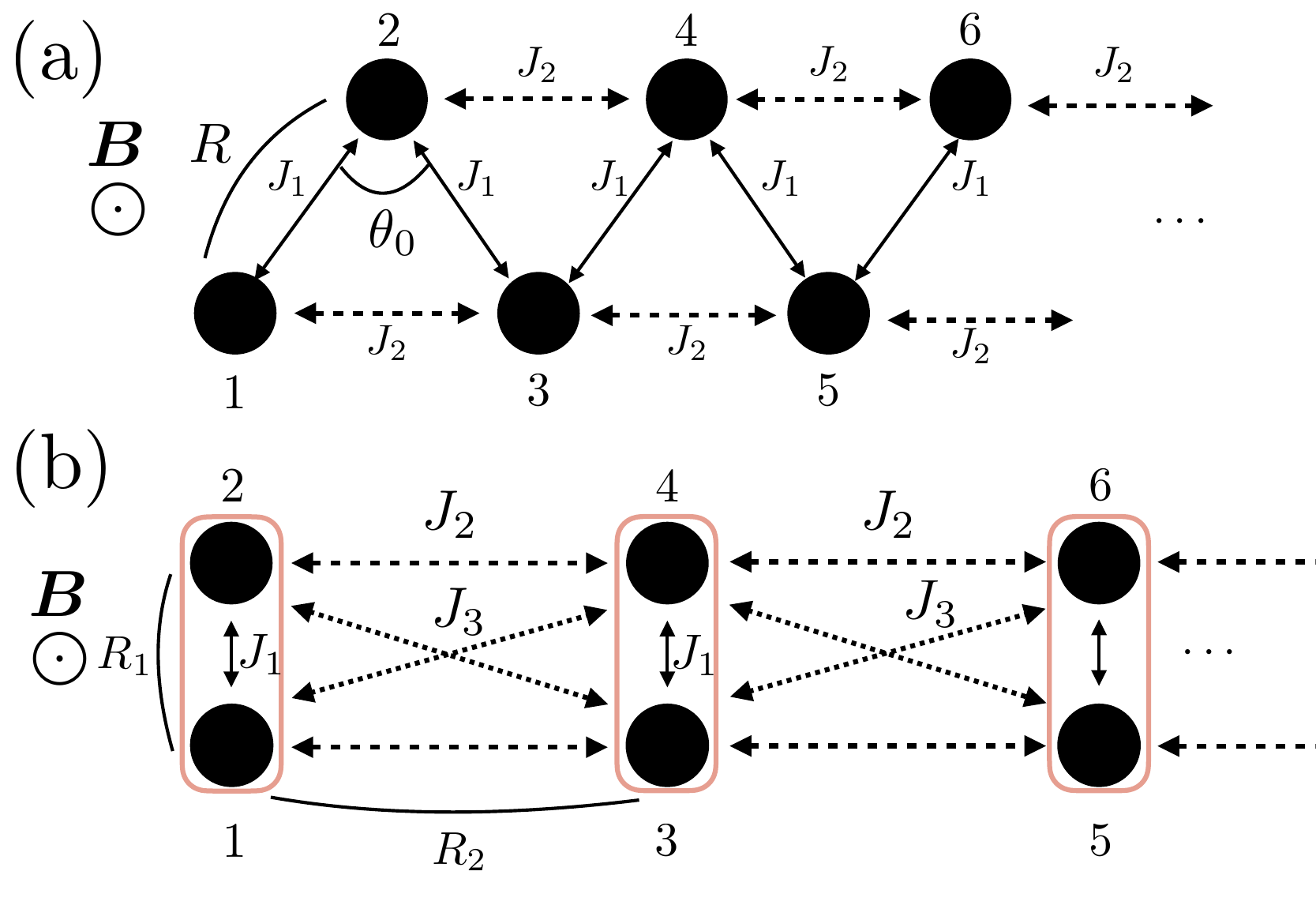}
\caption{(a) Atom configuration for the spin-1/2 $J_1$-$J_2$ model. Black circles represent the positions of Rydberg atoms. Here, $\theta_0$ denotes the angle between the vertices of the same length. (b) Atom configuration for the spin-1 Heisenberg model. Two atoms enclosed by the red solid line represent an effective spin-1 degree of freedom.
}
\label{fig:position_of_Rydberg_atoms_ladder}
\vspace{-0.75em}
\end{figure}

As an application of the above results, we propose an experimental realization of a tunable spin-1/2 $J_1$-$J_2$ Heisenberg model using Rydberg atoms. In the following, we fix the anisotropy parameter at the Heisenberg point, $\delta =1$, and neglect the nonuniformity arising from the term proportional to $\hat{S}_1^z+\hat{S}_2^z$ in Eq.~(\ref{eq:effective_Hamltonian_in_spin_language}) because the nonuniformity appears only at the edges of the system \footnote{If one desires a more accurate Hamiltonian, the nonuniformity can be compensated by applying an additional ac Stark shift at the edges of the system.}. After neglecting it, this term reduces to a uniform-magnetic-field term, which can be removed by a unitary transformation. We consider a zigzag ladder configuration of atoms, as shown in Fig.~\ref{fig:position_of_Rydberg_atoms_ladder}(a). The atoms are arranged in the $xy$ plane, and a uniform magnetic field is applied along the positive $z$ axis. The distance between the nearest-neighbor (NN) atoms is denoted by $R$. The Hamiltonian is given by
\begin{align}
\hat{H}_{J_1-J_2}&=J_1\sum_{j=1}^{M-1}\hat{\bm{S}}_j\cdot\hat{\bm{S}}_{j+1}+J_2\sum_{j=1}^{M-2}\hat{\bm{S}}_j\cdot\hat{\bm{S}}_{j+2}+\cdots
\end{align}
where the NN interaction $J_1$ is given by $J_1=2h C_6/R^6$ and $M$ is the number of lattice sites. The second-, third-, and fourth-neighbor interaction strengths $J_2$, $J_3$, and $J_4$ can be written as
\begin{align}
J_{2}&=\frac{1}{64\sin^6(\theta_0/2)}J_1,\label{eq:expression_of_J2}\\
J_3&=\frac{1}{\left[9\sin^2\left(\theta_0/2\right)+\cos^2\left(\theta_0/2\right)\right]^3}J_1,\label{eq:expression_of_J3}\\
J_4&=\frac{1}{4096\sin^6(\theta_0/2)}J_1,\label{eq:expression_of_J4}
\end{align}
where the angle $\theta_0$ is defined in Fig.~\ref{fig:position_of_Rydberg_atoms_ladder}(a). Here, the next-NN (NNN) interaction can be tuned by changing the angle $\theta_0$. For $0 \le \theta_0 \le \pi/3$, the range of $J_2$ is given by $J_1/64 \le J_2 \le J_1$. The linear chain case ($\theta_0=\pi$) corresponds to the minimum $J_2$ and the equilateral triangle case ($\theta_0=\pi/3$) corresponds to the maximum $J_2$.

By tuning the angle such that $\sin(\theta_0/2) = 2^{-5/6}\;(\theta_0\simeq 68.3^{\circ})$, we obtain the Majumdar-Ghosh model ($J_2 = J_1/2$)~\cite{majumdar1969next,majumdar1969next2}, with small longer-range interactions. In this case, the third- and fourth-neighbor interactions are given by $J_3 / J_1 = (1 + 2^{4/3})^{-3} \simeq 0.02$ and $J_4 / J_1 = 1/128 \simeq 0.008$, respectively. See details in the Supplemental Material \footnotemark[1].

As another application, we propose an experimental realization of a spin-1 Heisenberg model. In the previous work, the spin-1 model can be experimentally realized in the Rydberg systems using three different Rydberg states~\cite{Qiao2025} (see also the theoretical proposal in Ref.~\cite{mogerle2025spin1}). In these works, they encode three different Rydberg states to spin-1 systems. For example, in Ref.~\cite{Qiao2025}, the following encoding is used: $\cket{+}=\cket{61S_{1/2},m_J=1/2}, \cket{0}=\cket{60P_{3/2},m_J=-1/2}$, and $\cket{-}=\cket{60S_{1/2},m_J=1/2}$. Our strategy is different from this work. We construct the spin-1 degrees of freedom from two spin-1/2 degrees of freedom. To do this, we consider a two-leg ladder configuration, as shown in Fig.~\ref{fig:position_of_Rydberg_atoms_ladder}(b). This configuration is known as the Gelfand ladder~\cite{gelfand1991linked,honecker2000magnetization,chandra2006exact,kohshiro2021multiple}. Here, we tune the inter- and intra-leg NN distances, denoted by $R_1$ and $R_2$, respectively. The Hamiltonian is given by
\begin{align}
\hat{H}_{\rm Gelfand}&=J_1\sum_{j=1}^{M}\hat{\bm{S}}_{2j-1}\cdot\hat{\bm{S}}_{2j}\notag \\
&+J_2\sum_{j=1}^{M-1}\left(\hat{\bm{S}}_{2j-1}\cdot\hat{\bm{S}}_{2j+1}+\hat{\bm{S}}_{2j}\cdot\hat{\bm{S}}_{2j+2}\right)\notag \\
&+J_3\sum_{j=1}^{M-1}\left(\hat{\bm{S}}_{2j-1}\cdot\hat{\bm{S}}_{2j+2}+\hat{\bm{S}}_{2j}\cdot\hat{\bm{S}}_{2j+1}\right)+\cdots,\label{eq:definition_of_Gelfand_Ladder_Hamiltonian}
\end{align}
where $M$ is the number of the atoms in each ladder, and the indices of the atoms are defined in Fig~\ref{fig:position_of_Rydberg_atoms_ladder}(b). The interaction strengths can be written as $J_1=2h C_6/R^6_1$, $J_2=2h C_6/R_2^6$, and $J_3=2h C_6/(R_1^2+R_2^2)^3$.

Here, we assume the condition $|J_1| \gg |J_2|, |J_3|$. In this case, the energy difference between the triplet states and the singlet state of the $(2j - 1)$th and $2j$th atoms is large. Therefore, we can apply perturbation theory~\cite{hida1992crossover,hung2005numerical} and obtain the spin-1 Heisenberg chain as an effective Hamiltonian:
\begin{align}
\hat{H}_{S=1} &= J_1^{S=1} \sum_{j=1}^{M-1} \hat{\bm{\tau}}_j \cdot \hat{\bm{\tau}}_{j+1}, \label{eq:S=1_Heisenberg_model}
\end{align}
where $J_1^{S=1} \equiv (J_2 + J_3)/2$, and $\hat{\tau}_j^\mu$ ($\mu = x, y, z$) denotes the spin-1 operator at site $j$. Here, the $j$th site of the spin-1 system means the triplet formed by $(2j-1)$th and $2j$th atoms [see the red solid lines in Fig.~\ref{fig:position_of_Rydberg_atoms_ladder}(b)]. A detailed calculation of the perturbation theory, a discussion of the second-order perturbation, and the effects of the NNN interactions are provided in the Supplemental Material~\footnotemark[1].
 
Finally, we discuss the validity of the perturbation theory in the spin-1 model. Since our derivation of the spin-1 Hamiltonian relies on perturbation theory, the NN interaction in the spin-1 system is much smaller than that in the spin-1/2 case. In fact, $J_1^{S=1}$ can be expressed as a function of $J_2/J_1 = (R_1/R_2)^6$:
\begin{align}
\frac{J_1^{S=1}}{J_1} = \frac{1}{2} \frac{J_2}{J_1} + \frac{1}{2} \frac{J_2/J_1}{\left[1 + (J_2/J_1)^{1/3}\right]^3}. \label{eq:expression_J1_S=1}
\end{align}
For example, when $J_2/J_1 = 0.2$ (corresponding to $R_2 \simeq 1.31 R_1$), the NN interaction in the spin-1 system becomes $J_1^{S=1}/J_1 \simeq 0.13$. We adopted an interaction strength of $500~\mathrm{kHz}$ as a criterion for safely conducting the experiment in order to avoid decoherence. Because the typical decoherence time of a Rydberg-atom platform is on the order of a few microseconds, the characteristic timescale of $500~\mathrm{kHz}$ is about ten times shorter than the decoherence time. When $J_2/J_1 = 0.2$, $J_1 \gtrsim h \times 3.8~{\rm MHz}$ is necessary for satisfying the above condition. In addition to this condition, the derivation of the spin-1/2 {\it XXZ} model requires that $R_1^3 \gg R_{\rm c}^3$. Since these conditions are mutually competing, the choice of parameters, such as the atomic species, the pair states, and the magnetic field, must be made with care. We find some suitable parameters for realizing the spin-1 Heisenberg model, such as the ${}^{23}$Na atom for $n=75, m_J=-1/2, B=78.25~{\rm G}, \delta\simeq 1.00, R_{\rm c}=5.2~\mu{\rm m}$, $J_1=h\times 4.02~{\rm MHz}$ at $R_1=2R_{\rm c}$, and $d\delta/dB\simeq 5.5\times 10^{-7}~{\rm mG}^{-1}$. See also the list of the Heisenberg points shown in the Supplemental Material~\footnotemark[1].

{\it Summary.} In this Letter, we investigated the magnetic-field dependence of the interaction strength between the Rydberg states $|nS_{1/2},m_J\rangle$ and $|(n+1)S_{1/2},m_J\rangle$. In this setting, the effective spin Hamiltonian takes the form of a spin-1/2 {\it XXZ} model. We found that the anisotropy parameter $\delta$ changes drastically near the F\"orster resonance point. Exploiting this behavior, we proposed experimental realizations of the $J_1$-$J_2$ Heisenberg model and the spin-1 Heisenberg model.

Our work opens a route toward realizing Heisenberg-type quantum spin models for spin-1/2 and spin-1 systems on the Rydberg platform. The Heisenberg model exhibits a variety of nontrivial quantum many-body phenomena, such as the Haldane phase~\cite{Haldane1983nonlinear,haldane1983continuum,affleck1987rigorous,affleck1988valence}, spin transport~\cite{sekino2024thermomagnetic,sekino2025thermomagnetic,fujimoto2024quantum}, and the Kardar-Parisi-Zhang universality class~\cite{Ljubotina2017,Ljubotina2019,takeuchi2025partial}. Our results provide a basis for experimental exploration of such quantum many-body phenomena using Rydberg atom platforms. Although we focus on one-dimensional spin chains in this Letter, our approach can be readily extended to two-dimensional systems. Another direction is to extend our calculations to alkaline-earth-like atoms, such as Sr and Yb~\cite{vaillant2014multichannel,robertson2021arc,hummel2024engineering}, circular Rydberg states~\cite{ravon2023array,Holzl2024,mehaignerie2025interacting,nguyen2018towards,dobrzyniecki2025tunable}, and dual-species or -isotope systems~\cite{sheng2022defect,anand2024dual,nakamura2024hybrid,wei2024dual}.

Although we focus on the Heisenberg point $(\delta=1)$ in this Letter, several interesting cases arise for $\delta\not=1$. For example, the ground state of the {\it XXZ} zigzag chain with $\delta=-1/2$ can be obtained analytically due to the frustration-free nature of the Hamiltonian~\cite{gerhardt1998metamagnetism,batista2009canted}. Another notable case is the {\it XXZ} model with additional edge-magnetic-field terms for $J<0$ and $\delta=-1$~\cite{sable2025toward}, which is related to supersymmetry.

{\it Acknowledgments.}
The authors thank H. Katsura, I. Danshita, S. de L\'es\'eleuc, K. Takasan, K. Fujimoto, H. Tamura, and S. Nakajima for their helpful discussions. This work was supported by JSPS KAKENHI under Grant No.~JP25K00215 (M.K.), JST ASPIRE under Grant No.~JPMJAP24C2 (M.K.), MEXT Quantum Leap Flagship Program (MEXT Q-LEAP) under Grant No. JPMXS0118069021 (T.T.), and JST Moonshot R\&D Program under Grant No. JPMJMS2269 (T.T.).

{\it Data availability.}
The data that support the findings of this article are openly available \cite{kunimi_2025_16349138}.


\newpage
\bibliography{references}

\begin{thebibliography}{125}%
\makeatletter
\providecommand \@ifxundefined [1]{%
 \@ifx{#1\undefined}
}%
\providecommand \@ifnum [1]{%
 \ifnum #1\expandafter \@firstoftwo
 \else \expandafter \@secondoftwo
 \fi
}%
\providecommand \@ifx [1]{%
 \ifx #1\expandafter \@firstoftwo
 \else \expandafter \@secondoftwo
 \fi
}%
\providecommand \natexlab [1]{#1}%
\providecommand \enquote  [1]{``#1''}%
\providecommand \bibnamefont  [1]{#1}%
\providecommand \bibfnamefont [1]{#1}%
\providecommand \citenamefont [1]{#1}%
\providecommand \href@noop [0]{\@secondoftwo}%
\providecommand \href [0]{\begingroup \@sanitize@url \@href}%
\providecommand \@href[1]{\@@startlink{#1}\@@href}%
\providecommand \@@href[1]{\endgroup#1\@@endlink}%
\providecommand \@sanitize@url [0]{\catcode `\\12\catcode `\$12\catcode
  `\&12\catcode `\#12\catcode `\^12\catcode `\_12\catcode `\%12\relax}%
\providecommand \@@startlink[1]{}%
\providecommand \@@endlink[0]{}%
\providecommand \url  [0]{\begingroup\@sanitize@url \@url }%
\providecommand \@url [1]{\endgroup\@href {#1}{\urlprefix }}%
\providecommand \urlprefix  [0]{URL }%
\providecommand \Eprint [0]{\href }%
\providecommand \doibase [0]{https://doi.org/}%
\providecommand \selectlanguage [0]{\@gobble}%
\providecommand \bibinfo  [0]{\@secondoftwo}%
\providecommand \bibfield  [0]{\@secondoftwo}%
\providecommand \translation [1]{[#1]}%
\providecommand \BibitemOpen [0]{}%
\providecommand \bibitemStop [0]{}%
\providecommand \bibitemNoStop [0]{.\EOS\space}%
\providecommand \EOS [0]{\spacefactor3000\relax}%
\providecommand \BibitemShut  [1]{\csname bibitem#1\endcsname}%
\let\auto@bib@innerbib\@empty
\bibitem [{\citenamefont {Auerbach}(1998)}]{Auerbach1998interacting}%
  \BibitemOpen
  \bibfield  {author} {\bibinfo {author} {\bibfnamefont {A.}~\bibnamefont
  {Auerbach}},\ }\href@noop {} {\emph {\bibinfo {title} {Interacting Electrons
  and Quantum Magnetism}}}\ (\bibinfo  {publisher} {Springer},\ \bibinfo
  {address} {Berlin},\ \bibinfo {year} {1998})\BibitemShut {NoStop}%
\bibitem [{\citenamefont {Lee}\ \emph {et~al.}(2006)\citenamefont {Lee},
  \citenamefont {Nagaosa},\ and\ \citenamefont {Wen}}]{Lee2006doping}%
  \BibitemOpen
  \bibfield  {author} {\bibinfo {author} {\bibfnamefont {P.~A.}\ \bibnamefont
  {Lee}}, \bibinfo {author} {\bibfnamefont {N.}~\bibnamefont {Nagaosa}},\ and\
  \bibinfo {author} {\bibfnamefont {X.-G.}\ \bibnamefont {Wen}},\ }\bibfield
  {title} {\bibinfo {title} {Doping a {Mott} insulator: {Physics} of
  high-temperature superconductivity},\ }\href@noop {} {\bibfield  {journal}
  {\bibinfo  {journal} {Rev. Mod. Phys.}\ }\textbf {\bibinfo {volume} {78}},\
  \bibinfo {pages} {17} (\bibinfo {year} {2006})}\BibitemShut {NoStop}%
\bibitem [{\citenamefont {Balents}(2010)}]{Balents2010}%
  \BibitemOpen
  \bibfield  {author} {\bibinfo {author} {\bibfnamefont {L.}~\bibnamefont
  {Balents}},\ }\bibfield  {title} {\bibinfo {title} {Spin liquids in
  frustrated magnets},\ }\href@noop {} {\bibfield  {journal} {\bibinfo
  {journal} {Nature}\ }\textbf {\bibinfo {volume} {464}},\ \bibinfo {pages}
  {199} (\bibinfo {year} {2010})}\BibitemShut {NoStop}%
\bibitem [{\citenamefont {Takahashi}(1999)}]{Takahashi1999thermodynamics}%
  \BibitemOpen
  \bibfield  {author} {\bibinfo {author} {\bibfnamefont {M.}~\bibnamefont
  {Takahashi}},\ }\href@noop {} {\emph {\bibinfo {title} {Thermodynamics of
  One-Dimensional Solvable Models}}}\ (\bibinfo  {publisher} {Cambridge
  University},\ \bibinfo {address} {Cambridge, England},\ \bibinfo {year}
  {1999})\BibitemShut {NoStop}%
\bibitem [{\citenamefont {Sutherland}(2004)}]{Sutherland2004beautiful}%
  \BibitemOpen
  \bibfield  {author} {\bibinfo {author} {\bibfnamefont {B.}~\bibnamefont
  {Sutherland}},\ }\href@noop {} {\emph {\bibinfo {title} {Beautiful Models: 70
  Years of Exactly Solved Quantum Many-Body Problems}}}\ (\bibinfo  {publisher}
  {World Scientific},\ \bibinfo {address} {Singapore},\ \bibinfo {year}
  {2004})\BibitemShut {NoStop}%
\bibitem [{\citenamefont {Tasaki}(2020)}]{tasaki2020physics}%
  \BibitemOpen
  \bibfield  {author} {\bibinfo {author} {\bibfnamefont {H.}~\bibnamefont
  {Tasaki}},\ }\href@noop {} {\emph {\bibinfo {title} {Physics and Mathematics
  of Quantum Many-Body Systems}}}\ (\bibinfo  {publisher} {Springer},\ \bibinfo
  {address} {Berlin},\ \bibinfo {year} {2020})\BibitemShut {NoStop}%
\bibitem [{\citenamefont {Eisert}\ \emph {et~al.}(2010)\citenamefont {Eisert},
  \citenamefont {Cramer},\ and\ \citenamefont {Plenio}}]{Eisert2010}%
  \BibitemOpen
  \bibfield  {author} {\bibinfo {author} {\bibfnamefont {J.}~\bibnamefont
  {Eisert}}, \bibinfo {author} {\bibfnamefont {M.}~\bibnamefont {Cramer}},\
  and\ \bibinfo {author} {\bibfnamefont {M.~B.}\ \bibnamefont {Plenio}},\
  }\bibfield  {title} {\bibinfo {title} {Colloquium: {Area} laws for the
  entanglement entropy},\ }\href@noop {} {\bibfield  {journal} {\bibinfo
  {journal} {Rev. Mod. Phys.}\ }\textbf {\bibinfo {volume} {82}},\ \bibinfo
  {pages} {277} (\bibinfo {year} {2010})}\BibitemShut {NoStop}%
\bibitem [{\citenamefont {Bethe}(1931)}]{Bethe1931}%
  \BibitemOpen
  \bibfield  {author} {\bibinfo {author} {\bibfnamefont {H.}~\bibnamefont
  {Bethe}},\ }\bibfield  {title} {\bibinfo {title} {Zur {Theorie} der
  {Metalle}},\ }\href@noop {} {\bibfield  {journal} {\bibinfo  {journal} {Z.
  Phys.}\ }\textbf {\bibinfo {volume} {71}},\ \bibinfo {pages} {205} (\bibinfo
  {year} {1931})}\BibitemShut {NoStop}%
\bibitem [{\citenamefont {Mazurenko}\ \emph {et~al.}(2017)\citenamefont
  {Mazurenko}, \citenamefont {Chiu}, \citenamefont {Ji}, \citenamefont
  {Parsons}, \citenamefont {Kan{\'a}sz-Nagy}, \citenamefont {Schmidt},
  \citenamefont {Grusdt}, \citenamefont {Demler}, \citenamefont {Greif},\ and\
  \citenamefont {Greiner}}]{Mazurenko2017}%
  \BibitemOpen
  \bibfield  {author} {\bibinfo {author} {\bibfnamefont {A.}~\bibnamefont
  {Mazurenko}}, \bibinfo {author} {\bibfnamefont {C.~S.}\ \bibnamefont {Chiu}},
  \bibinfo {author} {\bibfnamefont {G.}~\bibnamefont {Ji}}, \bibinfo {author}
  {\bibfnamefont {M.~F.}\ \bibnamefont {Parsons}}, \bibinfo {author}
  {\bibfnamefont {M.}~\bibnamefont {Kan{\'a}sz-Nagy}}, \bibinfo {author}
  {\bibfnamefont {R.}~\bibnamefont {Schmidt}}, \bibinfo {author} {\bibfnamefont
  {F.}~\bibnamefont {Grusdt}}, \bibinfo {author} {\bibfnamefont
  {E.}~\bibnamefont {Demler}}, \bibinfo {author} {\bibfnamefont
  {D.}~\bibnamefont {Greif}},\ and\ \bibinfo {author} {\bibfnamefont
  {M.}~\bibnamefont {Greiner}},\ }\bibfield  {title} {\bibinfo {title} {A
  cold-atom {Fermi}--{Hubbard} antiferromagnet},\ }\href@noop {} {\bibfield
  {journal} {\bibinfo  {journal} {Nature}\ }\textbf {\bibinfo {volume} {545}},\
  \bibinfo {pages} {462} (\bibinfo {year} {2017})}\BibitemShut {NoStop}%
\bibitem [{\citenamefont {Brown}\ \emph {et~al.}(2017)\citenamefont {Brown},
  \citenamefont {Mitra}, \citenamefont {Guardado-Sanchez}, \citenamefont
  {Schau{\ss}}, \citenamefont {Kondov}, \citenamefont {Khatami}, \citenamefont
  {Paiva}, \citenamefont {Trivedi}, \citenamefont {Huse},\ and\ \citenamefont
  {Bakr}}]{Brown2017}%
  \BibitemOpen
  \bibfield  {author} {\bibinfo {author} {\bibfnamefont {P.~T.}\ \bibnamefont
  {Brown}}, \bibinfo {author} {\bibfnamefont {D.}~\bibnamefont {Mitra}},
  \bibinfo {author} {\bibfnamefont {E.}~\bibnamefont {Guardado-Sanchez}},
  \bibinfo {author} {\bibfnamefont {P.}~\bibnamefont {Schau{\ss}}}, \bibinfo
  {author} {\bibfnamefont {S.~S.}\ \bibnamefont {Kondov}}, \bibinfo {author}
  {\bibfnamefont {E.}~\bibnamefont {Khatami}}, \bibinfo {author} {\bibfnamefont
  {T.}~\bibnamefont {Paiva}}, \bibinfo {author} {\bibfnamefont
  {N.}~\bibnamefont {Trivedi}}, \bibinfo {author} {\bibfnamefont {D.~A.}\
  \bibnamefont {Huse}},\ and\ \bibinfo {author} {\bibfnamefont {W.~S.}\
  \bibnamefont {Bakr}},\ }\bibfield  {title} {\bibinfo {title} {Spin-imbalance
  in a {2D} {Fermi-Hubbard} system},\ }\href@noop {} {\bibfield  {journal}
  {\bibinfo  {journal} {Science}\ }\textbf {\bibinfo {volume} {357}},\ \bibinfo
  {pages} {1385} (\bibinfo {year} {2017})}\BibitemShut {NoStop}%
\bibitem [{\citenamefont {Xu}\ \emph {et~al.}(2023)\citenamefont {Xu},
  \citenamefont {Kendrick}, \citenamefont {Kale}, \citenamefont {Gang},
  \citenamefont {Ji}, \citenamefont {Scalettar}, \citenamefont {Lebrat},\ and\
  \citenamefont {Greiner}}]{Xu2023}%
  \BibitemOpen
  \bibfield  {author} {\bibinfo {author} {\bibfnamefont {M.}~\bibnamefont
  {Xu}}, \bibinfo {author} {\bibfnamefont {L.~H.}\ \bibnamefont {Kendrick}},
  \bibinfo {author} {\bibfnamefont {A.}~\bibnamefont {Kale}}, \bibinfo {author}
  {\bibfnamefont {Y.}~\bibnamefont {Gang}}, \bibinfo {author} {\bibfnamefont
  {G.}~\bibnamefont {Ji}}, \bibinfo {author} {\bibfnamefont {R.~T.}\
  \bibnamefont {Scalettar}}, \bibinfo {author} {\bibfnamefont {M.}~\bibnamefont
  {Lebrat}},\ and\ \bibinfo {author} {\bibfnamefont {M.}~\bibnamefont
  {Greiner}},\ }\bibfield  {title} {\bibinfo {title} {Frustration-and
  doping-induced magnetism in a {Fermi--Hubbard} simulator},\ }\href@noop {}
  {\bibfield  {journal} {\bibinfo  {journal} {Nature}\ }\textbf {\bibinfo
  {volume} {620}},\ \bibinfo {pages} {971} (\bibinfo {year}
  {2023})}\BibitemShut {NoStop}%
\bibitem [{\citenamefont {Shao}\ \emph {et~al.}(2024)\citenamefont {Shao},
  \citenamefont {Wang}, \citenamefont {Zhu}, \citenamefont {Zhu}, \citenamefont
  {Sun}, \citenamefont {Chen}, \citenamefont {Zhang}, \citenamefont {Fan},
  \citenamefont {Deng}, \citenamefont {Yao}, \citenamefont {Chen},\ and\
  \citenamefont {Pan}}]{Shao2024}%
  \BibitemOpen
  \bibfield  {author} {\bibinfo {author} {\bibfnamefont {H.-J.}\ \bibnamefont
  {Shao}}, \bibinfo {author} {\bibfnamefont {Y.-X.}\ \bibnamefont {Wang}},
  \bibinfo {author} {\bibfnamefont {D.-Z.}\ \bibnamefont {Zhu}}, \bibinfo
  {author} {\bibfnamefont {Y.-S.}\ \bibnamefont {Zhu}}, \bibinfo {author}
  {\bibfnamefont {H.-N.}\ \bibnamefont {Sun}}, \bibinfo {author} {\bibfnamefont
  {S.-Y.}\ \bibnamefont {Chen}}, \bibinfo {author} {\bibfnamefont
  {C.}~\bibnamefont {Zhang}}, \bibinfo {author} {\bibfnamefont {Z.-J.}\
  \bibnamefont {Fan}}, \bibinfo {author} {\bibfnamefont {Y.}~\bibnamefont
  {Deng}}, \bibinfo {author} {\bibfnamefont {X.-C.}\ \bibnamefont {Yao}},
  \bibinfo {author} {\bibfnamefont {Y.-A.}\ \bibnamefont {Chen}},\ and\
  \bibinfo {author} {\bibfnamefont {J.-W.}\ \bibnamefont {Pan}},\ }\bibfield
  {title} {\bibinfo {title} {Antiferromagnetic phase transition in a {3D}
  fermionic {Hubbard} model},\ }\href@noop {} {\bibfield  {journal} {\bibinfo
  {journal} {Nature}\ }\textbf {\bibinfo {volume} {632}},\ \bibinfo {pages}
  {267} (\bibinfo {year} {2024})}\BibitemShut {NoStop}%
\bibitem [{\citenamefont {Fukuhara}\ \emph
  {et~al.}(2013{\natexlab{a}})\citenamefont {Fukuhara}, \citenamefont
  {Kantian}, \citenamefont {Endres}, \citenamefont {Cheneau}, \citenamefont
  {Schau{\ss}}, \citenamefont {Hild}, \citenamefont {Bellem}, \citenamefont
  {Schollw{\"o}ck}, \citenamefont {Giamarchi}, \citenamefont {Gross},
  \citenamefont {Bloch},\ and\ \citenamefont {Kuhr}}]{Fukuhara2013}%
  \BibitemOpen
  \bibfield  {author} {\bibinfo {author} {\bibfnamefont {T.}~\bibnamefont
  {Fukuhara}}, \bibinfo {author} {\bibfnamefont {A.}~\bibnamefont {Kantian}},
  \bibinfo {author} {\bibfnamefont {M.}~\bibnamefont {Endres}}, \bibinfo
  {author} {\bibfnamefont {M.}~\bibnamefont {Cheneau}}, \bibinfo {author}
  {\bibfnamefont {P.}~\bibnamefont {Schau{\ss}}}, \bibinfo {author}
  {\bibfnamefont {S.}~\bibnamefont {Hild}}, \bibinfo {author} {\bibfnamefont
  {D.}~\bibnamefont {Bellem}}, \bibinfo {author} {\bibfnamefont
  {U.}~\bibnamefont {Schollw{\"o}ck}}, \bibinfo {author} {\bibfnamefont
  {T.}~\bibnamefont {Giamarchi}}, \bibinfo {author} {\bibfnamefont
  {C.}~\bibnamefont {Gross}}, \bibinfo {author} {\bibfnamefont
  {I.}~\bibnamefont {Bloch}},\ and\ \bibinfo {author} {\bibfnamefont
  {S.}~\bibnamefont {Kuhr}},\ }\bibfield  {title} {\bibinfo {title} {Quantum
  dynamics of a mobile spin impurity},\ }\href@noop {} {\bibfield  {journal}
  {\bibinfo  {journal} {Nat. Phys.}\ }\textbf {\bibinfo {volume} {9}},\
  \bibinfo {pages} {235} (\bibinfo {year} {2013}{\natexlab{a}})}\BibitemShut
  {NoStop}%
\bibitem [{\citenamefont {Fukuhara}\ \emph
  {et~al.}(2013{\natexlab{b}})\citenamefont {Fukuhara}, \citenamefont
  {Schau{\ss}}, \citenamefont {Endres}, \citenamefont {Hild}, \citenamefont
  {Cheneau}, \citenamefont {Bloch},\ and\ \citenamefont
  {Gross}}]{Fukuhara2013microscopic}%
  \BibitemOpen
  \bibfield  {author} {\bibinfo {author} {\bibfnamefont {T.}~\bibnamefont
  {Fukuhara}}, \bibinfo {author} {\bibfnamefont {P.}~\bibnamefont
  {Schau{\ss}}}, \bibinfo {author} {\bibfnamefont {M.}~\bibnamefont {Endres}},
  \bibinfo {author} {\bibfnamefont {S.}~\bibnamefont {Hild}}, \bibinfo {author}
  {\bibfnamefont {M.}~\bibnamefont {Cheneau}}, \bibinfo {author} {\bibfnamefont
  {I.}~\bibnamefont {Bloch}},\ and\ \bibinfo {author} {\bibfnamefont
  {C.}~\bibnamefont {Gross}},\ }\bibfield  {title} {\bibinfo {title}
  {Microscopic observation of magnon bound states and their dynamics},\
  }\href@noop {} {\bibfield  {journal} {\bibinfo  {journal} {Nature}\ }\textbf
  {\bibinfo {volume} {502}},\ \bibinfo {pages} {76} (\bibinfo {year}
  {2013}{\natexlab{b}})}\BibitemShut {NoStop}%
\bibitem [{\citenamefont {Hild}\ \emph {et~al.}(2014)\citenamefont {Hild},
  \citenamefont {Fukuhara}, \citenamefont {Schau{\ss}}, \citenamefont {Zeiher},
  \citenamefont {Knap}, \citenamefont {Demler}, \citenamefont {Bloch},\ and\
  \citenamefont {Gross}}]{Hild2014}%
  \BibitemOpen
  \bibfield  {author} {\bibinfo {author} {\bibfnamefont {S.}~\bibnamefont
  {Hild}}, \bibinfo {author} {\bibfnamefont {T.}~\bibnamefont {Fukuhara}},
  \bibinfo {author} {\bibfnamefont {P.}~\bibnamefont {Schau{\ss}}}, \bibinfo
  {author} {\bibfnamefont {J.}~\bibnamefont {Zeiher}}, \bibinfo {author}
  {\bibfnamefont {M.}~\bibnamefont {Knap}}, \bibinfo {author} {\bibfnamefont
  {E.}~\bibnamefont {Demler}}, \bibinfo {author} {\bibfnamefont
  {I.}~\bibnamefont {Bloch}},\ and\ \bibinfo {author} {\bibfnamefont
  {C.}~\bibnamefont {Gross}},\ }\bibfield  {title} {\bibinfo {title}
  {Far-from-equilibrium spin transport in {Heisenberg} quantum magnets},\
  }\href@noop {} {\bibfield  {journal} {\bibinfo  {journal} {Phys. Rev. Lett.}\
  }\textbf {\bibinfo {volume} {113}},\ \bibinfo {pages} {147205} (\bibinfo
  {year} {2014})}\BibitemShut {NoStop}%
\bibitem [{\citenamefont {Jepsen}\ \emph {et~al.}(2020)\citenamefont {Jepsen},
  \citenamefont {Amato-Grill}, \citenamefont {Dimitrova}, \citenamefont {Ho},
  \citenamefont {Demler},\ and\ \citenamefont {Ketterle}}]{Jepsen2020}%
  \BibitemOpen
  \bibfield  {author} {\bibinfo {author} {\bibfnamefont {P.~N.}\ \bibnamefont
  {Jepsen}}, \bibinfo {author} {\bibfnamefont {J.}~\bibnamefont {Amato-Grill}},
  \bibinfo {author} {\bibfnamefont {I.}~\bibnamefont {Dimitrova}}, \bibinfo
  {author} {\bibfnamefont {W.~W.}\ \bibnamefont {Ho}}, \bibinfo {author}
  {\bibfnamefont {E.}~\bibnamefont {Demler}},\ and\ \bibinfo {author}
  {\bibfnamefont {W.}~\bibnamefont {Ketterle}},\ }\bibfield  {title} {\bibinfo
  {title} {Spin transport in a tunable {Heisenberg} model realized with
  ultracold atoms},\ }\href@noop {} {\bibfield  {journal} {\bibinfo  {journal}
  {Nature}\ }\textbf {\bibinfo {volume} {588}},\ \bibinfo {pages} {403}
  (\bibinfo {year} {2020})}\BibitemShut {NoStop}%
\bibitem [{\citenamefont {Jepsen}\ \emph {et~al.}(2021)\citenamefont {Jepsen},
  \citenamefont {Ho}, \citenamefont {Amato-Grill}, \citenamefont {Dimitrova},
  \citenamefont {Demler},\ and\ \citenamefont {Ketterle}}]{Jepsen2021}%
  \BibitemOpen
  \bibfield  {author} {\bibinfo {author} {\bibfnamefont {P.~N.}\ \bibnamefont
  {Jepsen}}, \bibinfo {author} {\bibfnamefont {W.~W.}\ \bibnamefont {Ho}},
  \bibinfo {author} {\bibfnamefont {J.}~\bibnamefont {Amato-Grill}}, \bibinfo
  {author} {\bibfnamefont {I.}~\bibnamefont {Dimitrova}}, \bibinfo {author}
  {\bibfnamefont {E.}~\bibnamefont {Demler}},\ and\ \bibinfo {author}
  {\bibfnamefont {W.}~\bibnamefont {Ketterle}},\ }\bibfield  {title} {\bibinfo
  {title} {Transverse spin dynamics in the anisotropic {Heisenberg} model
  realized with ultracold atoms},\ }\href@noop {} {\bibfield  {journal}
  {\bibinfo  {journal} {Phys. Rev. X}\ }\textbf {\bibinfo {volume} {11}},\
  \bibinfo {pages} {041054} (\bibinfo {year} {2021})}\BibitemShut {NoStop}%
\bibitem [{\citenamefont {Sun}\ \emph {et~al.}(2021)\citenamefont {Sun},
  \citenamefont {Yang}, \citenamefont {Wang}, \citenamefont {Zhou},
  \citenamefont {Su}, \citenamefont {Dai}, \citenamefont {Yuan},\ and\
  \citenamefont {Pan}}]{Sun2021}%
  \BibitemOpen
  \bibfield  {author} {\bibinfo {author} {\bibfnamefont {H.}~\bibnamefont
  {Sun}}, \bibinfo {author} {\bibfnamefont {B.}~\bibnamefont {Yang}}, \bibinfo
  {author} {\bibfnamefont {H.-Y.}\ \bibnamefont {Wang}}, \bibinfo {author}
  {\bibfnamefont {Z.-Y.}\ \bibnamefont {Zhou}}, \bibinfo {author}
  {\bibfnamefont {G.-X.}\ \bibnamefont {Su}}, \bibinfo {author} {\bibfnamefont
  {H.-N.}\ \bibnamefont {Dai}}, \bibinfo {author} {\bibfnamefont {Z.-S.}\
  \bibnamefont {Yuan}},\ and\ \bibinfo {author} {\bibfnamefont {J.-W.}\
  \bibnamefont {Pan}},\ }\bibfield  {title} {\bibinfo {title} {Realization of a
  bosonic antiferromagnet},\ }\href@noop {} {\bibfield  {journal} {\bibinfo
  {journal} {Nature Physics}\ }\textbf {\bibinfo {volume} {17}},\ \bibinfo
  {pages} {990} (\bibinfo {year} {2021})}\BibitemShut {NoStop}%
\bibitem [{\citenamefont {Jepsen}\ \emph {et~al.}(2022)\citenamefont {Jepsen},
  \citenamefont {Lee}, \citenamefont {Lin}, \citenamefont {Dimitrova},
  \citenamefont {Margalit}, \citenamefont {Ho},\ and\ \citenamefont
  {Ketterle}}]{Jepsen2022}%
  \BibitemOpen
  \bibfield  {author} {\bibinfo {author} {\bibfnamefont {P.~N.}\ \bibnamefont
  {Jepsen}}, \bibinfo {author} {\bibfnamefont {Y.~K.~E.}\ \bibnamefont {Lee}},
  \bibinfo {author} {\bibfnamefont {H.}~\bibnamefont {Lin}}, \bibinfo {author}
  {\bibfnamefont {I.}~\bibnamefont {Dimitrova}}, \bibinfo {author}
  {\bibfnamefont {Y.}~\bibnamefont {Margalit}}, \bibinfo {author}
  {\bibfnamefont {W.~W.}\ \bibnamefont {Ho}},\ and\ \bibinfo {author}
  {\bibfnamefont {W.}~\bibnamefont {Ketterle}},\ }\bibfield  {title} {\bibinfo
  {title} {Long-lived phantom helix states in {Heisenberg} quantum magnets},\
  }\href@noop {} {\bibfield  {journal} {\bibinfo  {journal} {Nat. Phys.}\
  }\textbf {\bibinfo {volume} {18}},\ \bibinfo {pages} {899} (\bibinfo {year}
  {2022})}\BibitemShut {NoStop}%
\bibitem [{\citenamefont {Wei}\ \emph {et~al.}(2022)\citenamefont {Wei},
  \citenamefont {Rubio-Abadal}, \citenamefont {Ye}, \citenamefont {Machado},
  \citenamefont {Kemp}, \citenamefont {Srakaew}, \citenamefont {Hollerith},
  \citenamefont {Rui}, \citenamefont {Gopalakrishnan}, \citenamefont {Yao},
  \citenamefont {Bloch},\ and\ \citenamefont {Zeiher}}]{Wei2022}%
  \BibitemOpen
  \bibfield  {author} {\bibinfo {author} {\bibfnamefont {D.}~\bibnamefont
  {Wei}}, \bibinfo {author} {\bibfnamefont {A.}~\bibnamefont {Rubio-Abadal}},
  \bibinfo {author} {\bibfnamefont {B.}~\bibnamefont {Ye}}, \bibinfo {author}
  {\bibfnamefont {F.}~\bibnamefont {Machado}}, \bibinfo {author} {\bibfnamefont
  {J.}~\bibnamefont {Kemp}}, \bibinfo {author} {\bibfnamefont {K.}~\bibnamefont
  {Srakaew}}, \bibinfo {author} {\bibfnamefont {S.}~\bibnamefont {Hollerith}},
  \bibinfo {author} {\bibfnamefont {J.}~\bibnamefont {Rui}}, \bibinfo {author}
  {\bibfnamefont {S.}~\bibnamefont {Gopalakrishnan}}, \bibinfo {author}
  {\bibfnamefont {N.~Y.}\ \bibnamefont {Yao}}, \bibinfo {author} {\bibfnamefont
  {I.}~\bibnamefont {Bloch}},\ and\ \bibinfo {author} {\bibfnamefont
  {J.}~\bibnamefont {Zeiher}},\ }\bibfield  {title} {\bibinfo {title} {Quantum
  gas microscopy of {Kardar-Parisi-Zhang} superdiffusion},\ }\href@noop {}
  {\bibfield  {journal} {\bibinfo  {journal} {Science}\ }\textbf {\bibinfo
  {volume} {376}},\ \bibinfo {pages} {716} (\bibinfo {year}
  {2022})}\BibitemShut {NoStop}%
\bibitem [{\citenamefont {Browaeys}\ and\ \citenamefont
  {Lahaye}(2020)}]{Browaeys2020}%
  \BibitemOpen
  \bibfield  {author} {\bibinfo {author} {\bibfnamefont {A.}~\bibnamefont
  {Browaeys}}\ and\ \bibinfo {author} {\bibfnamefont {T.}~\bibnamefont
  {Lahaye}},\ }\bibfield  {title} {\bibinfo {title} {Many-body physics with
  individually controlled {Rydberg} atoms},\ }\href@noop {} {\bibfield
  {journal} {\bibinfo  {journal} {Nat. Phys}\ }\textbf {\bibinfo {volume}
  {16}},\ \bibinfo {pages} {132} (\bibinfo {year} {2020})}\BibitemShut
  {NoStop}%
\bibitem [{\citenamefont {Geier}\ \emph {et~al.}(2021)\citenamefont {Geier},
  \citenamefont {Thaicharoen}, \citenamefont {Hainaut}, \citenamefont {Franz},
  \citenamefont {Salzinger}, \citenamefont {Tebben}, \citenamefont
  {Grimshandl}, \citenamefont {Z{\"u}rn},\ and\ \citenamefont
  {Weidem{\"u}ller}}]{Geier2021}%
  \BibitemOpen
  \bibfield  {author} {\bibinfo {author} {\bibfnamefont {S.}~\bibnamefont
  {Geier}}, \bibinfo {author} {\bibfnamefont {N.}~\bibnamefont {Thaicharoen}},
  \bibinfo {author} {\bibfnamefont {C.}~\bibnamefont {Hainaut}}, \bibinfo
  {author} {\bibfnamefont {T.}~\bibnamefont {Franz}}, \bibinfo {author}
  {\bibfnamefont {A.}~\bibnamefont {Salzinger}}, \bibinfo {author}
  {\bibfnamefont {A.}~\bibnamefont {Tebben}}, \bibinfo {author} {\bibfnamefont
  {D.}~\bibnamefont {Grimshandl}}, \bibinfo {author} {\bibfnamefont
  {G.}~\bibnamefont {Z{\"u}rn}},\ and\ \bibinfo {author} {\bibfnamefont
  {M.}~\bibnamefont {Weidem{\"u}ller}},\ }\bibfield  {title} {\bibinfo {title}
  {Floquet {Hamiltonian} engineering of an isolated many-body spin system},\
  }\href@noop {} {\bibfield  {journal} {\bibinfo  {journal} {Science}\ }\textbf
  {\bibinfo {volume} {374}},\ \bibinfo {pages} {1149} (\bibinfo {year}
  {2021})}\BibitemShut {NoStop}%
\bibitem [{\citenamefont {Scholl}\ \emph {et~al.}(2022)\citenamefont {Scholl},
  \citenamefont {Williams}, \citenamefont {Bornet}, \citenamefont {Wallner},
  \citenamefont {Barredo}, \citenamefont {Henriet}, \citenamefont {Signoles},
  \citenamefont {Hainaut}, \citenamefont {Franz}, \citenamefont {Geier},
  \citenamefont {Tebben}, \citenamefont {Salzinger}, \citenamefont {Z\"{u}rn},
  \citenamefont {Lahaye}, \citenamefont {Weidem\"{u}ller},\ and\ \citenamefont
  {Browaeys}}]{Scholl2022}%
  \BibitemOpen
  \bibfield  {author} {\bibinfo {author} {\bibfnamefont {P.}~\bibnamefont
  {Scholl}}, \bibinfo {author} {\bibfnamefont {H.~J.}\ \bibnamefont
  {Williams}}, \bibinfo {author} {\bibfnamefont {G.}~\bibnamefont {Bornet}},
  \bibinfo {author} {\bibfnamefont {F.}~\bibnamefont {Wallner}}, \bibinfo
  {author} {\bibfnamefont {D.}~\bibnamefont {Barredo}}, \bibinfo {author}
  {\bibfnamefont {L.}~\bibnamefont {Henriet}}, \bibinfo {author} {\bibfnamefont
  {A.}~\bibnamefont {Signoles}}, \bibinfo {author} {\bibfnamefont
  {C.}~\bibnamefont {Hainaut}}, \bibinfo {author} {\bibfnamefont
  {T.}~\bibnamefont {Franz}}, \bibinfo {author} {\bibfnamefont
  {S.}~\bibnamefont {Geier}}, \bibinfo {author} {\bibfnamefont
  {A.}~\bibnamefont {Tebben}}, \bibinfo {author} {\bibfnamefont
  {A.}~\bibnamefont {Salzinger}}, \bibinfo {author} {\bibfnamefont
  {G.}~\bibnamefont {Z\"{u}rn}}, \bibinfo {author} {\bibfnamefont
  {T.}~\bibnamefont {Lahaye}}, \bibinfo {author} {\bibfnamefont
  {M.}~\bibnamefont {Weidem\"{u}ller}},\ and\ \bibinfo {author} {\bibfnamefont
  {A.}~\bibnamefont {Browaeys}},\ }\bibfield  {title} {\bibinfo {title}
  {Microwave engineering of programmable {XXZ} {Hamiltonians} in arrays of
  {Rydberg} atoms},\ }\href@noop {} {\bibfield  {journal} {\bibinfo  {journal}
  {PRX Quantum}\ }\textbf {\bibinfo {volume} {3}},\ \bibinfo {pages} {020303}
  (\bibinfo {year} {2022})}\BibitemShut {NoStop}%
\bibitem [{\citenamefont {Morong}\ \emph {et~al.}(2023)\citenamefont {Morong},
  \citenamefont {Collins}, \citenamefont {De}, \citenamefont {Stavropoulos},
  \citenamefont {You},\ and\ \citenamefont {Monroe}}]{Morong2023}%
  \BibitemOpen
  \bibfield  {author} {\bibinfo {author} {\bibfnamefont {W.}~\bibnamefont
  {Morong}}, \bibinfo {author} {\bibfnamefont {K.}~\bibnamefont {Collins}},
  \bibinfo {author} {\bibfnamefont {A.}~\bibnamefont {De}}, \bibinfo {author}
  {\bibfnamefont {E.}~\bibnamefont {Stavropoulos}}, \bibinfo {author}
  {\bibfnamefont {T.}~\bibnamefont {You}},\ and\ \bibinfo {author}
  {\bibfnamefont {C.}~\bibnamefont {Monroe}},\ }\bibfield  {title} {\bibinfo
  {title} {Engineering dynamically decoupled quantum simulations with trapped
  ions},\ }\href@noop {} {\bibfield  {journal} {\bibinfo  {journal} {PRX
  Quantum}\ }\textbf {\bibinfo {volume} {4}},\ \bibinfo {pages} {010334}
  (\bibinfo {year} {2023})}\BibitemShut {NoStop}%
\bibitem [{\citenamefont {Kranzl}\ \emph
  {et~al.}(2023{\natexlab{a}})\citenamefont {Kranzl}, \citenamefont {Lasek},
  \citenamefont {Joshi}, \citenamefont {Kalev}, \citenamefont {Blatt},
  \citenamefont {Roos},\ and\ \citenamefont
  {Yunger~Halpern}}]{Kranzl2023experimental}%
  \BibitemOpen
  \bibfield  {author} {\bibinfo {author} {\bibfnamefont {F.}~\bibnamefont
  {Kranzl}}, \bibinfo {author} {\bibfnamefont {A.}~\bibnamefont {Lasek}},
  \bibinfo {author} {\bibfnamefont {M.~K.}\ \bibnamefont {Joshi}}, \bibinfo
  {author} {\bibfnamefont {A.}~\bibnamefont {Kalev}}, \bibinfo {author}
  {\bibfnamefont {R.}~\bibnamefont {Blatt}}, \bibinfo {author} {\bibfnamefont
  {C.~F.}\ \bibnamefont {Roos}},\ and\ \bibinfo {author} {\bibfnamefont
  {N.}~\bibnamefont {Yunger~Halpern}},\ }\bibfield  {title} {\bibinfo {title}
  {Experimental observation of thermalization with noncommuting charges},\
  }\href@noop {} {\bibfield  {journal} {\bibinfo  {journal} {PRX Quantum}\
  }\textbf {\bibinfo {volume} {4}},\ \bibinfo {pages} {020318} (\bibinfo {year}
  {2023}{\natexlab{a}})}\BibitemShut {NoStop}%
\bibitem [{\citenamefont {Kranzl}\ \emph
  {et~al.}(2023{\natexlab{b}})\citenamefont {Kranzl}, \citenamefont
  {Birnkammer}, \citenamefont {Joshi}, \citenamefont {Bastianello},
  \citenamefont {Blatt}, \citenamefont {Knap},\ and\ \citenamefont
  {Roos}}]{Kranzl2023}%
  \BibitemOpen
  \bibfield  {author} {\bibinfo {author} {\bibfnamefont {F.}~\bibnamefont
  {Kranzl}}, \bibinfo {author} {\bibfnamefont {S.}~\bibnamefont {Birnkammer}},
  \bibinfo {author} {\bibfnamefont {M.~K.}\ \bibnamefont {Joshi}}, \bibinfo
  {author} {\bibfnamefont {A.}~\bibnamefont {Bastianello}}, \bibinfo {author}
  {\bibfnamefont {R.}~\bibnamefont {Blatt}}, \bibinfo {author} {\bibfnamefont
  {M.}~\bibnamefont {Knap}},\ and\ \bibinfo {author} {\bibfnamefont {C.~F.}\
  \bibnamefont {Roos}},\ }\bibfield  {title} {\bibinfo {title} {Observation of
  magnon bound states in the long-range, anisotropic {Heisenberg} model},\
  }\href@noop {} {\bibfield  {journal} {\bibinfo  {journal} {Phys. Rev. X}\
  }\textbf {\bibinfo {volume} {13}},\ \bibinfo {pages} {031017} (\bibinfo
  {year} {2023}{\natexlab{b}})}\BibitemShut {NoStop}%
\bibitem [{\citenamefont {Christakis}\ \emph {et~al.}(2023)\citenamefont
  {Christakis}, \citenamefont {Rosenberg}, \citenamefont {Raj}, \citenamefont
  {Chi}, \citenamefont {Morningstar}, \citenamefont {Huse}, \citenamefont
  {Yan},\ and\ \citenamefont {Bakr}}]{christakis2023probing}%
  \BibitemOpen
  \bibfield  {author} {\bibinfo {author} {\bibfnamefont {L.}~\bibnamefont
  {Christakis}}, \bibinfo {author} {\bibfnamefont {J.~S.}\ \bibnamefont
  {Rosenberg}}, \bibinfo {author} {\bibfnamefont {R.}~\bibnamefont {Raj}},
  \bibinfo {author} {\bibfnamefont {S.}~\bibnamefont {Chi}}, \bibinfo {author}
  {\bibfnamefont {A.}~\bibnamefont {Morningstar}}, \bibinfo {author}
  {\bibfnamefont {D.~A.}\ \bibnamefont {Huse}}, \bibinfo {author}
  {\bibfnamefont {Z.~Z.}\ \bibnamefont {Yan}},\ and\ \bibinfo {author}
  {\bibfnamefont {W.~S.}\ \bibnamefont {Bakr}},\ }\bibfield  {title} {\bibinfo
  {title} {Probing site-resolved correlations in a spin system of ultracold
  molecules},\ }\href@noop {} {\bibfield  {journal} {\bibinfo  {journal}
  {Nature}\ }\textbf {\bibinfo {volume} {614}},\ \bibinfo {pages} {64}
  (\bibinfo {year} {2023})}\BibitemShut {NoStop}%
\bibitem [{\citenamefont {Miller}\ \emph {et~al.}(2024)\citenamefont {Miller},
  \citenamefont {Carroll}, \citenamefont {Lin}, \citenamefont {Hirzler},
  \citenamefont {Gao}, \citenamefont {Zhou}, \citenamefont {Lukin},\ and\
  \citenamefont {Ye}}]{miller2024two}%
  \BibitemOpen
  \bibfield  {author} {\bibinfo {author} {\bibfnamefont {C.}~\bibnamefont
  {Miller}}, \bibinfo {author} {\bibfnamefont {A.~N.}\ \bibnamefont {Carroll}},
  \bibinfo {author} {\bibfnamefont {J.}~\bibnamefont {Lin}}, \bibinfo {author}
  {\bibfnamefont {H.}~\bibnamefont {Hirzler}}, \bibinfo {author} {\bibfnamefont
  {H.}~\bibnamefont {Gao}}, \bibinfo {author} {\bibfnamefont {H.}~\bibnamefont
  {Zhou}}, \bibinfo {author} {\bibfnamefont {M.~D.}\ \bibnamefont {Lukin}},\
  and\ \bibinfo {author} {\bibfnamefont {J.}~\bibnamefont {Ye}},\ }\bibfield
  {title} {\bibinfo {title} {Two-axis twisting using {Floquet-engineered} {XYZ}
  spin models with polar molecules},\ }\href@noop {} {\bibfield  {journal}
  {\bibinfo  {journal} {Nature}\ }\textbf {\bibinfo {volume} {633}},\ \bibinfo
  {pages} {332} (\bibinfo {year} {2024})}\BibitemShut {NoStop}%
\bibitem [{\citenamefont {Carroll}\ \emph {et~al.}(2025)\citenamefont
  {Carroll}, \citenamefont {Hirzler}, \citenamefont {Miller}, \citenamefont
  {Wellnitz}, \citenamefont {Muleady}, \citenamefont {Lin}, \citenamefont
  {Zamarski}, \citenamefont {Wang}, \citenamefont {Bohn}, \citenamefont {Rey},\
  and\ \citenamefont {Ye}}]{carroll2025observation}%
  \BibitemOpen
  \bibfield  {author} {\bibinfo {author} {\bibfnamefont {A.~N.}\ \bibnamefont
  {Carroll}}, \bibinfo {author} {\bibfnamefont {H.}~\bibnamefont {Hirzler}},
  \bibinfo {author} {\bibfnamefont {C.}~\bibnamefont {Miller}}, \bibinfo
  {author} {\bibfnamefont {D.}~\bibnamefont {Wellnitz}}, \bibinfo {author}
  {\bibfnamefont {S.~R.}\ \bibnamefont {Muleady}}, \bibinfo {author}
  {\bibfnamefont {J.}~\bibnamefont {Lin}}, \bibinfo {author} {\bibfnamefont
  {K.~P.}\ \bibnamefont {Zamarski}}, \bibinfo {author} {\bibfnamefont {R.~R.}\
  \bibnamefont {Wang}}, \bibinfo {author} {\bibfnamefont {J.~L.}\ \bibnamefont
  {Bohn}}, \bibinfo {author} {\bibfnamefont {A.~M.}\ \bibnamefont {Rey}},\ and\
  \bibinfo {author} {\bibfnamefont {J.}~\bibnamefont {Ye}},\ }\bibfield
  {title} {\bibinfo {title} {Observation of generalized {$t$-$J$} spin dynamics
  with tunable dipolar interactions},\ }\href@noop {} {\bibfield  {journal}
  {\bibinfo  {journal} {Science}\ }\textbf {\bibinfo {volume} {388}},\ \bibinfo
  {pages} {381} (\bibinfo {year} {2025})}\BibitemShut {NoStop}%
\bibitem [{\citenamefont {Nguyen}\ \emph {et~al.}(2024)\citenamefont {Nguyen},
  \citenamefont {Kim}, \citenamefont {Hashim}, \citenamefont {Goss},
  \citenamefont {Marinelli}, \citenamefont {Bhandari}, \citenamefont {Das},
  \citenamefont {Naik}, \citenamefont {Kreikebaum}, \citenamefont {Jordan},
  \citenamefont {Santiago},\ and\ \citenamefont {Siddiqi}}]{Nguyen2024}%
  \BibitemOpen
  \bibfield  {author} {\bibinfo {author} {\bibfnamefont {L.~B.}\ \bibnamefont
  {Nguyen}}, \bibinfo {author} {\bibfnamefont {Y.}~\bibnamefont {Kim}},
  \bibinfo {author} {\bibfnamefont {A.}~\bibnamefont {Hashim}}, \bibinfo
  {author} {\bibfnamefont {N.}~\bibnamefont {Goss}}, \bibinfo {author}
  {\bibfnamefont {B.}~\bibnamefont {Marinelli}}, \bibinfo {author}
  {\bibfnamefont {B.}~\bibnamefont {Bhandari}}, \bibinfo {author}
  {\bibfnamefont {D.}~\bibnamefont {Das}}, \bibinfo {author} {\bibfnamefont
  {R.~K.}\ \bibnamefont {Naik}}, \bibinfo {author} {\bibfnamefont {J.~M.}\
  \bibnamefont {Kreikebaum}}, \bibinfo {author} {\bibfnamefont {A.~N.}\
  \bibnamefont {Jordan}}, \bibinfo {author} {\bibfnamefont {D.~I.}\
  \bibnamefont {Santiago}},\ and\ \bibinfo {author} {\bibfnamefont
  {I.}~\bibnamefont {Siddiqi}},\ }\bibfield  {title} {\bibinfo {title}
  {Programmable {Heisenberg} interactions between {Floquet} qubits},\
  }\href@noop {} {\bibfield  {journal} {\bibinfo  {journal} {Nat. Phys.}\
  }\textbf {\bibinfo {volume} {20}},\ \bibinfo {pages} {240} (\bibinfo {year}
  {2024})}\BibitemShut {NoStop}%
\bibitem [{\citenamefont {Chowdhury}\ \emph {et~al.}(2024)\citenamefont
  {Chowdhury}, \citenamefont {Yu}, \citenamefont {Shamim}, \citenamefont
  {Kabir},\ and\ \citenamefont {Sufian}}]{Chowdhury2024}%
  \BibitemOpen
  \bibfield  {author} {\bibinfo {author} {\bibfnamefont {T.~A.}\ \bibnamefont
  {Chowdhury}}, \bibinfo {author} {\bibfnamefont {K.}~\bibnamefont {Yu}},
  \bibinfo {author} {\bibfnamefont {M.~A.}\ \bibnamefont {Shamim}}, \bibinfo
  {author} {\bibfnamefont {M.}~\bibnamefont {Kabir}},\ and\ \bibinfo {author}
  {\bibfnamefont {R.~S.}\ \bibnamefont {Sufian}},\ }\bibfield  {title}
  {\bibinfo {title} {Enhancing quantum utility: {Simulating} large-scale
  quantum spin chains on superconducting quantum computers},\ }\href@noop {}
  {\bibfield  {journal} {\bibinfo  {journal} {Phys. Rev. Res.}\ }\textbf
  {\bibinfo {volume} {6}},\ \bibinfo {pages} {033107} (\bibinfo {year}
  {2024})}\BibitemShut {NoStop}%
\bibitem [{\citenamefont {Rosenberg}\ \emph {et~al.}(2024)\citenamefont
  {Rosenberg}, \citenamefont {Andersen}, \citenamefont {Samajdar},
  \citenamefont {Petukhov}, \citenamefont {Hoke}, \citenamefont {Abanin},
  \citenamefont {Bengtsson}, \citenamefont {Drozdov}, \citenamefont {Erickson},
  \citenamefont {Klimov}, \citenamefont {Mi}, \citenamefont {Morvan},
  \citenamefont {Neeley}, \citenamefont {Neill}, \citenamefont {Acharya},
  \citenamefont {Allen}, \citenamefont {Anderson}, \citenamefont {Ansmann},
  \citenamefont {Arute}, \citenamefont {Arya}, \citenamefont {Asfaw},
  \citenamefont {Atalaya}, \citenamefont {Bardin}, \citenamefont {Bilmes},
  \citenamefont {Bortoli}, \citenamefont {Bourassa}, \citenamefont {Bovaird},
  \citenamefont {Brill}, \citenamefont {Broughton}, \citenamefont {Buckley},
  \citenamefont {Buell}, \citenamefont {Burger}, \citenamefont {Burkett},
  \citenamefont {Bushnell}, \citenamefont {Campero}, \citenamefont {Chang},
  \citenamefont {Chen}, \citenamefont {Chiaro}, \citenamefont {Chik},
  \citenamefont {Cogan}, \citenamefont {Collins}, \citenamefont {Conner},
  \citenamefont {Courtney}, \citenamefont {Crook}, \citenamefont {Curtin},
  \citenamefont {Debroy}, \citenamefont {Barba}, \citenamefont {Demura},
  \citenamefont {Paolo}, \citenamefont {Dunsworth}, \citenamefont {Earle},
  \citenamefont {Faoro}, \citenamefont {Farhi}, \citenamefont {Fatemi},
  \citenamefont {Ferreira}, \citenamefont {Burgos}, \citenamefont {Forati},
  \citenamefont {Fowler}, \citenamefont {Foxen}, \citenamefont {Garcia},
  \citenamefont {Genois}, \citenamefont {Giang}, \citenamefont {Gidney},
  \citenamefont {Gilboa}, \citenamefont {Giustina}, \citenamefont {Gosula},
  \citenamefont {Dau}, \citenamefont {Gross}, \citenamefont {Habegger},
  \citenamefont {Hamilton}, \citenamefont {Hansen}, \citenamefont {Harrigan},
  \citenamefont {Harrington}, \citenamefont {Heu}, \citenamefont {Hill},
  \citenamefont {Hoffmann}, \citenamefont {Hong}, \citenamefont {Huang},
  \citenamefont {Huff}, \citenamefont {Huggins}, \citenamefont {Ioffe},
  \citenamefont {Isakov}, \citenamefont {Iveland}, \citenamefont {Jeffrey},
  \citenamefont {Jiang}, \citenamefont {Jones}, \citenamefont {Juhas},
  \citenamefont {Kafri}, \citenamefont {Khattar}, \citenamefont {Khezri},
  \citenamefont {Kieferov\'a}, \citenamefont {Kim}, \citenamefont {Kitaev},
  \citenamefont {Klots}, \citenamefont {Korotkov}, \citenamefont {Kostritsa},
  \citenamefont {Kreikebaum}, \citenamefont {Landhuis}, \citenamefont {Laptev},
  \citenamefont {Lau}, \citenamefont {Laws}, \citenamefont {Lee}, \citenamefont
  {Lee}, \citenamefont {Lensky}, \citenamefont {Lester}, \citenamefont {Lill},
  \citenamefont {Liu}, \citenamefont {Locharla}, \citenamefont {Mandr\`a},
  \citenamefont {Martin}, \citenamefont {Martin}, \citenamefont {McClean},
  \citenamefont {McEwen}, \citenamefont {Meeks}, \citenamefont {Miao},
  \citenamefont {Mieszala}, \citenamefont {Montazeri}, \citenamefont
  {Movassagh}, \citenamefont {Mruczkiewicz}, \citenamefont {Nersisyan},
  \citenamefont {Newman}, \citenamefont {Ng}, \citenamefont {Nguyen},
  \citenamefont {Nguyen}, \citenamefont {Niu}, \citenamefont {O’Brien},
  \citenamefont {Omonije}, \citenamefont {Opremcak}, \citenamefont {Potter},
  \citenamefont {Pryadko}, \citenamefont {Quintana}, \citenamefont {Rhodes},
  \citenamefont {Rocque}, \citenamefont {Rubin}, \citenamefont {Saei},
  \citenamefont {Sank}, \citenamefont {Sankaragomathi}, \citenamefont
  {Satzinger}, \citenamefont {Schurkus}, \citenamefont {Schuster},
  \citenamefont {Shearn}, \citenamefont {Shorter}, \citenamefont {Shutty},
  \citenamefont {Shvarts}, \citenamefont {Sivak}, \citenamefont {Skruzny},
  \citenamefont {Smith}, \citenamefont {Somma}, \citenamefont {Sterling},
  \citenamefont {Strain}, \citenamefont {Szalay}, \citenamefont {Thor},
  \citenamefont {Torres}, \citenamefont {Vidal}, \citenamefont {Villalonga},
  \citenamefont {Heidweiller}, \citenamefont {White}, \citenamefont {Woo},
  \citenamefont {Xing}, \citenamefont {Yao}, \citenamefont {Yeh}, \citenamefont
  {Yoo}, \citenamefont {Young}, \citenamefont {Zalcman}, \citenamefont {Zhang},
  \citenamefont {Zhu}, \citenamefont {Zobrist}, \citenamefont {Neven},
  \citenamefont {Babbush}, \citenamefont {Bacon}, \citenamefont {Boixo},
  \citenamefont {Hilton}, \citenamefont {Lucero}, \citenamefont {Megrant},
  \citenamefont {Kelly}, \citenamefont {Chen}, \citenamefont {Smelyanskiy},
  \citenamefont {Khemani}, \citenamefont {Gopalakrishnan}, \citenamefont
  {Prosen},\ and\ \citenamefont {Roushan}}]{Rosenberg2024}%
  \BibitemOpen
  \bibfield  {author} {\bibinfo {author} {\bibfnamefont {E.}~\bibnamefont
  {Rosenberg}}, \bibinfo {author} {\bibfnamefont {T.~I.}\ \bibnamefont
  {Andersen}}, \bibinfo {author} {\bibfnamefont {R.}~\bibnamefont {Samajdar}},
  \bibinfo {author} {\bibfnamefont {A.}~\bibnamefont {Petukhov}}, \bibinfo
  {author} {\bibfnamefont {J.~C.}\ \bibnamefont {Hoke}}, \bibinfo {author}
  {\bibfnamefont {D.}~\bibnamefont {Abanin}}, \bibinfo {author} {\bibfnamefont
  {A.}~\bibnamefont {Bengtsson}}, \bibinfo {author} {\bibfnamefont {I.~K.}\
  \bibnamefont {Drozdov}}, \bibinfo {author} {\bibfnamefont {C.}~\bibnamefont
  {Erickson}}, \bibinfo {author} {\bibfnamefont {P.~V.}\ \bibnamefont
  {Klimov}}, \bibinfo {author} {\bibfnamefont {X.}~\bibnamefont {Mi}}, \bibinfo
  {author} {\bibfnamefont {A.}~\bibnamefont {Morvan}}, \bibinfo {author}
  {\bibfnamefont {M.}~\bibnamefont {Neeley}}, \bibinfo {author} {\bibfnamefont
  {C.}~\bibnamefont {Neill}}, \bibinfo {author} {\bibfnamefont
  {R.}~\bibnamefont {Acharya}}, \bibinfo {author} {\bibfnamefont
  {R.}~\bibnamefont {Allen}}, \bibinfo {author} {\bibfnamefont
  {K.}~\bibnamefont {Anderson}}, \bibinfo {author} {\bibfnamefont
  {M.}~\bibnamefont {Ansmann}}, \bibinfo {author} {\bibfnamefont
  {F.}~\bibnamefont {Arute}}, \bibinfo {author} {\bibfnamefont
  {K.}~\bibnamefont {Arya}}, \bibinfo {author} {\bibfnamefont {A.}~\bibnamefont
  {Asfaw}}, \bibinfo {author} {\bibfnamefont {J.}~\bibnamefont {Atalaya}},
  \bibinfo {author} {\bibfnamefont {J.~C.}\ \bibnamefont {Bardin}}, \bibinfo
  {author} {\bibfnamefont {A.}~\bibnamefont {Bilmes}}, \bibinfo {author}
  {\bibfnamefont {G.}~\bibnamefont {Bortoli}}, \bibinfo {author} {\bibfnamefont
  {A.}~\bibnamefont {Bourassa}}, \bibinfo {author} {\bibfnamefont
  {J.}~\bibnamefont {Bovaird}}, \bibinfo {author} {\bibfnamefont
  {L.}~\bibnamefont {Brill}}, \bibinfo {author} {\bibfnamefont
  {M.}~\bibnamefont {Broughton}}, \bibinfo {author} {\bibfnamefont {B.~B.}\
  \bibnamefont {Buckley}}, \bibinfo {author} {\bibfnamefont {D.~A.}\
  \bibnamefont {Buell}}, \bibinfo {author} {\bibfnamefont {T.}~\bibnamefont
  {Burger}}, \bibinfo {author} {\bibfnamefont {B.}~\bibnamefont {Burkett}},
  \bibinfo {author} {\bibfnamefont {N.}~\bibnamefont {Bushnell}}, \bibinfo
  {author} {\bibfnamefont {J.}~\bibnamefont {Campero}}, \bibinfo {author}
  {\bibfnamefont {H.-S.}\ \bibnamefont {Chang}}, \bibinfo {author}
  {\bibfnamefont {Z.}~\bibnamefont {Chen}}, \bibinfo {author} {\bibfnamefont
  {B.}~\bibnamefont {Chiaro}}, \bibinfo {author} {\bibfnamefont
  {D.}~\bibnamefont {Chik}}, \bibinfo {author} {\bibfnamefont {J.}~\bibnamefont
  {Cogan}}, \bibinfo {author} {\bibfnamefont {R.}~\bibnamefont {Collins}},
  \bibinfo {author} {\bibfnamefont {P.}~\bibnamefont {Conner}}, \bibinfo
  {author} {\bibfnamefont {W.}~\bibnamefont {Courtney}}, \bibinfo {author}
  {\bibfnamefont {A.~L.}\ \bibnamefont {Crook}}, \bibinfo {author}
  {\bibfnamefont {B.}~\bibnamefont {Curtin}}, \bibinfo {author} {\bibfnamefont
  {D.~M.}\ \bibnamefont {Debroy}}, \bibinfo {author} {\bibfnamefont {A.~D.~T.}\
  \bibnamefont {Barba}}, \bibinfo {author} {\bibfnamefont {S.}~\bibnamefont
  {Demura}}, \bibinfo {author} {\bibfnamefont {A.~D.}\ \bibnamefont {Paolo}},
  \bibinfo {author} {\bibfnamefont {A.}~\bibnamefont {Dunsworth}}, \bibinfo
  {author} {\bibfnamefont {C.}~\bibnamefont {Earle}}, \bibinfo {author}
  {\bibfnamefont {L.}~\bibnamefont {Faoro}}, \bibinfo {author} {\bibfnamefont
  {E.}~\bibnamefont {Farhi}}, \bibinfo {author} {\bibfnamefont
  {R.}~\bibnamefont {Fatemi}}, \bibinfo {author} {\bibfnamefont {V.~S.}\
  \bibnamefont {Ferreira}}, \bibinfo {author} {\bibfnamefont {L.~F.}\
  \bibnamefont {Burgos}}, \bibinfo {author} {\bibfnamefont {E.}~\bibnamefont
  {Forati}}, \bibinfo {author} {\bibfnamefont {A.~G.}\ \bibnamefont {Fowler}},
  \bibinfo {author} {\bibfnamefont {B.}~\bibnamefont {Foxen}}, \bibinfo
  {author} {\bibfnamefont {G.}~\bibnamefont {Garcia}}, \bibinfo {author}
  {\bibfnamefont {U.}~\bibnamefont {Genois}}, \bibinfo {author} {\bibfnamefont
  {W.}~\bibnamefont {Giang}}, \bibinfo {author} {\bibfnamefont
  {C.}~\bibnamefont {Gidney}}, \bibinfo {author} {\bibfnamefont
  {D.}~\bibnamefont {Gilboa}}, \bibinfo {author} {\bibfnamefont
  {M.}~\bibnamefont {Giustina}}, \bibinfo {author} {\bibfnamefont
  {R.}~\bibnamefont {Gosula}}, \bibinfo {author} {\bibfnamefont {A.~G.}\
  \bibnamefont {Dau}}, \bibinfo {author} {\bibfnamefont {J.~A.}\ \bibnamefont
  {Gross}}, \bibinfo {author} {\bibfnamefont {S.}~\bibnamefont {Habegger}},
  \bibinfo {author} {\bibfnamefont {M.~C.}\ \bibnamefont {Hamilton}}, \bibinfo
  {author} {\bibfnamefont {M.}~\bibnamefont {Hansen}}, \bibinfo {author}
  {\bibfnamefont {M.~P.}\ \bibnamefont {Harrigan}}, \bibinfo {author}
  {\bibfnamefont {S.~D.}\ \bibnamefont {Harrington}}, \bibinfo {author}
  {\bibfnamefont {P.}~\bibnamefont {Heu}}, \bibinfo {author} {\bibfnamefont
  {G.}~\bibnamefont {Hill}}, \bibinfo {author} {\bibfnamefont {M.~R.}\
  \bibnamefont {Hoffmann}}, \bibinfo {author} {\bibfnamefont {S.}~\bibnamefont
  {Hong}}, \bibinfo {author} {\bibfnamefont {T.}~\bibnamefont {Huang}},
  \bibinfo {author} {\bibfnamefont {A.}~\bibnamefont {Huff}}, \bibinfo {author}
  {\bibfnamefont {W.~J.}\ \bibnamefont {Huggins}}, \bibinfo {author}
  {\bibfnamefont {L.~B.}\ \bibnamefont {Ioffe}}, \bibinfo {author}
  {\bibfnamefont {S.~V.}\ \bibnamefont {Isakov}}, \bibinfo {author}
  {\bibfnamefont {J.}~\bibnamefont {Iveland}}, \bibinfo {author} {\bibfnamefont
  {E.}~\bibnamefont {Jeffrey}}, \bibinfo {author} {\bibfnamefont
  {Z.}~\bibnamefont {Jiang}}, \bibinfo {author} {\bibfnamefont
  {C.}~\bibnamefont {Jones}}, \bibinfo {author} {\bibfnamefont
  {P.}~\bibnamefont {Juhas}}, \bibinfo {author} {\bibfnamefont
  {D.}~\bibnamefont {Kafri}}, \bibinfo {author} {\bibfnamefont
  {T.}~\bibnamefont {Khattar}}, \bibinfo {author} {\bibfnamefont
  {M.}~\bibnamefont {Khezri}}, \bibinfo {author} {\bibfnamefont
  {M.}~\bibnamefont {Kieferov\'a}}, \bibinfo {author} {\bibfnamefont
  {S.}~\bibnamefont {Kim}}, \bibinfo {author} {\bibfnamefont {A.}~\bibnamefont
  {Kitaev}}, \bibinfo {author} {\bibfnamefont {A.~R.}\ \bibnamefont {Klots}},
  \bibinfo {author} {\bibfnamefont {A.~N.}\ \bibnamefont {Korotkov}}, \bibinfo
  {author} {\bibfnamefont {F.}~\bibnamefont {Kostritsa}}, \bibinfo {author}
  {\bibfnamefont {J.~M.}\ \bibnamefont {Kreikebaum}}, \bibinfo {author}
  {\bibfnamefont {D.}~\bibnamefont {Landhuis}}, \bibinfo {author}
  {\bibfnamefont {P.}~\bibnamefont {Laptev}}, \bibinfo {author} {\bibfnamefont
  {K.-M.}\ \bibnamefont {Lau}}, \bibinfo {author} {\bibfnamefont
  {L.}~\bibnamefont {Laws}}, \bibinfo {author} {\bibfnamefont {J.}~\bibnamefont
  {Lee}}, \bibinfo {author} {\bibfnamefont {K.~W.}\ \bibnamefont {Lee}},
  \bibinfo {author} {\bibfnamefont {Y.~D.}\ \bibnamefont {Lensky}}, \bibinfo
  {author} {\bibfnamefont {B.~J.}\ \bibnamefont {Lester}}, \bibinfo {author}
  {\bibfnamefont {A.~T.}\ \bibnamefont {Lill}}, \bibinfo {author}
  {\bibfnamefont {W.}~\bibnamefont {Liu}}, \bibinfo {author} {\bibfnamefont
  {A.}~\bibnamefont {Locharla}}, \bibinfo {author} {\bibfnamefont
  {S.}~\bibnamefont {Mandr\`a}}, \bibinfo {author} {\bibfnamefont
  {O.}~\bibnamefont {Martin}}, \bibinfo {author} {\bibfnamefont
  {S.}~\bibnamefont {Martin}}, \bibinfo {author} {\bibfnamefont {J.~R.}\
  \bibnamefont {McClean}}, \bibinfo {author} {\bibfnamefont {M.}~\bibnamefont
  {McEwen}}, \bibinfo {author} {\bibfnamefont {S.}~\bibnamefont {Meeks}},
  \bibinfo {author} {\bibfnamefont {K.~C.}\ \bibnamefont {Miao}}, \bibinfo
  {author} {\bibfnamefont {A.}~\bibnamefont {Mieszala}}, \bibinfo {author}
  {\bibfnamefont {S.}~\bibnamefont {Montazeri}}, \bibinfo {author}
  {\bibfnamefont {R.}~\bibnamefont {Movassagh}}, \bibinfo {author}
  {\bibfnamefont {W.}~\bibnamefont {Mruczkiewicz}}, \bibinfo {author}
  {\bibfnamefont {A.}~\bibnamefont {Nersisyan}}, \bibinfo {author}
  {\bibfnamefont {M.}~\bibnamefont {Newman}}, \bibinfo {author} {\bibfnamefont
  {J.~H.}\ \bibnamefont {Ng}}, \bibinfo {author} {\bibfnamefont
  {A.}~\bibnamefont {Nguyen}}, \bibinfo {author} {\bibfnamefont
  {M.}~\bibnamefont {Nguyen}}, \bibinfo {author} {\bibfnamefont {M.~Y.}\
  \bibnamefont {Niu}}, \bibinfo {author} {\bibfnamefont {T.~E.}\ \bibnamefont
  {O’Brien}}, \bibinfo {author} {\bibfnamefont {S.}~\bibnamefont {Omonije}},
  \bibinfo {author} {\bibfnamefont {A.}~\bibnamefont {Opremcak}}, \bibinfo
  {author} {\bibfnamefont {R.}~\bibnamefont {Potter}}, \bibinfo {author}
  {\bibfnamefont {L.~P.}\ \bibnamefont {Pryadko}}, \bibinfo {author}
  {\bibfnamefont {C.}~\bibnamefont {Quintana}}, \bibinfo {author}
  {\bibfnamefont {D.~M.}\ \bibnamefont {Rhodes}}, \bibinfo {author}
  {\bibfnamefont {C.}~\bibnamefont {Rocque}}, \bibinfo {author} {\bibfnamefont
  {N.~C.}\ \bibnamefont {Rubin}}, \bibinfo {author} {\bibfnamefont
  {N.}~\bibnamefont {Saei}}, \bibinfo {author} {\bibfnamefont {D.}~\bibnamefont
  {Sank}}, \bibinfo {author} {\bibfnamefont {K.}~\bibnamefont
  {Sankaragomathi}}, \bibinfo {author} {\bibfnamefont {K.~J.}\ \bibnamefont
  {Satzinger}}, \bibinfo {author} {\bibfnamefont {H.~F.}\ \bibnamefont
  {Schurkus}}, \bibinfo {author} {\bibfnamefont {C.}~\bibnamefont {Schuster}},
  \bibinfo {author} {\bibfnamefont {M.~J.}\ \bibnamefont {Shearn}}, \bibinfo
  {author} {\bibfnamefont {A.}~\bibnamefont {Shorter}}, \bibinfo {author}
  {\bibfnamefont {N.}~\bibnamefont {Shutty}}, \bibinfo {author} {\bibfnamefont
  {V.}~\bibnamefont {Shvarts}}, \bibinfo {author} {\bibfnamefont
  {V.}~\bibnamefont {Sivak}}, \bibinfo {author} {\bibfnamefont
  {J.}~\bibnamefont {Skruzny}}, \bibinfo {author} {\bibfnamefont {W.~C.}\
  \bibnamefont {Smith}}, \bibinfo {author} {\bibfnamefont {R.~D.}\ \bibnamefont
  {Somma}}, \bibinfo {author} {\bibfnamefont {G.}~\bibnamefont {Sterling}},
  \bibinfo {author} {\bibfnamefont {D.}~\bibnamefont {Strain}}, \bibinfo
  {author} {\bibfnamefont {M.}~\bibnamefont {Szalay}}, \bibinfo {author}
  {\bibfnamefont {D.}~\bibnamefont {Thor}}, \bibinfo {author} {\bibfnamefont
  {A.}~\bibnamefont {Torres}}, \bibinfo {author} {\bibfnamefont
  {G.}~\bibnamefont {Vidal}}, \bibinfo {author} {\bibfnamefont
  {B.}~\bibnamefont {Villalonga}}, \bibinfo {author} {\bibfnamefont {C.~V.}\
  \bibnamefont {Heidweiller}}, \bibinfo {author} {\bibfnamefont
  {T.}~\bibnamefont {White}}, \bibinfo {author} {\bibfnamefont {B.~W.~K.}\
  \bibnamefont {Woo}}, \bibinfo {author} {\bibfnamefont {C.}~\bibnamefont
  {Xing}}, \bibinfo {author} {\bibfnamefont {Z.~J.}\ \bibnamefont {Yao}},
  \bibinfo {author} {\bibfnamefont {P.}~\bibnamefont {Yeh}}, \bibinfo {author}
  {\bibfnamefont {J.}~\bibnamefont {Yoo}}, \bibinfo {author} {\bibfnamefont
  {G.}~\bibnamefont {Young}}, \bibinfo {author} {\bibfnamefont
  {A.}~\bibnamefont {Zalcman}}, \bibinfo {author} {\bibfnamefont
  {Y.}~\bibnamefont {Zhang}}, \bibinfo {author} {\bibfnamefont
  {N.}~\bibnamefont {Zhu}}, \bibinfo {author} {\bibfnamefont {N.}~\bibnamefont
  {Zobrist}}, \bibinfo {author} {\bibfnamefont {H.}~\bibnamefont {Neven}},
  \bibinfo {author} {\bibfnamefont {R.}~\bibnamefont {Babbush}}, \bibinfo
  {author} {\bibfnamefont {D.}~\bibnamefont {Bacon}}, \bibinfo {author}
  {\bibfnamefont {S.}~\bibnamefont {Boixo}}, \bibinfo {author} {\bibfnamefont
  {J.}~\bibnamefont {Hilton}}, \bibinfo {author} {\bibfnamefont
  {E.}~\bibnamefont {Lucero}}, \bibinfo {author} {\bibfnamefont
  {A.}~\bibnamefont {Megrant}}, \bibinfo {author} {\bibfnamefont
  {J.}~\bibnamefont {Kelly}}, \bibinfo {author} {\bibfnamefont
  {Y.}~\bibnamefont {Chen}}, \bibinfo {author} {\bibfnamefont {V.}~\bibnamefont
  {Smelyanskiy}}, \bibinfo {author} {\bibfnamefont {V.}~\bibnamefont
  {Khemani}}, \bibinfo {author} {\bibfnamefont {S.}~\bibnamefont
  {Gopalakrishnan}}, \bibinfo {author} {\bibfnamefont {T.}~\bibnamefont
  {Prosen}},\ and\ \bibinfo {author} {\bibfnamefont {P.}~\bibnamefont
  {Roushan}},\ }\bibfield  {title} {\bibinfo {title} {Dynamics of magnetization
  at infinite temperature in a {Heisenberg} spin chain},\ }\href@noop {}
  {\bibfield  {journal} {\bibinfo  {journal} {Science}\ }\textbf {\bibinfo
  {volume} {384}},\ \bibinfo {pages} {48} (\bibinfo {year} {2024})}\BibitemShut
  {NoStop}%
\bibitem [{\citenamefont {Chung}\ \emph {et~al.}(2021)\citenamefont {Chung},
  \citenamefont {de~Hond}, \citenamefont {Xiang}, \citenamefont
  {Cruz-Col{\'o}n},\ and\ \citenamefont {Ketterle}}]{Chung2021}%
  \BibitemOpen
  \bibfield  {author} {\bibinfo {author} {\bibfnamefont {W.~C.}\ \bibnamefont
  {Chung}}, \bibinfo {author} {\bibfnamefont {J.}~\bibnamefont {de~Hond}},
  \bibinfo {author} {\bibfnamefont {J.}~\bibnamefont {Xiang}}, \bibinfo
  {author} {\bibfnamefont {E.}~\bibnamefont {Cruz-Col{\'o}n}},\ and\ \bibinfo
  {author} {\bibfnamefont {W.}~\bibnamefont {Ketterle}},\ }\bibfield  {title}
  {\bibinfo {title} {Tunable single-ion anisotropy in spin-1 models realized
  with ultracold atoms},\ }\href@noop {} {\bibfield  {journal} {\bibinfo
  {journal} {Phys. Rev. Lett.}\ }\textbf {\bibinfo {volume} {126}},\ \bibinfo
  {pages} {163203} (\bibinfo {year} {2021})}\BibitemShut {NoStop}%
\bibitem [{\citenamefont {Luo}\ \emph {et~al.}(2024)\citenamefont {Luo},
  \citenamefont {Zheng}, \citenamefont {Zhang}, \citenamefont {He},
  \citenamefont {Shen}, \citenamefont {Zhu}, \citenamefont {Yuan},\ and\
  \citenamefont {Pan}}]{Luo2024}%
  \BibitemOpen
  \bibfield  {author} {\bibinfo {author} {\bibfnamefont {A.}~\bibnamefont
  {Luo}}, \bibinfo {author} {\bibfnamefont {Y.-G.}\ \bibnamefont {Zheng}},
  \bibinfo {author} {\bibfnamefont {W.-Y.}\ \bibnamefont {Zhang}}, \bibinfo
  {author} {\bibfnamefont {M.-G.}\ \bibnamefont {He}}, \bibinfo {author}
  {\bibfnamefont {Y.-C.}\ \bibnamefont {Shen}}, \bibinfo {author}
  {\bibfnamefont {Z.-H.}\ \bibnamefont {Zhu}}, \bibinfo {author} {\bibfnamefont
  {Z.-S.}\ \bibnamefont {Yuan}},\ and\ \bibinfo {author} {\bibfnamefont
  {J.-W.}\ \bibnamefont {Pan}},\ }\bibfield  {title} {\bibinfo {title}
  {Microscopic study on superexchange dynamics of composite spin-1 bosons},\
  }\href@noop {} {\bibfield  {journal} {\bibinfo  {journal} {Phys. Rev. Lett.}\
  }\textbf {\bibinfo {volume} {133}},\ \bibinfo {pages} {043401} (\bibinfo
  {year} {2024})}\BibitemShut {NoStop}%
\bibitem [{\citenamefont {Labuhn}\ \emph {et~al.}(2016)\citenamefont {Labuhn},
  \citenamefont {Barredo}, \citenamefont {Ravets}, \citenamefont
  {de~L{\'e}s{\'e}leuc}, \citenamefont {Macr{\`\i}}, \citenamefont {Lahaye},\
  and\ \citenamefont {Browaeys}}]{Labuhn2016}%
  \BibitemOpen
  \bibfield  {author} {\bibinfo {author} {\bibfnamefont {H.}~\bibnamefont
  {Labuhn}}, \bibinfo {author} {\bibfnamefont {D.}~\bibnamefont {Barredo}},
  \bibinfo {author} {\bibfnamefont {S.}~\bibnamefont {Ravets}}, \bibinfo
  {author} {\bibfnamefont {S.}~\bibnamefont {de~L{\'e}s{\'e}leuc}}, \bibinfo
  {author} {\bibfnamefont {T.}~\bibnamefont {Macr{\`\i}}}, \bibinfo {author}
  {\bibfnamefont {T.}~\bibnamefont {Lahaye}},\ and\ \bibinfo {author}
  {\bibfnamefont {A.}~\bibnamefont {Browaeys}},\ }\bibfield  {title} {\bibinfo
  {title} {Tunable two-dimensional arrays of single {Rydberg} atoms for
  realizing quantum {Ising} models},\ }\href@noop {} {\bibfield  {journal}
  {\bibinfo  {journal} {Nature}\ }\textbf {\bibinfo {volume} {534}},\ \bibinfo
  {pages} {667} (\bibinfo {year} {2016})}\BibitemShut {NoStop}%
\bibitem [{\citenamefont {Zeiher}\ \emph {et~al.}(2017)\citenamefont {Zeiher},
  \citenamefont {Choi}, \citenamefont {Rubio-Abadal}, \citenamefont {Pohl},
  \citenamefont {Van~Bijnen}, \citenamefont {Bloch},\ and\ \citenamefont
  {Gross}}]{Zeiher2017}%
  \BibitemOpen
  \bibfield  {author} {\bibinfo {author} {\bibfnamefont {J.}~\bibnamefont
  {Zeiher}}, \bibinfo {author} {\bibfnamefont {J.-y.}\ \bibnamefont {Choi}},
  \bibinfo {author} {\bibfnamefont {A.}~\bibnamefont {Rubio-Abadal}}, \bibinfo
  {author} {\bibfnamefont {T.}~\bibnamefont {Pohl}}, \bibinfo {author}
  {\bibfnamefont {R.}~\bibnamefont {Van~Bijnen}}, \bibinfo {author}
  {\bibfnamefont {I.}~\bibnamefont {Bloch}},\ and\ \bibinfo {author}
  {\bibfnamefont {C.}~\bibnamefont {Gross}},\ }\bibfield  {title} {\bibinfo
  {title} {Coherent many-body spin dynamics in a long-range interacting {Ising}
  chain},\ }\href@noop {} {\bibfield  {journal} {\bibinfo  {journal} {Phys.
  Rev. X}\ }\textbf {\bibinfo {volume} {7}},\ \bibinfo {pages} {041063}
  (\bibinfo {year} {2017})}\BibitemShut {NoStop}%
\bibitem [{\citenamefont {Bernien}\ \emph {et~al.}(2017)\citenamefont
  {Bernien}, \citenamefont {Schwartz}, \citenamefont {Keesling}, \citenamefont
  {Levine}, \citenamefont {Omran}, \citenamefont {Pichler}, \citenamefont
  {Choi}, \citenamefont {Zibrov}, \citenamefont {Endres}, \citenamefont
  {Greiner}, \citenamefont {Vuleti\'c},\ and\ \citenamefont
  {Lukin}}]{Bernien2017}%
  \BibitemOpen
  \bibfield  {author} {\bibinfo {author} {\bibfnamefont {H.}~\bibnamefont
  {Bernien}}, \bibinfo {author} {\bibfnamefont {S.}~\bibnamefont {Schwartz}},
  \bibinfo {author} {\bibfnamefont {A.}~\bibnamefont {Keesling}}, \bibinfo
  {author} {\bibfnamefont {H.}~\bibnamefont {Levine}}, \bibinfo {author}
  {\bibfnamefont {A.}~\bibnamefont {Omran}}, \bibinfo {author} {\bibfnamefont
  {H.}~\bibnamefont {Pichler}}, \bibinfo {author} {\bibfnamefont
  {S.}~\bibnamefont {Choi}}, \bibinfo {author} {\bibfnamefont {A.~S.}\
  \bibnamefont {Zibrov}}, \bibinfo {author} {\bibfnamefont {M.}~\bibnamefont
  {Endres}}, \bibinfo {author} {\bibfnamefont {M.}~\bibnamefont {Greiner}},
  \bibinfo {author} {\bibfnamefont {V.}~\bibnamefont {Vuleti\'c}},\ and\
  \bibinfo {author} {\bibfnamefont {M.~D.}\ \bibnamefont {Lukin}},\ }\bibfield
  {title} {\bibinfo {title} {Probing many-body dynamics on a 51-atom quantum
  simulator},\ }\href@noop {} {\bibfield  {journal} {\bibinfo  {journal}
  {Nature}\ }\textbf {\bibinfo {volume} {551}},\ \bibinfo {pages} {579}
  (\bibinfo {year} {2017})}\BibitemShut {NoStop}%
\bibitem [{\citenamefont {de~L{\'e}s{\'e}leuc}\ \emph
  {et~al.}(2018)\citenamefont {de~L{\'e}s{\'e}leuc}, \citenamefont {Weber},
  \citenamefont {Lienhard}, \citenamefont {Barredo}, \citenamefont
  {B{\"u}chler}, \citenamefont {Lahaye},\ and\ \citenamefont
  {Browaeys}}]{Leseleuc2018}%
  \BibitemOpen
  \bibfield  {author} {\bibinfo {author} {\bibfnamefont {S.}~\bibnamefont
  {de~L{\'e}s{\'e}leuc}}, \bibinfo {author} {\bibfnamefont {S.}~\bibnamefont
  {Weber}}, \bibinfo {author} {\bibfnamefont {V.}~\bibnamefont {Lienhard}},
  \bibinfo {author} {\bibfnamefont {D.}~\bibnamefont {Barredo}}, \bibinfo
  {author} {\bibfnamefont {H.~P.}\ \bibnamefont {B{\"u}chler}}, \bibinfo
  {author} {\bibfnamefont {T.}~\bibnamefont {Lahaye}},\ and\ \bibinfo {author}
  {\bibfnamefont {A.}~\bibnamefont {Browaeys}},\ }\bibfield  {title} {\bibinfo
  {title} {Accurate mapping of multilevel {Rydberg} atoms on interacting
  spin-1/2 particles for the quantum simulation of {Ising} models},\
  }\href@noop {} {\bibfield  {journal} {\bibinfo  {journal} {Phys. Rev. Lett.}\
  }\textbf {\bibinfo {volume} {120}},\ \bibinfo {pages} {113602} (\bibinfo
  {year} {2018})}\BibitemShut {NoStop}%
\bibitem [{\citenamefont {Lienhard}\ \emph {et~al.}(2018)\citenamefont
  {Lienhard}, \citenamefont {de~L{\'e}s{\'e}leuc}, \citenamefont {Barredo},
  \citenamefont {Lahaye}, \citenamefont {Browaeys}, \citenamefont {Schuler},
  \citenamefont {Henry},\ and\ \citenamefont {L{\"a}uchli}}]{Lienhard2018}%
  \BibitemOpen
  \bibfield  {author} {\bibinfo {author} {\bibfnamefont {V.}~\bibnamefont
  {Lienhard}}, \bibinfo {author} {\bibfnamefont {S.}~\bibnamefont
  {de~L{\'e}s{\'e}leuc}}, \bibinfo {author} {\bibfnamefont {D.}~\bibnamefont
  {Barredo}}, \bibinfo {author} {\bibfnamefont {T.}~\bibnamefont {Lahaye}},
  \bibinfo {author} {\bibfnamefont {A.}~\bibnamefont {Browaeys}}, \bibinfo
  {author} {\bibfnamefont {M.}~\bibnamefont {Schuler}}, \bibinfo {author}
  {\bibfnamefont {L.-P.}\ \bibnamefont {Henry}},\ and\ \bibinfo {author}
  {\bibfnamefont {A.~M.}\ \bibnamefont {L{\"a}uchli}},\ }\bibfield  {title}
  {\bibinfo {title} {Observing the space-and time-dependent growth of
  correlations in dynamically tuned synthetic {Ising} models with
  antiferromagnetic interactions},\ }\href@noop {} {\bibfield  {journal}
  {\bibinfo  {journal} {Phys. Rev. X}\ }\textbf {\bibinfo {volume} {8}},\
  \bibinfo {pages} {021070} (\bibinfo {year} {2018})}\BibitemShut {NoStop}%
\bibitem [{\citenamefont {Guardado-Sanchez}\ \emph {et~al.}(2018)\citenamefont
  {Guardado-Sanchez}, \citenamefont {Brown}, \citenamefont {Mitra},
  \citenamefont {Devakul}, \citenamefont {Huse}, \citenamefont {Schau{\ss}},\
  and\ \citenamefont {Bakr}}]{Guardado2018}%
  \BibitemOpen
  \bibfield  {author} {\bibinfo {author} {\bibfnamefont {E.}~\bibnamefont
  {Guardado-Sanchez}}, \bibinfo {author} {\bibfnamefont {P.~T.}\ \bibnamefont
  {Brown}}, \bibinfo {author} {\bibfnamefont {D.}~\bibnamefont {Mitra}},
  \bibinfo {author} {\bibfnamefont {T.}~\bibnamefont {Devakul}}, \bibinfo
  {author} {\bibfnamefont {D.~A.}\ \bibnamefont {Huse}}, \bibinfo {author}
  {\bibfnamefont {P.}~\bibnamefont {Schau{\ss}}},\ and\ \bibinfo {author}
  {\bibfnamefont {W.~S.}\ \bibnamefont {Bakr}},\ }\bibfield  {title} {\bibinfo
  {title} {Probing the quench dynamics of antiferromagnetic correlations in a
  {2D} quantum {Ising} spin system},\ }\href@noop {} {\bibfield  {journal}
  {\bibinfo  {journal} {Phys. Rev. X}\ }\textbf {\bibinfo {volume} {8}},\
  \bibinfo {pages} {021069} (\bibinfo {year} {2018})}\BibitemShut {NoStop}%
\bibitem [{\citenamefont {Keesling}\ \emph {et~al.}(2019)\citenamefont
  {Keesling}, \citenamefont {Omran}, \citenamefont {Levine}, \citenamefont
  {Bernien}, \citenamefont {Pichler}, \citenamefont {Choi}, \citenamefont
  {Samajdar}, \citenamefont {Schwartz}, \citenamefont {Silvi}, \citenamefont
  {Sachdev}, \citenamefont {Zoller}, \citenamefont {Endres}, \citenamefont
  {Greiner}, \citenamefont {Vuleti\'c},\ and\ \citenamefont
  {Lukin}}]{Keesling2019}%
  \BibitemOpen
  \bibfield  {author} {\bibinfo {author} {\bibfnamefont {A.}~\bibnamefont
  {Keesling}}, \bibinfo {author} {\bibfnamefont {A.}~\bibnamefont {Omran}},
  \bibinfo {author} {\bibfnamefont {H.}~\bibnamefont {Levine}}, \bibinfo
  {author} {\bibfnamefont {H.}~\bibnamefont {Bernien}}, \bibinfo {author}
  {\bibfnamefont {H.}~\bibnamefont {Pichler}}, \bibinfo {author} {\bibfnamefont
  {S.}~\bibnamefont {Choi}}, \bibinfo {author} {\bibfnamefont {R.}~\bibnamefont
  {Samajdar}}, \bibinfo {author} {\bibfnamefont {S.}~\bibnamefont {Schwartz}},
  \bibinfo {author} {\bibfnamefont {P.}~\bibnamefont {Silvi}}, \bibinfo
  {author} {\bibfnamefont {S.}~\bibnamefont {Sachdev}}, \bibinfo {author}
  {\bibfnamefont {P.}~\bibnamefont {Zoller}}, \bibinfo {author} {\bibfnamefont
  {M.}~\bibnamefont {Endres}}, \bibinfo {author} {\bibfnamefont
  {M.}~\bibnamefont {Greiner}}, \bibinfo {author} {\bibfnamefont
  {V.}~\bibnamefont {Vuleti\'c}},\ and\ \bibinfo {author} {\bibfnamefont
  {M.~D.}\ \bibnamefont {Lukin}},\ }\bibfield  {title} {\bibinfo {title}
  {Quantum {Kibble--Zurek} mechanism and critical dynamics on a programmable
  {Rydberg} simulator},\ }\href@noop {} {\bibfield  {journal} {\bibinfo
  {journal} {Nature}\ }\textbf {\bibinfo {volume} {568}},\ \bibinfo {pages}
  {207} (\bibinfo {year} {2019})}\BibitemShut {NoStop}%
\bibitem [{\citenamefont {Ebadi}\ \emph {et~al.}(2021)\citenamefont {Ebadi},
  \citenamefont {Wang}, \citenamefont {Levine}, \citenamefont {Keesling},
  \citenamefont {Semeghini}, \citenamefont {Omran}, \citenamefont {Bluvstein},
  \citenamefont {Samajdar}, \citenamefont {Pichler}, \citenamefont {Ho},
  \citenamefont {Choi}, \citenamefont {Sachdev}, \citenamefont {Greiner},
  \citenamefont {Vuleti\'c},\ and\ \citenamefont {Lukin}}]{Ebadi2021}%
  \BibitemOpen
  \bibfield  {author} {\bibinfo {author} {\bibfnamefont {S.}~\bibnamefont
  {Ebadi}}, \bibinfo {author} {\bibfnamefont {T.~T.}\ \bibnamefont {Wang}},
  \bibinfo {author} {\bibfnamefont {H.}~\bibnamefont {Levine}}, \bibinfo
  {author} {\bibfnamefont {A.}~\bibnamefont {Keesling}}, \bibinfo {author}
  {\bibfnamefont {G.}~\bibnamefont {Semeghini}}, \bibinfo {author}
  {\bibfnamefont {A.}~\bibnamefont {Omran}}, \bibinfo {author} {\bibfnamefont
  {D.}~\bibnamefont {Bluvstein}}, \bibinfo {author} {\bibfnamefont
  {R.}~\bibnamefont {Samajdar}}, \bibinfo {author} {\bibfnamefont
  {H.}~\bibnamefont {Pichler}}, \bibinfo {author} {\bibfnamefont {W.~W.}\
  \bibnamefont {Ho}}, \bibinfo {author} {\bibfnamefont {S.}~\bibnamefont
  {Choi}}, \bibinfo {author} {\bibfnamefont {S.}~\bibnamefont {Sachdev}},
  \bibinfo {author} {\bibfnamefont {M.}~\bibnamefont {Greiner}}, \bibinfo
  {author} {\bibfnamefont {V.}~\bibnamefont {Vuleti\'c}},\ and\ \bibinfo
  {author} {\bibfnamefont {M.~D.}\ \bibnamefont {Lukin}},\ }\bibfield  {title}
  {\bibinfo {title} {Quantum phases of matter on a 256-atom programmable
  quantum simulator},\ }\href@noop {} {\bibfield  {journal} {\bibinfo
  {journal} {Nature}\ }\textbf {\bibinfo {volume} {595}},\ \bibinfo {pages}
  {227} (\bibinfo {year} {2021})}\BibitemShut {NoStop}%
\bibitem [{\citenamefont {Semeghini}\ \emph {et~al.}(2021)\citenamefont
  {Semeghini}, \citenamefont {Levine}, \citenamefont {Keesling}, \citenamefont
  {Ebadi}, \citenamefont {Wang}, \citenamefont {Bluvstein}, \citenamefont
  {Verresen}, \citenamefont {Pichler}, \citenamefont {Kalinowski},
  \citenamefont {Samajdar}, \citenamefont {Omran}, \citenamefont {Sachdev},
  \citenamefont {Vishwanath}, \citenamefont {Greiner}, \citenamefont
  {Vuleti\'c},\ and\ \citenamefont {Lukin}}]{Semeghini2021}%
  \BibitemOpen
  \bibfield  {author} {\bibinfo {author} {\bibfnamefont {G.}~\bibnamefont
  {Semeghini}}, \bibinfo {author} {\bibfnamefont {H.}~\bibnamefont {Levine}},
  \bibinfo {author} {\bibfnamefont {A.}~\bibnamefont {Keesling}}, \bibinfo
  {author} {\bibfnamefont {S.}~\bibnamefont {Ebadi}}, \bibinfo {author}
  {\bibfnamefont {T.~T.}\ \bibnamefont {Wang}}, \bibinfo {author}
  {\bibfnamefont {D.}~\bibnamefont {Bluvstein}}, \bibinfo {author}
  {\bibfnamefont {R.}~\bibnamefont {Verresen}}, \bibinfo {author}
  {\bibfnamefont {H.}~\bibnamefont {Pichler}}, \bibinfo {author} {\bibfnamefont
  {M.}~\bibnamefont {Kalinowski}}, \bibinfo {author} {\bibfnamefont
  {R.}~\bibnamefont {Samajdar}}, \bibinfo {author} {\bibfnamefont
  {A.}~\bibnamefont {Omran}}, \bibinfo {author} {\bibfnamefont
  {S.}~\bibnamefont {Sachdev}}, \bibinfo {author} {\bibfnamefont
  {A.}~\bibnamefont {Vishwanath}}, \bibinfo {author} {\bibfnamefont
  {M.}~\bibnamefont {Greiner}}, \bibinfo {author} {\bibfnamefont
  {V.}~\bibnamefont {Vuleti\'c}},\ and\ \bibinfo {author} {\bibfnamefont
  {M.~D.}\ \bibnamefont {Lukin}},\ }\bibfield  {title} {\bibinfo {title}
  {Probing topological spin liquids on a programmable quantum simulator},\
  }\href@noop {} {\bibfield  {journal} {\bibinfo  {journal} {Science}\ }\textbf
  {\bibinfo {volume} {374}},\ \bibinfo {pages} {1242} (\bibinfo {year}
  {2021})}\BibitemShut {NoStop}%
\bibitem [{\citenamefont {Bluvstein}\ \emph {et~al.}(2021)\citenamefont
  {Bluvstein}, \citenamefont {Omran}, \citenamefont {Levine}, \citenamefont
  {Keesling}, \citenamefont {Semeghini}, \citenamefont {Ebadi}, \citenamefont
  {Wang}, \citenamefont {Michailidis}, \citenamefont {Maskara}, \citenamefont
  {Ho}, \citenamefont {Choi}, \citenamefont {Serbyn}, \citenamefont {Greiner},
  \citenamefont {Vuleti\'c},\ and\ \citenamefont {Lukin}}]{Bluvstein2021}%
  \BibitemOpen
  \bibfield  {author} {\bibinfo {author} {\bibfnamefont {D.}~\bibnamefont
  {Bluvstein}}, \bibinfo {author} {\bibfnamefont {A.}~\bibnamefont {Omran}},
  \bibinfo {author} {\bibfnamefont {H.}~\bibnamefont {Levine}}, \bibinfo
  {author} {\bibfnamefont {A.}~\bibnamefont {Keesling}}, \bibinfo {author}
  {\bibfnamefont {G.}~\bibnamefont {Semeghini}}, \bibinfo {author}
  {\bibfnamefont {S.}~\bibnamefont {Ebadi}}, \bibinfo {author} {\bibfnamefont
  {T.~T.}\ \bibnamefont {Wang}}, \bibinfo {author} {\bibfnamefont {A.~A.}\
  \bibnamefont {Michailidis}}, \bibinfo {author} {\bibfnamefont
  {N.}~\bibnamefont {Maskara}}, \bibinfo {author} {\bibfnamefont {W.~W.}\
  \bibnamefont {Ho}}, \bibinfo {author} {\bibfnamefont {S.}~\bibnamefont
  {Choi}}, \bibinfo {author} {\bibfnamefont {M.}~\bibnamefont {Serbyn}},
  \bibinfo {author} {\bibfnamefont {M.}~\bibnamefont {Greiner}}, \bibinfo
  {author} {\bibfnamefont {V.}~\bibnamefont {Vuleti\'c}},\ and\ \bibinfo
  {author} {\bibfnamefont {M.~D.}\ \bibnamefont {Lukin}},\ }\bibfield  {title}
  {\bibinfo {title} {Controlling quantum many-body dynamics in driven {Rydberg}
  atom arrays},\ }\href@noop {} {\bibfield  {journal} {\bibinfo  {journal}
  {Science}\ }\textbf {\bibinfo {volume} {371}},\ \bibinfo {pages} {1355}
  (\bibinfo {year} {2021})}\BibitemShut {NoStop}%
\bibitem [{\citenamefont {Scholl}\ \emph {et~al.}(2021)\citenamefont {Scholl},
  \citenamefont {Schuler}, \citenamefont {Williams}, \citenamefont
  {Eberharter}, \citenamefont {Barredo}, \citenamefont {Schymik}, \citenamefont
  {Lienhard}, \citenamefont {Henry}, \citenamefont {Lang}, \citenamefont
  {Lahaye}, \citenamefont {L\"auchli},\ and\ \citenamefont
  {Browaeys}}]{Scholl2021}%
  \BibitemOpen
  \bibfield  {author} {\bibinfo {author} {\bibfnamefont {P.}~\bibnamefont
  {Scholl}}, \bibinfo {author} {\bibfnamefont {M.}~\bibnamefont {Schuler}},
  \bibinfo {author} {\bibfnamefont {H.~J.}\ \bibnamefont {Williams}}, \bibinfo
  {author} {\bibfnamefont {A.~A.}\ \bibnamefont {Eberharter}}, \bibinfo
  {author} {\bibfnamefont {D.}~\bibnamefont {Barredo}}, \bibinfo {author}
  {\bibfnamefont {K.-N.}\ \bibnamefont {Schymik}}, \bibinfo {author}
  {\bibfnamefont {V.}~\bibnamefont {Lienhard}}, \bibinfo {author}
  {\bibfnamefont {L.-P.}\ \bibnamefont {Henry}}, \bibinfo {author}
  {\bibfnamefont {T.~C.}\ \bibnamefont {Lang}}, \bibinfo {author}
  {\bibfnamefont {T.}~\bibnamefont {Lahaye}}, \bibinfo {author} {\bibfnamefont
  {A.~M.}\ \bibnamefont {L\"auchli}},\ and\ \bibinfo {author} {\bibfnamefont
  {A.}~\bibnamefont {Browaeys}},\ }\bibfield  {title} {\bibinfo {title}
  {Quantum simulation of {2D} antiferromagnets with hundreds of {Rydberg}
  atoms},\ }\href@noop {} {\bibfield  {journal} {\bibinfo  {journal} {Nature}\
  }\textbf {\bibinfo {volume} {595}},\ \bibinfo {pages} {233} (\bibinfo {year}
  {2021})}\BibitemShut {NoStop}%
\bibitem [{\citenamefont {Hollerith}\ \emph {et~al.}(2022)\citenamefont
  {Hollerith}, \citenamefont {Srakaew}, \citenamefont {Wei}, \citenamefont
  {Rubio-Abadal}, \citenamefont {Adler}, \citenamefont {Weckesser},
  \citenamefont {Kruckenhauser}, \citenamefont {Walther}, \citenamefont {van
  Bijnen}, \citenamefont {Rui}, \citenamefont {Gross}, \citenamefont {Bloch},\
  and\ \citenamefont {Zeiher}}]{Hollerith2022}%
  \BibitemOpen
  \bibfield  {author} {\bibinfo {author} {\bibfnamefont {S.}~\bibnamefont
  {Hollerith}}, \bibinfo {author} {\bibfnamefont {K.}~\bibnamefont {Srakaew}},
  \bibinfo {author} {\bibfnamefont {D.}~\bibnamefont {Wei}}, \bibinfo {author}
  {\bibfnamefont {A.}~\bibnamefont {Rubio-Abadal}}, \bibinfo {author}
  {\bibfnamefont {D.}~\bibnamefont {Adler}}, \bibinfo {author} {\bibfnamefont
  {P.}~\bibnamefont {Weckesser}}, \bibinfo {author} {\bibfnamefont
  {A.}~\bibnamefont {Kruckenhauser}}, \bibinfo {author} {\bibfnamefont
  {V.}~\bibnamefont {Walther}}, \bibinfo {author} {\bibfnamefont
  {R.}~\bibnamefont {van Bijnen}}, \bibinfo {author} {\bibfnamefont
  {J.}~\bibnamefont {Rui}}, \bibinfo {author} {\bibfnamefont {C.}~\bibnamefont
  {Gross}}, \bibinfo {author} {\bibfnamefont {I.}~\bibnamefont {Bloch}},\ and\
  \bibinfo {author} {\bibfnamefont {J.}~\bibnamefont {Zeiher}},\ }\bibfield
  {title} {\bibinfo {title} {Realizing distance-selective interactions in a
  {Rydberg-dressed} atom array},\ }\href@noop {} {\bibfield  {journal}
  {\bibinfo  {journal} {Phys. Rev. Lett.}\ }\textbf {\bibinfo {volume} {128}},\
  \bibinfo {pages} {113602} (\bibinfo {year} {2022})}\BibitemShut {NoStop}%
\bibitem [{\citenamefont {Zhao}\ \emph {et~al.}(2023)\citenamefont {Zhao},
  \citenamefont {Lee}, \citenamefont {Aliyu},\ and\ \citenamefont
  {Loh}}]{Zhao2023}%
  \BibitemOpen
  \bibfield  {author} {\bibinfo {author} {\bibfnamefont {L.}~\bibnamefont
  {Zhao}}, \bibinfo {author} {\bibfnamefont {M.~D.~K.}\ \bibnamefont {Lee}},
  \bibinfo {author} {\bibfnamefont {M.~M.}\ \bibnamefont {Aliyu}},\ and\
  \bibinfo {author} {\bibfnamefont {H.}~\bibnamefont {Loh}},\ }\bibfield
  {title} {\bibinfo {title} {Floquet-tailored {Rydberg} interactions},\
  }\href@noop {} {\bibfield  {journal} {\bibinfo  {journal} {Nat. Commun.}\
  }\textbf {\bibinfo {volume} {14}},\ \bibinfo {pages} {7128} (\bibinfo {year}
  {2023})}\BibitemShut {NoStop}%
\bibitem [{\citenamefont {Bharti}\ \emph {et~al.}(2023)\citenamefont {Bharti},
  \citenamefont {Sugawa}, \citenamefont {Mizoguchi}, \citenamefont {Kunimi},
  \citenamefont {Zhang}, \citenamefont {de~L{\'e}s{\'e}leuc}, \citenamefont
  {Tomita}, \citenamefont {Franz}, \citenamefont {Weidem{\"u}ller},\ and\
  \citenamefont {Ohmori}}]{Bharti2023}%
  \BibitemOpen
  \bibfield  {author} {\bibinfo {author} {\bibfnamefont {V.}~\bibnamefont
  {Bharti}}, \bibinfo {author} {\bibfnamefont {S.}~\bibnamefont {Sugawa}},
  \bibinfo {author} {\bibfnamefont {M.}~\bibnamefont {Mizoguchi}}, \bibinfo
  {author} {\bibfnamefont {M.}~\bibnamefont {Kunimi}}, \bibinfo {author}
  {\bibfnamefont {Y.}~\bibnamefont {Zhang}}, \bibinfo {author} {\bibfnamefont
  {S.}~\bibnamefont {de~L{\'e}s{\'e}leuc}}, \bibinfo {author} {\bibfnamefont
  {T.}~\bibnamefont {Tomita}}, \bibinfo {author} {\bibfnamefont
  {T.}~\bibnamefont {Franz}}, \bibinfo {author} {\bibfnamefont
  {M.}~\bibnamefont {Weidem{\"u}ller}},\ and\ \bibinfo {author} {\bibfnamefont
  {K.}~\bibnamefont {Ohmori}},\ }\bibfield  {title} {\bibinfo {title}
  {Picosecond-scale ultrafast many-body dynamics in an ultracold
  {Rydberg-excited} atomic {Mott} insulator},\ }\href@noop {} {\bibfield
  {journal} {\bibinfo  {journal} {Phys. Rev. Lett.}\ }\textbf {\bibinfo
  {volume} {131}},\ \bibinfo {pages} {123201} (\bibinfo {year}
  {2023})}\BibitemShut {NoStop}%
\bibitem [{\citenamefont {Franz}\ \emph {et~al.}(2024)\citenamefont {Franz},
  \citenamefont {Geier}, \citenamefont {Hainaut}, \citenamefont {Braemer},
  \citenamefont {Thaicharoen}, \citenamefont {Hornung}, \citenamefont {Braun},
  \citenamefont {G{\"a}rttner}, \citenamefont {Z{\"u}rn},\ and\ \citenamefont
  {Weidem{\"u}ller}}]{Franz2024}%
  \BibitemOpen
  \bibfield  {author} {\bibinfo {author} {\bibfnamefont {T.}~\bibnamefont
  {Franz}}, \bibinfo {author} {\bibfnamefont {S.}~\bibnamefont {Geier}},
  \bibinfo {author} {\bibfnamefont {C.}~\bibnamefont {Hainaut}}, \bibinfo
  {author} {\bibfnamefont {A.}~\bibnamefont {Braemer}}, \bibinfo {author}
  {\bibfnamefont {N.}~\bibnamefont {Thaicharoen}}, \bibinfo {author}
  {\bibfnamefont {M.}~\bibnamefont {Hornung}}, \bibinfo {author} {\bibfnamefont
  {E.}~\bibnamefont {Braun}}, \bibinfo {author} {\bibfnamefont
  {M.}~\bibnamefont {G{\"a}rttner}}, \bibinfo {author} {\bibfnamefont
  {G.}~\bibnamefont {Z{\"u}rn}},\ and\ \bibinfo {author} {\bibfnamefont
  {M.}~\bibnamefont {Weidem{\"u}ller}},\ }\bibfield  {title} {\bibinfo {title}
  {Observation of anisotropy-independent magnetization dynamics in spatially
  disordered {Heisenberg} spin systems},\ }\href@noop {} {\bibfield  {journal}
  {\bibinfo  {journal} {Phys. Rev. Res.}\ }\textbf {\bibinfo {volume} {6}},\
  \bibinfo {pages} {033131} (\bibinfo {year} {2024})}\BibitemShut {NoStop}%
\bibitem [{\citenamefont {Bharti}\ \emph {et~al.}(2024)\citenamefont {Bharti},
  \citenamefont {Sugawa}, \citenamefont {Kunimi}, \citenamefont {Chauhan},
  \citenamefont {Mahesh}, \citenamefont {Mizoguchi}, \citenamefont {Matsubara},
  \citenamefont {Tomita}, \citenamefont {de~L\'es\'eleuc},\ and\ \citenamefont
  {Ohmori}}]{Bharti2024}%
  \BibitemOpen
  \bibfield  {author} {\bibinfo {author} {\bibfnamefont {V.}~\bibnamefont
  {Bharti}}, \bibinfo {author} {\bibfnamefont {S.}~\bibnamefont {Sugawa}},
  \bibinfo {author} {\bibfnamefont {M.}~\bibnamefont {Kunimi}}, \bibinfo
  {author} {\bibfnamefont {V.}~\bibnamefont {Chauhan}}, \bibinfo {author}
  {\bibfnamefont {T.}~\bibnamefont {Mahesh}}, \bibinfo {author} {\bibfnamefont
  {M.}~\bibnamefont {Mizoguchi}}, \bibinfo {author} {\bibfnamefont
  {T.}~\bibnamefont {Matsubara}}, \bibinfo {author} {\bibfnamefont
  {T.}~\bibnamefont {Tomita}}, \bibinfo {author} {\bibfnamefont
  {S.}~\bibnamefont {de~L\'es\'eleuc}},\ and\ \bibinfo {author} {\bibfnamefont
  {K.}~\bibnamefont {Ohmori}},\ }\bibfield  {title} {\bibinfo {title} {Strong
  spin-motion coupling in the ultrafast dynamics of {Rydberg} atoms},\
  }\href@noop {} {\bibfield  {journal} {\bibinfo  {journal} {Phys. Rev. Lett.}\
  }\textbf {\bibinfo {volume} {133}},\ \bibinfo {pages} {093405} (\bibinfo
  {year} {2024})}\BibitemShut {NoStop}%
\bibitem [{\citenamefont {Zhao}\ \emph {et~al.}(2025)\citenamefont {Zhao},
  \citenamefont {Datla}, \citenamefont {Tian}, \citenamefont {Aliyu},\ and\
  \citenamefont {Loh}}]{Zhao2025}%
  \BibitemOpen
  \bibfield  {author} {\bibinfo {author} {\bibfnamefont {L.}~\bibnamefont
  {Zhao}}, \bibinfo {author} {\bibfnamefont {P.~R.}\ \bibnamefont {Datla}},
  \bibinfo {author} {\bibfnamefont {W.}~\bibnamefont {Tian}}, \bibinfo {author}
  {\bibfnamefont {M.~M.}\ \bibnamefont {Aliyu}},\ and\ \bibinfo {author}
  {\bibfnamefont {H.}~\bibnamefont {Loh}},\ }\bibfield  {title} {\bibinfo
  {title} {Observation of quantum thermalization restricted to {Hilbert} space
  fragments and $\mathbb{Z}_{2k}$ scars},\ }\href@noop {} {\bibfield  {journal}
  {\bibinfo  {journal} {Phys. Rev. X}\ }\textbf {\bibinfo {volume} {15}},\
  \bibinfo {pages} {011035} (\bibinfo {year} {2025})}\BibitemShut {NoStop}%
\bibitem [{\citenamefont {Manovitz}\ \emph {et~al.}(2025)\citenamefont
  {Manovitz}, \citenamefont {Li}, \citenamefont {Ebadi}, \citenamefont
  {Samajdar}, \citenamefont {Geim}, \citenamefont {Evered}, \citenamefont
  {Bluvstein}, \citenamefont {Zhou}, \citenamefont {Koyluoglu}, \citenamefont
  {Feldmeier}, \citenamefont {Dolgirev}, \citenamefont {Maskara}, \citenamefont
  {Kalinowski}, \citenamefont {Sachdev}, \citenamefont {Huse}, \citenamefont
  {Greiner}, \citenamefont {Vuleti\'c},\ and\ \citenamefont
  {Lukin}}]{manovitz2025quantum}%
  \BibitemOpen
  \bibfield  {author} {\bibinfo {author} {\bibfnamefont {T.}~\bibnamefont
  {Manovitz}}, \bibinfo {author} {\bibfnamefont {S.~H.}\ \bibnamefont {Li}},
  \bibinfo {author} {\bibfnamefont {S.}~\bibnamefont {Ebadi}}, \bibinfo
  {author} {\bibfnamefont {R.}~\bibnamefont {Samajdar}}, \bibinfo {author}
  {\bibfnamefont {A.~A.}\ \bibnamefont {Geim}}, \bibinfo {author}
  {\bibfnamefont {S.~J.}\ \bibnamefont {Evered}}, \bibinfo {author}
  {\bibfnamefont {D.}~\bibnamefont {Bluvstein}}, \bibinfo {author}
  {\bibfnamefont {H.}~\bibnamefont {Zhou}}, \bibinfo {author} {\bibfnamefont
  {N.~U.}\ \bibnamefont {Koyluoglu}}, \bibinfo {author} {\bibfnamefont
  {J.}~\bibnamefont {Feldmeier}}, \bibinfo {author} {\bibfnamefont {P.~E.}\
  \bibnamefont {Dolgirev}}, \bibinfo {author} {\bibfnamefont {N.}~\bibnamefont
  {Maskara}}, \bibinfo {author} {\bibfnamefont {M.}~\bibnamefont {Kalinowski}},
  \bibinfo {author} {\bibfnamefont {S.}~\bibnamefont {Sachdev}}, \bibinfo
  {author} {\bibfnamefont {D.~A.}\ \bibnamefont {Huse}}, \bibinfo {author}
  {\bibfnamefont {M.}~\bibnamefont {Greiner}}, \bibinfo {author} {\bibfnamefont
  {V.}~\bibnamefont {Vuleti\'c}},\ and\ \bibinfo {author} {\bibfnamefont
  {M.~D.}\ \bibnamefont {Lukin}},\ }\bibfield  {title} {\bibinfo {title}
  {Quantum coarsening and collective dynamics on a programmable simulator},\
  }\href@noop {} {\bibfield  {journal} {\bibinfo  {journal} {Nature}\ }\textbf
  {\bibinfo {volume} {638}},\ \bibinfo {pages} {86} (\bibinfo {year}
  {2025})}\BibitemShut {NoStop}%
\bibitem [{\citenamefont {Datla}\ \emph {et~al.}(2025)\citenamefont {Datla},
  \citenamefont {Zhao}, \citenamefont {Ho}, \citenamefont {Klco},\ and\
  \citenamefont {Loh}}]{datla2025statistical}%
  \BibitemOpen
  \bibfield  {author} {\bibinfo {author} {\bibfnamefont {P.~R.}\ \bibnamefont
  {Datla}}, \bibinfo {author} {\bibfnamefont {L.}~\bibnamefont {Zhao}},
  \bibinfo {author} {\bibfnamefont {W.~W.}\ \bibnamefont {Ho}}, \bibinfo
  {author} {\bibfnamefont {N.}~\bibnamefont {Klco}},\ and\ \bibinfo {author}
  {\bibfnamefont {H.}~\bibnamefont {Loh}},\ }\bibfield  {title} {\bibinfo
  {title} {Statistical localization in a {Rydberg} simulator of {$U(1)$}
  lattice gauge theory},\ }\href@noop {} {\bibfield  {journal} {\bibinfo
  {journal} {arXiv:2505.18143}\ } (\bibinfo {year} {2025})}\BibitemShut
  {NoStop}%
\bibitem [{\citenamefont {Ravets}\ \emph {et~al.}(2014)\citenamefont {Ravets},
  \citenamefont {Labuhn}, \citenamefont {Barredo}, \citenamefont {B{\'e}guin},
  \citenamefont {Lahaye},\ and\ \citenamefont {Browaeys}}]{Ravets2014}%
  \BibitemOpen
  \bibfield  {author} {\bibinfo {author} {\bibfnamefont {S.}~\bibnamefont
  {Ravets}}, \bibinfo {author} {\bibfnamefont {H.}~\bibnamefont {Labuhn}},
  \bibinfo {author} {\bibfnamefont {D.}~\bibnamefont {Barredo}}, \bibinfo
  {author} {\bibfnamefont {L.}~\bibnamefont {B{\'e}guin}}, \bibinfo {author}
  {\bibfnamefont {T.}~\bibnamefont {Lahaye}},\ and\ \bibinfo {author}
  {\bibfnamefont {A.}~\bibnamefont {Browaeys}},\ }\bibfield  {title} {\bibinfo
  {title} {Coherent dipole--dipole coupling between two single {Rydberg} atoms
  at an electrically-tuned {F{\"o}rster} resonance},\ }\href@noop {} {\bibfield
   {journal} {\bibinfo  {journal} {Nat. Phys.}\ }\textbf {\bibinfo {volume}
  {10}},\ \bibinfo {pages} {914} (\bibinfo {year} {2014})}\BibitemShut
  {NoStop}%
\bibitem [{\citenamefont {Ravets}\ \emph {et~al.}(2015)\citenamefont {Ravets},
  \citenamefont {Labuhn}, \citenamefont {Barredo}, \citenamefont {Lahaye},\
  and\ \citenamefont {Browaeys}}]{Ravets2015}%
  \BibitemOpen
  \bibfield  {author} {\bibinfo {author} {\bibfnamefont {S.}~\bibnamefont
  {Ravets}}, \bibinfo {author} {\bibfnamefont {H.}~\bibnamefont {Labuhn}},
  \bibinfo {author} {\bibfnamefont {D.}~\bibnamefont {Barredo}}, \bibinfo
  {author} {\bibfnamefont {T.}~\bibnamefont {Lahaye}},\ and\ \bibinfo {author}
  {\bibfnamefont {A.}~\bibnamefont {Browaeys}},\ }\bibfield  {title} {\bibinfo
  {title} {Measurement of the angular dependence of the dipole-dipole
  interaction between two individual {Rydberg} atoms at a {F{\"o}rster
  resonance}},\ }\href@noop {} {\bibfield  {journal} {\bibinfo  {journal}
  {Phys. Rev. A}\ }\textbf {\bibinfo {volume} {92}},\ \bibinfo {pages} {020701}
  (\bibinfo {year} {2015})}\BibitemShut {NoStop}%
\bibitem [{\citenamefont {Orioli}\ \emph {et~al.}(2018)\citenamefont {Orioli},
  \citenamefont {Signoles}, \citenamefont {Wildhagen}, \citenamefont
  {G{\"u}nter}, \citenamefont {Berges}, \citenamefont {Whitlock},\ and\
  \citenamefont {Weidem{\"u}ller}}]{Orioli2018}%
  \BibitemOpen
  \bibfield  {author} {\bibinfo {author} {\bibfnamefont {A.~P.}\ \bibnamefont
  {Orioli}}, \bibinfo {author} {\bibfnamefont {A.}~\bibnamefont {Signoles}},
  \bibinfo {author} {\bibfnamefont {H.}~\bibnamefont {Wildhagen}}, \bibinfo
  {author} {\bibfnamefont {G.}~\bibnamefont {G{\"u}nter}}, \bibinfo {author}
  {\bibfnamefont {J.}~\bibnamefont {Berges}}, \bibinfo {author} {\bibfnamefont
  {S.}~\bibnamefont {Whitlock}},\ and\ \bibinfo {author} {\bibfnamefont
  {M.}~\bibnamefont {Weidem{\"u}ller}},\ }\bibfield  {title} {\bibinfo {title}
  {Relaxation of an isolated dipolar-interacting {Rydberg} quantum spin
  system},\ }\href@noop {} {\bibfield  {journal} {\bibinfo  {journal} {Phys.
  Rev. Lett.}\ }\textbf {\bibinfo {volume} {120}},\ \bibinfo {pages} {063601}
  (\bibinfo {year} {2018})}\BibitemShut {NoStop}%
\bibitem [{\citenamefont {de~L{\'e}s{\'e}leuc}\ \emph
  {et~al.}(2019)\citenamefont {de~L{\'e}s{\'e}leuc}, \citenamefont {Lienhard},
  \citenamefont {Scholl}, \citenamefont {Barredo}, \citenamefont {Weber},
  \citenamefont {Lang}, \citenamefont {B{\"u}chler}, \citenamefont {Lahaye},\
  and\ \citenamefont {Browaeys}}]{Leseleuc2019}%
  \BibitemOpen
  \bibfield  {author} {\bibinfo {author} {\bibfnamefont {S.}~\bibnamefont
  {de~L{\'e}s{\'e}leuc}}, \bibinfo {author} {\bibfnamefont {V.}~\bibnamefont
  {Lienhard}}, \bibinfo {author} {\bibfnamefont {P.}~\bibnamefont {Scholl}},
  \bibinfo {author} {\bibfnamefont {D.}~\bibnamefont {Barredo}}, \bibinfo
  {author} {\bibfnamefont {S.}~\bibnamefont {Weber}}, \bibinfo {author}
  {\bibfnamefont {N.}~\bibnamefont {Lang}}, \bibinfo {author} {\bibfnamefont
  {H.~P.}\ \bibnamefont {B{\"u}chler}}, \bibinfo {author} {\bibfnamefont
  {T.}~\bibnamefont {Lahaye}},\ and\ \bibinfo {author} {\bibfnamefont
  {A.}~\bibnamefont {Browaeys}},\ }\bibfield  {title} {\bibinfo {title}
  {Observation of a symmetry-protected topological phase of interacting bosons
  with {Rydberg} atoms},\ }\href@noop {} {\bibfield  {journal} {\bibinfo
  {journal} {Science}\ }\textbf {\bibinfo {volume} {365}},\ \bibinfo {pages}
  {775} (\bibinfo {year} {2019})}\BibitemShut {NoStop}%
\bibitem [{\citenamefont {Chew}\ \emph {et~al.}(2022)\citenamefont {Chew},
  \citenamefont {Tomita}, \citenamefont {Mahesh}, \citenamefont {Sugawa},
  \citenamefont {de~L{\'e}s{\'e}leuc},\ and\ \citenamefont
  {Ohmori}}]{Chew2022}%
  \BibitemOpen
  \bibfield  {author} {\bibinfo {author} {\bibfnamefont {Y.}~\bibnamefont
  {Chew}}, \bibinfo {author} {\bibfnamefont {T.}~\bibnamefont {Tomita}},
  \bibinfo {author} {\bibfnamefont {T.~P.}\ \bibnamefont {Mahesh}}, \bibinfo
  {author} {\bibfnamefont {S.}~\bibnamefont {Sugawa}}, \bibinfo {author}
  {\bibfnamefont {S.}~\bibnamefont {de~L{\'e}s{\'e}leuc}},\ and\ \bibinfo
  {author} {\bibfnamefont {K.}~\bibnamefont {Ohmori}},\ }\bibfield  {title}
  {\bibinfo {title} {Ultrafast energy exchange between two single {Rydberg}
  atoms on a nanosecond timescale},\ }\href@noop {} {\bibfield  {journal}
  {\bibinfo  {journal} {Nat. Photonics}\ }\textbf {\bibinfo {volume} {16}},\
  \bibinfo {pages} {724} (\bibinfo {year} {2022})}\BibitemShut {NoStop}%
\bibitem [{\citenamefont {Franz}\ \emph {et~al.}(2022)\citenamefont {Franz},
  \citenamefont {Geier}, \citenamefont {Hainaut}, \citenamefont {Signoles},
  \citenamefont {Thaicharoen}, \citenamefont {Tebben}, \citenamefont
  {Salzinger}, \citenamefont {Braemer}, \citenamefont {G{\"a}rttner},
  \citenamefont {Z{\"u}rn},\ and\ \citenamefont {Weidem\"uller}}]{Franz2022a}%
  \BibitemOpen
  \bibfield  {author} {\bibinfo {author} {\bibfnamefont {T.}~\bibnamefont
  {Franz}}, \bibinfo {author} {\bibfnamefont {S.}~\bibnamefont {Geier}},
  \bibinfo {author} {\bibfnamefont {C.}~\bibnamefont {Hainaut}}, \bibinfo
  {author} {\bibfnamefont {A.}~\bibnamefont {Signoles}}, \bibinfo {author}
  {\bibfnamefont {N.}~\bibnamefont {Thaicharoen}}, \bibinfo {author}
  {\bibfnamefont {A.}~\bibnamefont {Tebben}}, \bibinfo {author} {\bibfnamefont
  {A.}~\bibnamefont {Salzinger}}, \bibinfo {author} {\bibfnamefont
  {A.}~\bibnamefont {Braemer}}, \bibinfo {author} {\bibfnamefont
  {M.}~\bibnamefont {G{\"a}rttner}}, \bibinfo {author} {\bibfnamefont
  {G.}~\bibnamefont {Z{\"u}rn}},\ and\ \bibinfo {author} {\bibfnamefont
  {M.}~\bibnamefont {Weidem\"uller}},\ }\bibfield  {title} {\bibinfo {title}
  {Emergent pair localization in a many-body quantum spin system},\ }\href@noop
  {} {\bibfield  {journal} {\bibinfo  {journal} {arXiv:2207.14216}\ } (\bibinfo
  {year} {2022})}\BibitemShut {NoStop}%
\bibitem [{\citenamefont {Chen}\ \emph {et~al.}(2023)\citenamefont {Chen},
  \citenamefont {Bornet}, \citenamefont {Bintz}, \citenamefont {Emperauger},
  \citenamefont {Leclerc}, \citenamefont {Liu}, \citenamefont {Scholl},
  \citenamefont {Barredo}, \citenamefont {Hauschild}, \citenamefont
  {Chatterjee}, \citenamefont {Schuler}, \citenamefont {L\"auchli},
  \citenamefont {Zaletel}, \citenamefont {Lahaye}, \citenamefont {Yao},\ and\
  \citenamefont {Browaeys}}]{Chen2023}%
  \BibitemOpen
  \bibfield  {author} {\bibinfo {author} {\bibfnamefont {C.}~\bibnamefont
  {Chen}}, \bibinfo {author} {\bibfnamefont {G.}~\bibnamefont {Bornet}},
  \bibinfo {author} {\bibfnamefont {M.}~\bibnamefont {Bintz}}, \bibinfo
  {author} {\bibfnamefont {G.}~\bibnamefont {Emperauger}}, \bibinfo {author}
  {\bibfnamefont {L.}~\bibnamefont {Leclerc}}, \bibinfo {author} {\bibfnamefont
  {V.~S.}\ \bibnamefont {Liu}}, \bibinfo {author} {\bibfnamefont
  {P.}~\bibnamefont {Scholl}}, \bibinfo {author} {\bibfnamefont
  {D.}~\bibnamefont {Barredo}}, \bibinfo {author} {\bibfnamefont
  {J.}~\bibnamefont {Hauschild}}, \bibinfo {author} {\bibfnamefont
  {S.}~\bibnamefont {Chatterjee}}, \bibinfo {author} {\bibfnamefont
  {M.}~\bibnamefont {Schuler}}, \bibinfo {author} {\bibfnamefont {A.~M.}\
  \bibnamefont {L\"auchli}}, \bibinfo {author} {\bibfnamefont {M.~P.}\
  \bibnamefont {Zaletel}}, \bibinfo {author} {\bibfnamefont {T.}~\bibnamefont
  {Lahaye}}, \bibinfo {author} {\bibfnamefont {N.~Y.}\ \bibnamefont {Yao}},\
  and\ \bibinfo {author} {\bibfnamefont {A.}~\bibnamefont {Browaeys}},\
  }\bibfield  {title} {\bibinfo {title} {Continuous symmetry breaking in a
  two-dimensional {Rydberg} array},\ }\href@noop {} {\bibfield  {journal}
  {\bibinfo  {journal} {Nature}\ }\textbf {\bibinfo {volume} {616}},\ \bibinfo
  {pages} {691} (\bibinfo {year} {2023})}\BibitemShut {NoStop}%
\bibitem [{\citenamefont {Bornet}\ \emph {et~al.}(2023)\citenamefont {Bornet},
  \citenamefont {Emperauger}, \citenamefont {Chen}, \citenamefont {Ye},
  \citenamefont {Block}, \citenamefont {Bintz}, \citenamefont {Boyd},
  \citenamefont {Barredo}, \citenamefont {Comparin}, \citenamefont {Mezzacapo},
  \citenamefont {Roscilde}, \citenamefont {Lahaye}, \citenamefont {Yao},\ and\
  \citenamefont {Browaeys}}]{Bornet2023}%
  \BibitemOpen
  \bibfield  {author} {\bibinfo {author} {\bibfnamefont {G.}~\bibnamefont
  {Bornet}}, \bibinfo {author} {\bibfnamefont {G.}~\bibnamefont {Emperauger}},
  \bibinfo {author} {\bibfnamefont {C.}~\bibnamefont {Chen}}, \bibinfo {author}
  {\bibfnamefont {B.}~\bibnamefont {Ye}}, \bibinfo {author} {\bibfnamefont
  {M.}~\bibnamefont {Block}}, \bibinfo {author} {\bibfnamefont
  {M.}~\bibnamefont {Bintz}}, \bibinfo {author} {\bibfnamefont {J.~A.}\
  \bibnamefont {Boyd}}, \bibinfo {author} {\bibfnamefont {D.}~\bibnamefont
  {Barredo}}, \bibinfo {author} {\bibfnamefont {T.}~\bibnamefont {Comparin}},
  \bibinfo {author} {\bibfnamefont {F.}~\bibnamefont {Mezzacapo}}, \bibinfo
  {author} {\bibfnamefont {T.}~\bibnamefont {Roscilde}}, \bibinfo {author}
  {\bibfnamefont {T.}~\bibnamefont {Lahaye}}, \bibinfo {author} {\bibfnamefont
  {N.~Y.}\ \bibnamefont {Yao}},\ and\ \bibinfo {author} {\bibfnamefont
  {A.}~\bibnamefont {Browaeys}},\ }\bibfield  {title} {\bibinfo {title}
  {Scalable spin squeezing in a dipolar {Rydberg} atom array},\ }\href@noop {}
  {\bibfield  {journal} {\bibinfo  {journal} {Nature}\ }\textbf {\bibinfo
  {volume} {621}},\ \bibinfo {pages} {728} (\bibinfo {year}
  {2023})}\BibitemShut {NoStop}%
\bibitem [{\citenamefont {Bornet}\ \emph {et~al.}(2024)\citenamefont {Bornet},
  \citenamefont {Emperauger}, \citenamefont {Chen}, \citenamefont {Machado},
  \citenamefont {Chern}, \citenamefont {Leclerc}, \citenamefont {G{\'e}ly},
  \citenamefont {Chew}, \citenamefont {Barredo}, \citenamefont {Lahaye},
  \citenamefont {Yao},\ and\ \citenamefont {Browaeys}}]{Bornet2024}%
  \BibitemOpen
  \bibfield  {author} {\bibinfo {author} {\bibfnamefont {G.}~\bibnamefont
  {Bornet}}, \bibinfo {author} {\bibfnamefont {G.}~\bibnamefont {Emperauger}},
  \bibinfo {author} {\bibfnamefont {C.}~\bibnamefont {Chen}}, \bibinfo {author}
  {\bibfnamefont {F.}~\bibnamefont {Machado}}, \bibinfo {author} {\bibfnamefont
  {S.}~\bibnamefont {Chern}}, \bibinfo {author} {\bibfnamefont
  {L.}~\bibnamefont {Leclerc}}, \bibinfo {author} {\bibfnamefont
  {B.}~\bibnamefont {G{\'e}ly}}, \bibinfo {author} {\bibfnamefont {Y.~T.}\
  \bibnamefont {Chew}}, \bibinfo {author} {\bibfnamefont {D.}~\bibnamefont
  {Barredo}}, \bibinfo {author} {\bibfnamefont {T.}~\bibnamefont {Lahaye}},
  \bibinfo {author} {\bibfnamefont {N.~Y.}\ \bibnamefont {Yao}},\ and\ \bibinfo
  {author} {\bibfnamefont {A.}~\bibnamefont {Browaeys}},\ }\bibfield  {title}
  {\bibinfo {title} {Enhancing a many-body dipolar {Rydberg} tweezer array with
  arbitrary local controls},\ }\href@noop {} {\bibfield  {journal} {\bibinfo
  {journal} {Phys. Rev. Lett.}\ }\textbf {\bibinfo {volume} {132}},\ \bibinfo
  {pages} {263601} (\bibinfo {year} {2024})}\BibitemShut {NoStop}%
\bibitem [{\citenamefont {Emperauger}\ \emph
  {et~al.}(2025{\natexlab{a}})\citenamefont {Emperauger}, \citenamefont {Qiao},
  \citenamefont {Chen}, \citenamefont {Caleca}, \citenamefont {Bocini},
  \citenamefont {Bintz}, \citenamefont {Bornet}, \citenamefont {Martin},
  \citenamefont {G\'ely}, \citenamefont {Klein}, \citenamefont {Barredo},
  \citenamefont {Chatterjee}, \citenamefont {Yao}, \citenamefont {Mezzacapo},
  \citenamefont {Lahaye}, \citenamefont {Roscilde},\ and\ \citenamefont
  {Browaeys}}]{Emperauger2025}%
  \BibitemOpen
  \bibfield  {author} {\bibinfo {author} {\bibfnamefont {G.}~\bibnamefont
  {Emperauger}}, \bibinfo {author} {\bibfnamefont {M.}~\bibnamefont {Qiao}},
  \bibinfo {author} {\bibfnamefont {C.}~\bibnamefont {Chen}}, \bibinfo {author}
  {\bibfnamefont {F.}~\bibnamefont {Caleca}}, \bibinfo {author} {\bibfnamefont
  {S.}~\bibnamefont {Bocini}}, \bibinfo {author} {\bibfnamefont
  {M.}~\bibnamefont {Bintz}}, \bibinfo {author} {\bibfnamefont
  {G.}~\bibnamefont {Bornet}}, \bibinfo {author} {\bibfnamefont
  {R.}~\bibnamefont {Martin}}, \bibinfo {author} {\bibfnamefont
  {B.}~\bibnamefont {G\'ely}}, \bibinfo {author} {\bibfnamefont
  {L.}~\bibnamefont {Klein}}, \bibinfo {author} {\bibfnamefont
  {D.}~\bibnamefont {Barredo}}, \bibinfo {author} {\bibfnamefont
  {S.}~\bibnamefont {Chatterjee}}, \bibinfo {author} {\bibfnamefont {N.~Y.}\
  \bibnamefont {Yao}}, \bibinfo {author} {\bibfnamefont {F.}~\bibnamefont
  {Mezzacapo}}, \bibinfo {author} {\bibfnamefont {T.}~\bibnamefont {Lahaye}},
  \bibinfo {author} {\bibfnamefont {T.}~\bibnamefont {Roscilde}},\ and\
  \bibinfo {author} {\bibfnamefont {A.}~\bibnamefont {Browaeys}},\ }\bibfield
  {title} {\bibinfo {title} {{Tomonaga-Luttinger} liquid behavior in a
  {Rydberg-encoded} spin chain},\ }\href {https://doi.org/10.1103/qfnp-6dpz}
  {\bibfield  {journal} {\bibinfo  {journal} {Phys. Rev. X}\ }\textbf {\bibinfo
  {volume} {15}},\ \bibinfo {pages} {031021} (\bibinfo {year}
  {2025}{\natexlab{a}})}\BibitemShut {NoStop}%
\bibitem [{\citenamefont {Emperauger}\ \emph
  {et~al.}(2025{\natexlab{b}})\citenamefont {Emperauger}, \citenamefont {Qiao},
  \citenamefont {Bornet}, \citenamefont {Chen}, \citenamefont {Martin},
  \citenamefont {Chew}, \citenamefont {G\'ely}, \citenamefont {Klein},
  \citenamefont {Barredo}, \citenamefont {Browaeys},\ and\ \citenamefont
  {Lahaye}}]{Emperauger2025benchmarking}%
  \BibitemOpen
  \bibfield  {author} {\bibinfo {author} {\bibfnamefont {G.}~\bibnamefont
  {Emperauger}}, \bibinfo {author} {\bibfnamefont {M.}~\bibnamefont {Qiao}},
  \bibinfo {author} {\bibfnamefont {G.}~\bibnamefont {Bornet}}, \bibinfo
  {author} {\bibfnamefont {C.}~\bibnamefont {Chen}}, \bibinfo {author}
  {\bibfnamefont {R.}~\bibnamefont {Martin}}, \bibinfo {author} {\bibfnamefont
  {Y.~T.}\ \bibnamefont {Chew}}, \bibinfo {author} {\bibfnamefont
  {B.}~\bibnamefont {G\'ely}}, \bibinfo {author} {\bibfnamefont
  {L.}~\bibnamefont {Klein}}, \bibinfo {author} {\bibfnamefont
  {D.}~\bibnamefont {Barredo}}, \bibinfo {author} {\bibfnamefont
  {A.}~\bibnamefont {Browaeys}},\ and\ \bibinfo {author} {\bibfnamefont
  {T.}~\bibnamefont {Lahaye}},\ }\bibfield  {title} {\bibinfo {title}
  {Benchmarking direct and indirect dipolar spin-exchange interactions between
  two {Rydberg} atoms},\ }\href {https://doi.org/10.1103/PhysRevA.111.062806}
  {\bibfield  {journal} {\bibinfo  {journal} {Phys. Rev. A}\ }\textbf {\bibinfo
  {volume} {111}},\ \bibinfo {pages} {062806} (\bibinfo {year}
  {2025}{\natexlab{b}})}\BibitemShut {NoStop}%
\bibitem [{\citenamefont {Chen}\ \emph {et~al.}(2025)\citenamefont {Chen},
  \citenamefont {Emperauger}, \citenamefont {Bornet}, \citenamefont {Caleca},
  \citenamefont {G{\'e}ly}, \citenamefont {Bintz}, \citenamefont {Chatterjee},
  \citenamefont {Liu}, \citenamefont {Barredo}, \citenamefont {Yao},
  \citenamefont {Lahaye}, \citenamefont {Mezzacapo}, \citenamefont {Roscilde},\
  and\ \citenamefont {Browaeys}}]{chen2025spectroscopy}%
  \BibitemOpen
  \bibfield  {author} {\bibinfo {author} {\bibfnamefont {C.}~\bibnamefont
  {Chen}}, \bibinfo {author} {\bibfnamefont {G.}~\bibnamefont {Emperauger}},
  \bibinfo {author} {\bibfnamefont {G.}~\bibnamefont {Bornet}}, \bibinfo
  {author} {\bibfnamefont {F.}~\bibnamefont {Caleca}}, \bibinfo {author}
  {\bibfnamefont {B.}~\bibnamefont {G{\'e}ly}}, \bibinfo {author}
  {\bibfnamefont {M.}~\bibnamefont {Bintz}}, \bibinfo {author} {\bibfnamefont
  {S.}~\bibnamefont {Chatterjee}}, \bibinfo {author} {\bibfnamefont
  {V.}~\bibnamefont {Liu}}, \bibinfo {author} {\bibfnamefont {D.}~\bibnamefont
  {Barredo}}, \bibinfo {author} {\bibfnamefont {N.~Y.}\ \bibnamefont {Yao}},
  \bibinfo {author} {\bibfnamefont {T.}~\bibnamefont {Lahaye}}, \bibinfo
  {author} {\bibfnamefont {F.}~\bibnamefont {Mezzacapo}}, \bibinfo {author}
  {\bibfnamefont {T.}~\bibnamefont {Roscilde}},\ and\ \bibinfo {author}
  {\bibfnamefont {A.}~\bibnamefont {Browaeys}},\ }\bibfield  {title} {\bibinfo
  {title} {Spectroscopy of elementary excitations from quench dynamics in a
  dipolar {XY} {Rydberg} simulator},\ }\href
  {https://doi.org/10.1126/science.adn0618} {\bibfield  {journal} {\bibinfo
  {journal} {Science}\ }\textbf {\bibinfo {volume} {389}},\ \bibinfo {pages}
  {483} (\bibinfo {year} {2025})}\BibitemShut {NoStop}%
\bibitem [{\citenamefont {Signoles}\ \emph {et~al.}(2021)\citenamefont
  {Signoles}, \citenamefont {Franz}, \citenamefont {Ferracini~Alves},
  \citenamefont {G{\"a}rttner}, \citenamefont {Whitlock}, \citenamefont
  {Z{\"u}rn},\ and\ \citenamefont {Weidem{\"u}ller}}]{Signoles2021}%
  \BibitemOpen
  \bibfield  {author} {\bibinfo {author} {\bibfnamefont {A.}~\bibnamefont
  {Signoles}}, \bibinfo {author} {\bibfnamefont {T.}~\bibnamefont {Franz}},
  \bibinfo {author} {\bibfnamefont {R.}~\bibnamefont {Ferracini~Alves}},
  \bibinfo {author} {\bibfnamefont {M.}~\bibnamefont {G{\"a}rttner}}, \bibinfo
  {author} {\bibfnamefont {S.}~\bibnamefont {Whitlock}}, \bibinfo {author}
  {\bibfnamefont {G.}~\bibnamefont {Z{\"u}rn}},\ and\ \bibinfo {author}
  {\bibfnamefont {M.}~\bibnamefont {Weidem{\"u}ller}},\ }\bibfield  {title}
  {\bibinfo {title} {Glassy dynamics in a disordered {Heisenberg} quantum spin
  system},\ }\href@noop {} {\bibfield  {journal} {\bibinfo  {journal} {Phys.
  Rev. X}\ }\textbf {\bibinfo {volume} {11}},\ \bibinfo {pages} {011011}
  (\bibinfo {year} {2021})}\BibitemShut {NoStop}%
\bibitem [{\citenamefont {Steinert}\ \emph {et~al.}(2023)\citenamefont
  {Steinert}, \citenamefont {Osterholz}, \citenamefont {Eberhard},
  \citenamefont {Festa}, \citenamefont {Lorenz}, \citenamefont {Chen},
  \citenamefont {Trautmann},\ and\ \citenamefont {Gross}}]{Steinert2023}%
  \BibitemOpen
  \bibfield  {author} {\bibinfo {author} {\bibfnamefont {L.-M.}\ \bibnamefont
  {Steinert}}, \bibinfo {author} {\bibfnamefont {P.}~\bibnamefont {Osterholz}},
  \bibinfo {author} {\bibfnamefont {R.}~\bibnamefont {Eberhard}}, \bibinfo
  {author} {\bibfnamefont {L.}~\bibnamefont {Festa}}, \bibinfo {author}
  {\bibfnamefont {N.}~\bibnamefont {Lorenz}}, \bibinfo {author} {\bibfnamefont
  {Z.}~\bibnamefont {Chen}}, \bibinfo {author} {\bibfnamefont {A.}~\bibnamefont
  {Trautmann}},\ and\ \bibinfo {author} {\bibfnamefont {C.}~\bibnamefont
  {Gross}},\ }\bibfield  {title} {\bibinfo {title} {Spatially tunable spin
  interactions in neutral atom arrays},\ }\href@noop {} {\bibfield  {journal}
  {\bibinfo  {journal} {Phys. Rev. Lett.}\ }\textbf {\bibinfo {volume} {130}},\
  \bibinfo {pages} {243001} (\bibinfo {year} {2023})}\BibitemShut {NoStop}%
\bibitem [{\citenamefont {Lienhard}\ \emph {et~al.}(2020)\citenamefont
  {Lienhard}, \citenamefont {Scholl}, \citenamefont {Weber}, \citenamefont
  {Barredo}, \citenamefont {de~L{\'e}s{\'e}leuc}, \citenamefont {Bai},
  \citenamefont {Lang}, \citenamefont {Fleischhauer}, \citenamefont
  {B{\"u}chler}, \citenamefont {Lahaye},\ and\ \citenamefont
  {Browaeys}}]{Lienhard2020}%
  \BibitemOpen
  \bibfield  {author} {\bibinfo {author} {\bibfnamefont {V.}~\bibnamefont
  {Lienhard}}, \bibinfo {author} {\bibfnamefont {P.}~\bibnamefont {Scholl}},
  \bibinfo {author} {\bibfnamefont {S.}~\bibnamefont {Weber}}, \bibinfo
  {author} {\bibfnamefont {D.}~\bibnamefont {Barredo}}, \bibinfo {author}
  {\bibfnamefont {S.}~\bibnamefont {de~L{\'e}s{\'e}leuc}}, \bibinfo {author}
  {\bibfnamefont {R.}~\bibnamefont {Bai}}, \bibinfo {author} {\bibfnamefont
  {N.}~\bibnamefont {Lang}}, \bibinfo {author} {\bibfnamefont {M.}~\bibnamefont
  {Fleischhauer}}, \bibinfo {author} {\bibfnamefont {H.~P.}\ \bibnamefont
  {B{\"u}chler}}, \bibinfo {author} {\bibfnamefont {T.}~\bibnamefont
  {Lahaye}},\ and\ \bibinfo {author} {\bibfnamefont {A.}~\bibnamefont
  {Browaeys}},\ }\bibfield  {title} {\bibinfo {title} {Realization of a
  density-dependent peierls phase in a synthetic, spin-orbit coupled {Rydberg}
  system},\ }\href@noop {} {\bibfield  {journal} {\bibinfo  {journal} {Phys.
  Rev. X}\ }\textbf {\bibinfo {volume} {10}},\ \bibinfo {pages} {021031}
  (\bibinfo {year} {2020})}\BibitemShut {NoStop}%
\bibitem [{\citenamefont {Kanungo}\ \emph {et~al.}(2022)\citenamefont
  {Kanungo}, \citenamefont {Whalen}, \citenamefont {Lu}, \citenamefont {Yuan},
  \citenamefont {Dasgupta}, \citenamefont {Dunning}, \citenamefont {Hazzard},\
  and\ \citenamefont {Killian}}]{Kanungo2022}%
  \BibitemOpen
  \bibfield  {author} {\bibinfo {author} {\bibfnamefont {S.~K.}\ \bibnamefont
  {Kanungo}}, \bibinfo {author} {\bibfnamefont {J.~D.}\ \bibnamefont {Whalen}},
  \bibinfo {author} {\bibfnamefont {Y.}~\bibnamefont {Lu}}, \bibinfo {author}
  {\bibfnamefont {M.}~\bibnamefont {Yuan}}, \bibinfo {author} {\bibfnamefont
  {S.}~\bibnamefont {Dasgupta}}, \bibinfo {author} {\bibfnamefont {F.~B.}\
  \bibnamefont {Dunning}}, \bibinfo {author} {\bibfnamefont {K.~R.~A.}\
  \bibnamefont {Hazzard}},\ and\ \bibinfo {author} {\bibfnamefont {T.~C.}\
  \bibnamefont {Killian}},\ }\bibfield  {title} {\bibinfo {title} {Realizing
  topological edge states with {Rydberg-atom} synthetic dimensions},\
  }\href@noop {} {\bibfield  {journal} {\bibinfo  {journal} {Nat. commun.}\
  }\textbf {\bibinfo {volume} {13}},\ \bibinfo {pages} {972} (\bibinfo {year}
  {2022})}\BibitemShut {NoStop}%
\bibitem [{\citenamefont {Qiao}\ \emph {et~al.}(2025)\citenamefont {Qiao},
  \citenamefont {Emperauger}, \citenamefont {Chen}, \citenamefont {Homeier},
  \citenamefont {Hollerith}, \citenamefont {Bornet}, \citenamefont {Martin},
  \citenamefont {G{\'e}ly}, \citenamefont {Klein}, \citenamefont {Barredo},
  \citenamefont {Geier}, \citenamefont {Chiu}, \citenamefont {Grusdt},
  \citenamefont {Bohrdt}, \citenamefont {Lahaye},\ and\ \citenamefont
  {Browaeys}}]{Qiao2025}%
  \BibitemOpen
  \bibfield  {author} {\bibinfo {author} {\bibfnamefont {M.}~\bibnamefont
  {Qiao}}, \bibinfo {author} {\bibfnamefont {G.}~\bibnamefont {Emperauger}},
  \bibinfo {author} {\bibfnamefont {C.}~\bibnamefont {Chen}}, \bibinfo {author}
  {\bibfnamefont {L.}~\bibnamefont {Homeier}}, \bibinfo {author} {\bibfnamefont
  {S.}~\bibnamefont {Hollerith}}, \bibinfo {author} {\bibfnamefont
  {G.}~\bibnamefont {Bornet}}, \bibinfo {author} {\bibfnamefont
  {R.}~\bibnamefont {Martin}}, \bibinfo {author} {\bibfnamefont
  {B.}~\bibnamefont {G{\'e}ly}}, \bibinfo {author} {\bibfnamefont
  {L.}~\bibnamefont {Klein}}, \bibinfo {author} {\bibfnamefont
  {D.}~\bibnamefont {Barredo}}, \bibinfo {author} {\bibfnamefont
  {S.}~\bibnamefont {Geier}}, \bibinfo {author} {\bibfnamefont {N.-C.}\
  \bibnamefont {Chiu}}, \bibinfo {author} {\bibfnamefont {F.}~\bibnamefont
  {Grusdt}}, \bibinfo {author} {\bibfnamefont {A.}~\bibnamefont {Bohrdt}},
  \bibinfo {author} {\bibfnamefont {T.}~\bibnamefont {Lahaye}},\ and\ \bibinfo
  {author} {\bibfnamefont {A.}~\bibnamefont {Browaeys}},\ }\bibfield  {title}
  {\bibinfo {title} {Realization of a doped quantum antiferromagnet in a
  {Rydberg} tweezer array},\ }\href@noop {} {\bibfield  {journal} {\bibinfo
  {journal} {Nature}\ }\textbf {\bibinfo {volume} {644}},\ \bibinfo {pages}
  {889} (\bibinfo {year} {2025})}\BibitemShut {NoStop}%
\bibitem [{\citenamefont {Evered}\ \emph {et~al.}(2025)\citenamefont {Evered},
  \citenamefont {Kalinowski}, \citenamefont {Geim}, \citenamefont {Manovitz},
  \citenamefont {Bluvstein}, \citenamefont {Li}, \citenamefont {Maskara},
  \citenamefont {Zhou}, \citenamefont {Ebadi}, \citenamefont {Xu},
  \citenamefont {Campo}, \citenamefont {Cain}, \citenamefont {Ostermann},
  \citenamefont {Yelin}, \citenamefont {Sachev}, \citenamefont {Greiner},
  \citenamefont {Vuleti\'{c}},\ and\ \citenamefont
  {Lukin}}]{evered2025probing}%
  \BibitemOpen
  \bibfield  {author} {\bibinfo {author} {\bibfnamefont {S.~J.}\ \bibnamefont
  {Evered}}, \bibinfo {author} {\bibfnamefont {M.}~\bibnamefont {Kalinowski}},
  \bibinfo {author} {\bibfnamefont {A.~A.}\ \bibnamefont {Geim}}, \bibinfo
  {author} {\bibfnamefont {T.}~\bibnamefont {Manovitz}}, \bibinfo {author}
  {\bibfnamefont {D.}~\bibnamefont {Bluvstein}}, \bibinfo {author}
  {\bibfnamefont {S.~H.}\ \bibnamefont {Li}}, \bibinfo {author} {\bibfnamefont
  {N.}~\bibnamefont {Maskara}}, \bibinfo {author} {\bibfnamefont
  {H.}~\bibnamefont {Zhou}}, \bibinfo {author} {\bibfnamefont {S.}~\bibnamefont
  {Ebadi}}, \bibinfo {author} {\bibfnamefont {M.}~\bibnamefont {Xu}}, \bibinfo
  {author} {\bibfnamefont {J.}~\bibnamefont {Campo}}, \bibinfo {author}
  {\bibfnamefont {M.}~\bibnamefont {Cain}}, \bibinfo {author} {\bibfnamefont
  {S.}~\bibnamefont {Ostermann}}, \bibinfo {author} {\bibfnamefont {S.~F.}\
  \bibnamefont {Yelin}}, \bibinfo {author} {\bibfnamefont {S.}~\bibnamefont
  {Sachev}}, \bibinfo {author} {\bibfnamefont {M.}~\bibnamefont {Greiner}},
  \bibinfo {author} {\bibfnamefont {V.}~\bibnamefont {Vuleti\'{c}}},\ and\
  \bibinfo {author} {\bibfnamefont {M.~D.}\ \bibnamefont {Lukin}},\ }\bibfield
  {title} {\bibinfo {title} {Probing the {Kitaev} honeycomb model on a
  neutral-atom quantum computer},\ }\href@noop {} {\bibfield  {journal}
  {\bibinfo  {journal} {Nature}\ }\textbf {\bibinfo {volume} {645}},\ \bibinfo
  {pages} {341} (\bibinfo {year} {2025})}\BibitemShut {NoStop}%
\bibitem [{\citenamefont {Nishad}\ \emph {et~al.}(2023)\citenamefont {Nishad},
  \citenamefont {Keselman}, \citenamefont {Lahaye}, \citenamefont {Browaeys},\
  and\ \citenamefont {Tsesses}}]{nishad2023quantum}%
  \BibitemOpen
  \bibfield  {author} {\bibinfo {author} {\bibfnamefont {N.}~\bibnamefont
  {Nishad}}, \bibinfo {author} {\bibfnamefont {A.}~\bibnamefont {Keselman}},
  \bibinfo {author} {\bibfnamefont {T.}~\bibnamefont {Lahaye}}, \bibinfo
  {author} {\bibfnamefont {A.}~\bibnamefont {Browaeys}},\ and\ \bibinfo
  {author} {\bibfnamefont {S.}~\bibnamefont {Tsesses}},\ }\bibfield  {title}
  {\bibinfo {title} {Quantum simulation of generic spin-exchange models in
  {Floquet-engineered} {Rydberg-atom} arrays},\ }\href@noop {} {\bibfield
  {journal} {\bibinfo  {journal} {Phys. Rev. A}\ }\textbf {\bibinfo {volume}
  {108}},\ \bibinfo {pages} {053318} (\bibinfo {year} {2023})}\BibitemShut
  {NoStop}%
\bibitem [{\citenamefont {Kuji}\ \emph {et~al.}(2025)\citenamefont {Kuji},
  \citenamefont {Kunimi},\ and\ \citenamefont {Nikuni}}]{kuji2025proposal}%
  \BibitemOpen
  \bibfield  {author} {\bibinfo {author} {\bibfnamefont {H.}~\bibnamefont
  {Kuji}}, \bibinfo {author} {\bibfnamefont {M.}~\bibnamefont {Kunimi}},\ and\
  \bibinfo {author} {\bibfnamefont {T.}~\bibnamefont {Nikuni}},\ }\bibfield
  {title} {\bibinfo {title} {Proposal for realizing quantum-spin systems on a
  two-dimensional square lattice with {Dzyaloshinskii-Moriya} interaction by
  {Floquet} engineering using {Rydberg} atoms},\ }\href@noop {} {\bibfield
  {journal} {\bibinfo  {journal} {Phys. Rev. A}\ }\textbf {\bibinfo {volume}
  {112}},\ \bibinfo {pages} {022614} (\bibinfo {year} {2025})}\BibitemShut
  {NoStop}%
\bibitem [{\citenamefont {Tian}\ \emph {et~al.}(2025)\citenamefont {Tian},
  \citenamefont {Samajdar},\ and\ \citenamefont
  {Gadway}}]{tian2025engineering}%
  \BibitemOpen
  \bibfield  {author} {\bibinfo {author} {\bibfnamefont {M.}~\bibnamefont
  {Tian}}, \bibinfo {author} {\bibfnamefont {R.}~\bibnamefont {Samajdar}},\
  and\ \bibinfo {author} {\bibfnamefont {B.}~\bibnamefont {Gadway}},\
  }\bibfield  {title} {\bibinfo {title} {Engineering frustrated {Rydberg} spin
  models by graphical {Floquet} modulation},\ }\href@noop {} {\bibfield
  {journal} {\bibinfo  {journal} {arXiv:2505.01513}\ } (\bibinfo {year}
  {2025})}\BibitemShut {NoStop}%
\bibitem [{\citenamefont {Whitlock}\ \emph {et~al.}(2017)\citenamefont
  {Whitlock}, \citenamefont {Glaetzle},\ and\ \citenamefont
  {Hannaford}}]{Whitlock2017}%
  \BibitemOpen
  \bibfield  {author} {\bibinfo {author} {\bibfnamefont {S.}~\bibnamefont
  {Whitlock}}, \bibinfo {author} {\bibfnamefont {A.~W.}\ \bibnamefont
  {Glaetzle}},\ and\ \bibinfo {author} {\bibfnamefont {P.}~\bibnamefont
  {Hannaford}},\ }\bibfield  {title} {\bibinfo {title} {Simulating quantum spin
  models using {Rydberg-excited} atomic ensembles in magnetic microtrap
  arrays},\ }\href@noop {} {\bibfield  {journal} {\bibinfo  {journal} {J. Phys.
  B: At. Mol. Opt. Phys.}\ }\textbf {\bibinfo {volume} {50}},\ \bibinfo {pages}
  {074001} (\bibinfo {year} {2017})}\BibitemShut {NoStop}%
\bibitem [{\citenamefont {Majumdar}\ and\ \citenamefont
  {Ghosh}(1969{\natexlab{a}})}]{majumdar1969next}%
  \BibitemOpen
  \bibfield  {author} {\bibinfo {author} {\bibfnamefont {C.~K.}\ \bibnamefont
  {Majumdar}}\ and\ \bibinfo {author} {\bibfnamefont {D.~K.}\ \bibnamefont
  {Ghosh}},\ }\bibfield  {title} {\bibinfo {title} {On next-nearest-neighbor
  interaction in linear chain. {I}},\ }\href@noop {} {\bibfield  {journal}
  {\bibinfo  {journal} {J. Math. Phys.}\ }\textbf {\bibinfo {volume} {10}},\
  \bibinfo {pages} {1388} (\bibinfo {year} {1969}{\natexlab{a}})}\BibitemShut
  {NoStop}%
\bibitem [{\citenamefont {Majumdar}\ and\ \citenamefont
  {Ghosh}(1969{\natexlab{b}})}]{majumdar1969next2}%
  \BibitemOpen
  \bibfield  {author} {\bibinfo {author} {\bibfnamefont {C.~K.}\ \bibnamefont
  {Majumdar}}\ and\ \bibinfo {author} {\bibfnamefont {D.~K.}\ \bibnamefont
  {Ghosh}},\ }\bibfield  {title} {\bibinfo {title} {On next-nearest-neighbor
  interaction in linear chain. {II}},\ }\href@noop {} {\bibfield  {journal}
  {\bibinfo  {journal} {J. Math. Phys.}\ }\textbf {\bibinfo {volume} {10}},\
  \bibinfo {pages} {1399} (\bibinfo {year} {1969}{\natexlab{b}})}\BibitemShut
  {NoStop}%
\bibitem [{\citenamefont {{\v{S}}ibali{\'c}}\ \emph {et~al.}(2017)\citenamefont
  {{\v{S}}ibali{\'c}}, \citenamefont {Pritchard}, \citenamefont {Adams},\ and\
  \citenamefont {Weatherill}}]{Sibalic2017}%
  \BibitemOpen
  \bibfield  {author} {\bibinfo {author} {\bibfnamefont {N.}~\bibnamefont
  {{\v{S}}ibali{\'c}}}, \bibinfo {author} {\bibfnamefont {J.~D.}\ \bibnamefont
  {Pritchard}}, \bibinfo {author} {\bibfnamefont {C.~S.}\ \bibnamefont
  {Adams}},\ and\ \bibinfo {author} {\bibfnamefont {K.~J.}\ \bibnamefont
  {Weatherill}},\ }\bibfield  {title} {\bibinfo {title} {Arc: {An} open-source
  library for calculating properties of alkali {Rydberg} atoms},\ }\href@noop
  {} {\bibfield  {journal} {\bibinfo  {journal} {Computer Physics
  Communications}\ }\textbf {\bibinfo {volume} {220}},\ \bibinfo {pages} {319}
  (\bibinfo {year} {2017})}\BibitemShut {NoStop}%
\bibitem [{\citenamefont {Weber}\ \emph {et~al.}(2017)\citenamefont {Weber},
  \citenamefont {Tresp}, \citenamefont {Menke}, \citenamefont {Urvoy},
  \citenamefont {Firstenberg}, \citenamefont {B{\"u}chler},\ and\ \citenamefont
  {Hofferberth}}]{Weber2017}%
  \BibitemOpen
  \bibfield  {author} {\bibinfo {author} {\bibfnamefont {S.}~\bibnamefont
  {Weber}}, \bibinfo {author} {\bibfnamefont {C.}~\bibnamefont {Tresp}},
  \bibinfo {author} {\bibfnamefont {H.}~\bibnamefont {Menke}}, \bibinfo
  {author} {\bibfnamefont {A.}~\bibnamefont {Urvoy}}, \bibinfo {author}
  {\bibfnamefont {O.}~\bibnamefont {Firstenberg}}, \bibinfo {author}
  {\bibfnamefont {H.~P.}\ \bibnamefont {B{\"u}chler}},\ and\ \bibinfo {author}
  {\bibfnamefont {S.}~\bibnamefont {Hofferberth}},\ }\bibfield  {title}
  {\bibinfo {title} {Calculation of {Rydberg} interaction potentials},\
  }\href@noop {} {\bibfield  {journal} {\bibinfo  {journal} {J. Phys. B: At.
  Mol. Opt. Phys.}\ }\textbf {\bibinfo {volume} {50}},\ \bibinfo {pages}
  {133001} (\bibinfo {year} {2017})}\BibitemShut {NoStop}%
\bibitem [{\citenamefont {van Bijnen}(2013)}]{Bijnen_PhD_thesis}%
  \BibitemOpen
  \bibfield  {author} {\bibinfo {author} {\bibfnamefont {R.~M.~W.}\
  \bibnamefont {van Bijnen}},\ }\emph {\bibinfo {title} {Quantum engineering
  with ultracold atoms}},\ \href@noop {} {\bibinfo {type} {Ph.d. thesis}},\
  \bibinfo  {school} {Eindhoven University of Technology} (\bibinfo {year}
  {2013})\BibitemShut {NoStop}%
\bibitem [{\citenamefont {Kunimi}\ \emph {et~al.}(2024)\citenamefont {Kunimi},
  \citenamefont {Tomita}, \citenamefont {Katsura},\ and\ \citenamefont
  {Kato}}]{Kunimi2024}%
  \BibitemOpen
  \bibfield  {author} {\bibinfo {author} {\bibfnamefont {M.}~\bibnamefont
  {Kunimi}}, \bibinfo {author} {\bibfnamefont {T.}~\bibnamefont {Tomita}},
  \bibinfo {author} {\bibfnamefont {H.}~\bibnamefont {Katsura}},\ and\ \bibinfo
  {author} {\bibfnamefont {Y.}~\bibnamefont {Kato}},\ }\bibfield  {title}
  {\bibinfo {title} {Proposal for simulating quantum spin models with the
  {Dzyaloshinskii-Moriya} interaction using {Rydberg} atoms and the
  construction of asymptotic quantum many-body scar states},\ }\href@noop {}
  {\bibfield  {journal} {\bibinfo  {journal} {Phys. Rev. A}\ }\textbf {\bibinfo
  {volume} {110}},\ \bibinfo {pages} {043312} (\bibinfo {year}
  {2024})}\BibitemShut {NoStop}%
\bibitem [{\citenamefont {Wadenpfuhl}\ and\ \citenamefont
  {Adams}(2025)}]{Wadenpfuhl2025unraveling}%
  \BibitemOpen
  \bibfield  {author} {\bibinfo {author} {\bibfnamefont {K.}~\bibnamefont
  {Wadenpfuhl}}\ and\ \bibinfo {author} {\bibfnamefont {C.~S.}\ \bibnamefont
  {Adams}},\ }\bibfield  {title} {\bibinfo {title} {Unraveling the structures
  in the van der {Waals} interactions of alkali-metal {Rydberg} atoms},\ }\href
  {https://doi.org/10.1103/PhysRevA.111.062803} {\bibfield  {journal} {\bibinfo
   {journal} {Phys. Rev. A}\ }\textbf {\bibinfo {volume} {111}},\ \bibinfo
  {pages} {062803} (\bibinfo {year} {2025})}\BibitemShut {NoStop}%
\bibitem [{Note1()}]{Note1}%
  \BibitemOpen
  \bibinfo {note} {We can estimate the next-to-leading-order term, which is
  given by fourth-order perturbation due to the selection rule. The ratio
  between the fourth- and second-order terms scales as $(R_{\protect \rm
  c}/R)^6$. In this Letter, we adopt $R=2R_{\protect \rm c}~[(R_{\protect \rm
  c}/R)^3=1/8]$. Then, the contribution of the fourth-order term is roughly
  $1\%$ of the second-order term.}\BibitemShut {Stop}%
\bibitem [{Note2()}]{Note2}%
  \BibitemOpen
  \bibinfo {note} {See Supplemental Material for details on the results for
  other atomic species, a summary of Heisenberg points, details of the atom
  positions, and the perturbation theory for spin-1 systems.}\BibitemShut
  {Stop}%
\bibitem [{\citenamefont {Fuchs}\ \emph {et~al.}(2008)\citenamefont {Fuchs},
  \citenamefont {Ticknor}, \citenamefont {Dyke}, \citenamefont {Veeravalli},
  \citenamefont {Kuhnle}, \citenamefont {Rowlands}, \citenamefont {Hannaford},\
  and\ \citenamefont {Vale}}]{Fuchs2008binding}%
  \BibitemOpen
  \bibfield  {author} {\bibinfo {author} {\bibfnamefont {J.}~\bibnamefont
  {Fuchs}}, \bibinfo {author} {\bibfnamefont {C.}~\bibnamefont {Ticknor}},
  \bibinfo {author} {\bibfnamefont {P.}~\bibnamefont {Dyke}}, \bibinfo {author}
  {\bibfnamefont {G.}~\bibnamefont {Veeravalli}}, \bibinfo {author}
  {\bibfnamefont {E.}~\bibnamefont {Kuhnle}}, \bibinfo {author} {\bibfnamefont
  {W.}~\bibnamefont {Rowlands}}, \bibinfo {author} {\bibfnamefont
  {P.}~\bibnamefont {Hannaford}},\ and\ \bibinfo {author} {\bibfnamefont
  {C.~J.}\ \bibnamefont {Vale}},\ }\bibfield  {title} {\bibinfo {title}
  {Binding energies of $^{6}\text{L}\text{i}$ $p$-wave {Feshbach} molecules},\
  }\href {https://doi.org/10.1103/PhysRevA.77.053616} {\bibfield  {journal}
  {\bibinfo  {journal} {Phys. Rev. A}\ }\textbf {\bibinfo {volume} {77}},\
  \bibinfo {pages} {053616} (\bibinfo {year} {2008})}\BibitemShut {NoStop}%
\bibitem [{\citenamefont {Inada}\ \emph {et~al.}(2008)\citenamefont {Inada},
  \citenamefont {Horikoshi}, \citenamefont {Nakajima}, \citenamefont
  {Kuwata-Gonokami}, \citenamefont {Ueda},\ and\ \citenamefont
  {Mukaiyama}}]{Inada2008collisional}%
  \BibitemOpen
  \bibfield  {author} {\bibinfo {author} {\bibfnamefont {Y.}~\bibnamefont
  {Inada}}, \bibinfo {author} {\bibfnamefont {M.}~\bibnamefont {Horikoshi}},
  \bibinfo {author} {\bibfnamefont {S.}~\bibnamefont {Nakajima}}, \bibinfo
  {author} {\bibfnamefont {M.}~\bibnamefont {Kuwata-Gonokami}}, \bibinfo
  {author} {\bibfnamefont {M.}~\bibnamefont {Ueda}},\ and\ \bibinfo {author}
  {\bibfnamefont {T.}~\bibnamefont {Mukaiyama}},\ }\bibfield  {title} {\bibinfo
  {title} {Collisional properties of $p$-wave {Feshbach} molecules},\ }\href
  {https://doi.org/10.1103/PhysRevLett.101.100401} {\bibfield  {journal}
  {\bibinfo  {journal} {Phys. Rev. Lett.}\ }\textbf {\bibinfo {volume} {101}},\
  \bibinfo {pages} {100401} (\bibinfo {year} {2008})}\BibitemShut {NoStop}%
\bibitem [{\citenamefont {Chang}\ \emph {et~al.}(2020)\citenamefont {Chang},
  \citenamefont {Senaratne}, \citenamefont {Cavazos-Cavazos},\ and\
  \citenamefont {Hulet}}]{Chang2020collisional}%
  \BibitemOpen
  \bibfield  {author} {\bibinfo {author} {\bibfnamefont {Y.-T.}\ \bibnamefont
  {Chang}}, \bibinfo {author} {\bibfnamefont {R.}~\bibnamefont {Senaratne}},
  \bibinfo {author} {\bibfnamefont {D.}~\bibnamefont {Cavazos-Cavazos}},\ and\
  \bibinfo {author} {\bibfnamefont {R.~G.}\ \bibnamefont {Hulet}},\ }\bibfield
  {title} {\bibinfo {title} {Collisional loss of one-dimensional fermions near
  a $p$-wave {Feshbach} resonance},\ }\href
  {https://doi.org/10.1103/PhysRevLett.125.263402} {\bibfield  {journal}
  {\bibinfo  {journal} {Phys. Rev. Lett.}\ }\textbf {\bibinfo {volume} {125}},\
  \bibinfo {pages} {263402} (\bibinfo {year} {2020})}\BibitemShut {NoStop}%
\bibitem [{\citenamefont {Xie}\ \emph {et~al.}(2025)\citenamefont {Xie},
  \citenamefont {Li}, \citenamefont {Zhou}, \citenamefont {Luo}, \citenamefont
  {Wang}, \citenamefont {Nie}, \citenamefont {Shen}, \citenamefont {Chen},
  \citenamefont {Yao},\ and\ \citenamefont {Pan}}]{Xie2025Feshbach}%
  \BibitemOpen
  \bibfield  {author} {\bibinfo {author} {\bibfnamefont {K.}~\bibnamefont
  {Xie}}, \bibinfo {author} {\bibfnamefont {X.}~\bibnamefont {Li}}, \bibinfo
  {author} {\bibfnamefont {Y.-Y.}\ \bibnamefont {Zhou}}, \bibinfo {author}
  {\bibfnamefont {J.-H.}\ \bibnamefont {Luo}}, \bibinfo {author} {\bibfnamefont
  {S.}~\bibnamefont {Wang}}, \bibinfo {author} {\bibfnamefont {Y.-Z.}\
  \bibnamefont {Nie}}, \bibinfo {author} {\bibfnamefont {H.-C.}\ \bibnamefont
  {Shen}}, \bibinfo {author} {\bibfnamefont {Y.-A.}\ \bibnamefont {Chen}},
  \bibinfo {author} {\bibfnamefont {X.-C.}\ \bibnamefont {Yao}},\ and\ \bibinfo
  {author} {\bibfnamefont {J.-W.}\ \bibnamefont {Pan}},\ }\bibfield  {title}
  {\bibinfo {title} {Feshbach spectroscopy of ultracold mixtures of
  $^{6}\mathrm{Li}$ and $^{164}\mathrm{Dy}$ atoms},\ }\href
  {https://doi.org/10.1103/PhysRevA.111.023327} {\bibfield  {journal} {\bibinfo
   {journal} {Phys. Rev. A}\ }\textbf {\bibinfo {volume} {111}},\ \bibinfo
  {pages} {023327} (\bibinfo {year} {2025})}\BibitemShut {NoStop}%
\bibitem [{Note3()}]{Note3}%
  \BibitemOpen
  \bibinfo {note} {According to Refs.~\cite
  {merkel2019magnetic,borkowski2023active}, the ppm-level accuracy of the
  magnetic field has been achieved for fields in the range of
  $100$-$1000~{\protect \rm G}$. This means that the fluctuation of the
  magnetic field is about $0.01~{\protect \rm mG}$.}\BibitemShut {Stop}%
\bibitem [{Note4()}]{Note4}%
  \BibitemOpen
  \bibinfo {note} {If one desires a more accurate Hamiltonian, the
  nonuniformity can be compensated by applying an additional ac Stark shift at
  the edges of the system.}\BibitemShut {Stop}%
\bibitem [{\citenamefont {M\"ogerle}\ \emph {et~al.}(2025)\citenamefont
  {M\"ogerle}, \citenamefont {Brechtelsbauer}, \citenamefont {Gea-Caballero},
  \citenamefont {Prior}, \citenamefont {Emperauger}, \citenamefont {Bornet},
  \citenamefont {Chen}, \citenamefont {Lahaye}, \citenamefont {Browaeys},\ and\
  \citenamefont {B\"uchler}}]{mogerle2025spin1}%
  \BibitemOpen
  \bibfield  {author} {\bibinfo {author} {\bibfnamefont {J.}~\bibnamefont
  {M\"ogerle}}, \bibinfo {author} {\bibfnamefont {K.}~\bibnamefont
  {Brechtelsbauer}}, \bibinfo {author} {\bibfnamefont {A.}~\bibnamefont
  {Gea-Caballero}}, \bibinfo {author} {\bibfnamefont {J.}~\bibnamefont
  {Prior}}, \bibinfo {author} {\bibfnamefont {G.}~\bibnamefont {Emperauger}},
  \bibinfo {author} {\bibfnamefont {G.}~\bibnamefont {Bornet}}, \bibinfo
  {author} {\bibfnamefont {C.}~\bibnamefont {Chen}}, \bibinfo {author}
  {\bibfnamefont {T.}~\bibnamefont {Lahaye}}, \bibinfo {author} {\bibfnamefont
  {A.}~\bibnamefont {Browaeys}},\ and\ \bibinfo {author} {\bibfnamefont
  {H.}~\bibnamefont {B\"uchler}},\ }\bibfield  {title} {\bibinfo {title}
  {Spin-1 {Haldane} phase in a chain of {Rydberg} atoms},\ }\href
  {https://doi.org/10.1103/PRXQuantum.6.020332} {\bibfield  {journal} {\bibinfo
   {journal} {PRX Quantum}\ }\textbf {\bibinfo {volume} {6}},\ \bibinfo {pages}
  {020332} (\bibinfo {year} {2025})}\BibitemShut {NoStop}%
\bibitem [{\citenamefont {Gelfand}(1991)}]{gelfand1991linked}%
  \BibitemOpen
  \bibfield  {author} {\bibinfo {author} {\bibfnamefont {M.~P.}\ \bibnamefont
  {Gelfand}},\ }\bibfield  {title} {\bibinfo {title} {Linked-tetrahedra spin
  chain: {Exact} ground state and excitations},\ }\href@noop {} {\bibfield
  {journal} {\bibinfo  {journal} {Phys. Rev. B}\ }\textbf {\bibinfo {volume}
  {43}},\ \bibinfo {pages} {8644} (\bibinfo {year} {1991})}\BibitemShut
  {NoStop}%
\bibitem [{\citenamefont {Honecker}\ \emph {et~al.}(2000)\citenamefont
  {Honecker}, \citenamefont {Mila},\ and\ \citenamefont
  {Troyer}}]{honecker2000magnetization}%
  \BibitemOpen
  \bibfield  {author} {\bibinfo {author} {\bibfnamefont {A.}~\bibnamefont
  {Honecker}}, \bibinfo {author} {\bibfnamefont {F.}~\bibnamefont {Mila}},\
  and\ \bibinfo {author} {\bibfnamefont {M.}~\bibnamefont {Troyer}},\
  }\bibfield  {title} {\bibinfo {title} {Magnetization plateaux and jumps in a
  class of frustrated ladders: {A} simple route to a complex behaviour},\
  }\href@noop {} {\bibfield  {journal} {\bibinfo  {journal} {Eur. Phys. J. B}\
  }\textbf {\bibinfo {volume} {15}},\ \bibinfo {pages} {227} (\bibinfo {year}
  {2000})}\BibitemShut {NoStop}%
\bibitem [{\citenamefont {Chandra}\ and\ \citenamefont
  {Surendran}(2006)}]{chandra2006exact}%
  \BibitemOpen
  \bibfield  {author} {\bibinfo {author} {\bibfnamefont {V.~R.}\ \bibnamefont
  {Chandra}}\ and\ \bibinfo {author} {\bibfnamefont {N.}~\bibnamefont
  {Surendran}},\ }\bibfield  {title} {\bibinfo {title} {Exact magnetization
  plateaus and phase transitions in {spin-$S$} {Heisenberg} antiferromagnets in
  arbitrary dimensions},\ }\href@noop {} {\bibfield  {journal} {\bibinfo
  {journal} {Phys. Rev. B}\ }\textbf {\bibinfo {volume} {74}},\ \bibinfo
  {pages} {024421} (\bibinfo {year} {2006})}\BibitemShut {NoStop}%
\bibitem [{\citenamefont {Kohshiro}\ \emph {et~al.}(2021)\citenamefont
  {Kohshiro}, \citenamefont {Kaneko}, \citenamefont {Morita}, \citenamefont
  {Katsura},\ and\ \citenamefont {Kawashima}}]{kohshiro2021multiple}%
  \BibitemOpen
  \bibfield  {author} {\bibinfo {author} {\bibfnamefont {H.}~\bibnamefont
  {Kohshiro}}, \bibinfo {author} {\bibfnamefont {R.}~\bibnamefont {Kaneko}},
  \bibinfo {author} {\bibfnamefont {S.}~\bibnamefont {Morita}}, \bibinfo
  {author} {\bibfnamefont {H.}~\bibnamefont {Katsura}},\ and\ \bibinfo {author}
  {\bibfnamefont {N.}~\bibnamefont {Kawashima}},\ }\bibfield  {title} {\bibinfo
  {title} {Multiple magnetization plateaus induced by farther neighbor
  interactions in an {$S=1$} two-leg {Heisenberg} spin ladder},\ }\href@noop {}
  {\bibfield  {journal} {\bibinfo  {journal} {Phys. Rev. B}\ }\textbf {\bibinfo
  {volume} {104}},\ \bibinfo {pages} {214409} (\bibinfo {year}
  {2021})}\BibitemShut {NoStop}%
\bibitem [{\citenamefont {Hida}(1992)}]{hida1992crossover}%
  \BibitemOpen
  \bibfield  {author} {\bibinfo {author} {\bibfnamefont {K.}~\bibnamefont
  {Hida}},\ }\bibfield  {title} {\bibinfo {title} {Crossover between the
  {Haldane-gap} phase and the dimer phase in the spin-1/2 alternating
  {Heisenberg} chain},\ }\href@noop {} {\bibfield  {journal} {\bibinfo
  {journal} {Phys. Rev. B}\ }\textbf {\bibinfo {volume} {45}},\ \bibinfo
  {pages} {2207} (\bibinfo {year} {1992})}\BibitemShut {NoStop}%
\bibitem [{\citenamefont {Hung}\ and\ \citenamefont
  {Gong}(2005)}]{hung2005numerical}%
  \BibitemOpen
  \bibfield  {author} {\bibinfo {author} {\bibfnamefont {H.-H.}\ \bibnamefont
  {Hung}}\ and\ \bibinfo {author} {\bibfnamefont {C.-D.}\ \bibnamefont
  {Gong}},\ }\bibfield  {title} {\bibinfo {title} {Numerical evidence of a
  spin-1/2 chain approaching a spin-1 chain},\ }\href@noop {} {\bibfield
  {journal} {\bibinfo  {journal} {Phys. Rev. B}\ }\textbf {\bibinfo {volume}
  {71}},\ \bibinfo {pages} {054413} (\bibinfo {year} {2005})}\BibitemShut
  {NoStop}%
\bibitem [{\citenamefont {Haldane}(1983{\natexlab{a}})}]{Haldane1983nonlinear}%
  \BibitemOpen
  \bibfield  {author} {\bibinfo {author} {\bibfnamefont {F.~D.~M.}\
  \bibnamefont {Haldane}},\ }\bibfield  {title} {\bibinfo {title} {Nonlinear
  field theory of large-spin {Heisenberg} antiferromagnets: Semiclassically
  quantized solitons of the one-dimensional easy-axis {N\'eel} state},\
  }\href@noop {} {\bibfield  {journal} {\bibinfo  {journal} {Phys. Rev. Lett.}\
  }\textbf {\bibinfo {volume} {50}},\ \bibinfo {pages} {1153} (\bibinfo {year}
  {1983}{\natexlab{a}})}\BibitemShut {NoStop}%
\bibitem [{\citenamefont {Haldane}(1983{\natexlab{b}})}]{haldane1983continuum}%
  \BibitemOpen
  \bibfield  {author} {\bibinfo {author} {\bibfnamefont {F.~D.~M.}\
  \bibnamefont {Haldane}},\ }\bibfield  {title} {\bibinfo {title} {Continuum
  dynamics of the {1-D} {Heisenberg} antiferromagnet: Identification with the
  {O(3)} nonlinear sigma model},\ }\href@noop {} {\bibfield  {journal}
  {\bibinfo  {journal} {Phys. lett. A}\ }\textbf {\bibinfo {volume} {93}},\
  \bibinfo {pages} {464} (\bibinfo {year} {1983}{\natexlab{b}})}\BibitemShut
  {NoStop}%
\bibitem [{\citenamefont {Affleck}\ \emph {et~al.}(1987)\citenamefont
  {Affleck}, \citenamefont {Kennedy}, \citenamefont {Lieb},\ and\ \citenamefont
  {Tasaki}}]{affleck1987rigorous}%
  \BibitemOpen
  \bibfield  {author} {\bibinfo {author} {\bibfnamefont {I.}~\bibnamefont
  {Affleck}}, \bibinfo {author} {\bibfnamefont {T.}~\bibnamefont {Kennedy}},
  \bibinfo {author} {\bibfnamefont {E.~H.}\ \bibnamefont {Lieb}},\ and\
  \bibinfo {author} {\bibfnamefont {H.}~\bibnamefont {Tasaki}},\ }\bibfield
  {title} {\bibinfo {title} {Rigorous results on valence-bond ground states in
  antiferromagnets},\ }\href@noop {} {\bibfield  {journal} {\bibinfo  {journal}
  {Phys. Rev. Lett.}\ }\textbf {\bibinfo {volume} {59}},\ \bibinfo {pages}
  {799} (\bibinfo {year} {1987})}\BibitemShut {NoStop}%
\bibitem [{\citenamefont {Affleck}\ \emph {et~al.}(1988)\citenamefont
  {Affleck}, \citenamefont {Kennedy}, \citenamefont {Lieb},\ and\ \citenamefont
  {Tasaki}}]{affleck1988valence}%
  \BibitemOpen
  \bibfield  {author} {\bibinfo {author} {\bibfnamefont {I.}~\bibnamefont
  {Affleck}}, \bibinfo {author} {\bibfnamefont {T.}~\bibnamefont {Kennedy}},
  \bibinfo {author} {\bibfnamefont {E.~H.}\ \bibnamefont {Lieb}},\ and\
  \bibinfo {author} {\bibfnamefont {H.}~\bibnamefont {Tasaki}},\ }\bibfield
  {title} {\bibinfo {title} {Valence bond ground states in isotropic quantum
  antiferromagnets},\ }\href@noop {} {\bibfield  {journal} {\bibinfo  {journal}
  {Commun. Math. Phys.}\ }\textbf {\bibinfo {volume} {115}},\ \bibinfo {pages}
  {477} (\bibinfo {year} {1988})}\BibitemShut {NoStop}%
\bibitem [{\citenamefont {Sekino}\ \emph {et~al.}(2024)\citenamefont {Sekino},
  \citenamefont {Ominato}, \citenamefont {Tajima}, \citenamefont {Uchino},\
  and\ \citenamefont {Matsuo}}]{sekino2024thermomagnetic}%
  \BibitemOpen
  \bibfield  {author} {\bibinfo {author} {\bibfnamefont {Y.}~\bibnamefont
  {Sekino}}, \bibinfo {author} {\bibfnamefont {Y.}~\bibnamefont {Ominato}},
  \bibinfo {author} {\bibfnamefont {H.}~\bibnamefont {Tajima}}, \bibinfo
  {author} {\bibfnamefont {S.}~\bibnamefont {Uchino}},\ and\ \bibinfo {author}
  {\bibfnamefont {M.}~\bibnamefont {Matsuo}},\ }\bibfield  {title} {\bibinfo
  {title} {Thermomagnetic anomalies by magnonic criticality in ultracold atomic
  transport},\ }\href@noop {} {\bibfield  {journal} {\bibinfo  {journal} {Phys.
  Rev. Lett.}\ }\textbf {\bibinfo {volume} {133}},\ \bibinfo {pages} {163402}
  (\bibinfo {year} {2024})}\BibitemShut {NoStop}%
\bibitem [{\citenamefont {Sekino}\ \emph {et~al.}(2025)\citenamefont {Sekino},
  \citenamefont {Ominato}, \citenamefont {Tajima}, \citenamefont {Uchino},\
  and\ \citenamefont {Matsuo}}]{sekino2025thermomagnetic}%
  \BibitemOpen
  \bibfield  {author} {\bibinfo {author} {\bibfnamefont {Y.}~\bibnamefont
  {Sekino}}, \bibinfo {author} {\bibfnamefont {Y.}~\bibnamefont {Ominato}},
  \bibinfo {author} {\bibfnamefont {H.}~\bibnamefont {Tajima}}, \bibinfo
  {author} {\bibfnamefont {S.}~\bibnamefont {Uchino}},\ and\ \bibinfo {author}
  {\bibfnamefont {M.}~\bibnamefont {Matsuo}},\ }\bibfield  {title} {\bibinfo
  {title} {Thermomagnetic anomalies in quantum magnon transport caused by
  tunable junction geometries in cold atomic systems},\ }\href@noop {}
  {\bibfield  {journal} {\bibinfo  {journal} {Phys. Rev. A}\ }\textbf {\bibinfo
  {volume} {111}},\ \bibinfo {pages} {033312} (\bibinfo {year}
  {2025})}\BibitemShut {NoStop}%
\bibitem [{\citenamefont {Fujimoto}\ and\ \citenamefont
  {Sasamoto}(2024)}]{fujimoto2024quantum}%
  \BibitemOpen
  \bibfield  {author} {\bibinfo {author} {\bibfnamefont {K.}~\bibnamefont
  {Fujimoto}}\ and\ \bibinfo {author} {\bibfnamefont {T.}~\bibnamefont
  {Sasamoto}},\ }\bibfield  {title} {\bibinfo {title} {Quantum transport in
  interacting spin chains: {Exact} derivation of the {GUE} {Tracy-Widom}
  distribution},\ }\href@noop {} {\bibfield  {journal} {\bibinfo  {journal}
  {arXiv:2412.20147}\ } (\bibinfo {year} {2024})}\BibitemShut {NoStop}%
\bibitem [{\citenamefont {Ljubotina}\ \emph {et~al.}(2017)\citenamefont
  {Ljubotina}, \citenamefont {{\v{Z}}nidari{\v{c}}},\ and\ \citenamefont
  {Prosen}}]{Ljubotina2017}%
  \BibitemOpen
  \bibfield  {author} {\bibinfo {author} {\bibfnamefont {M.}~\bibnamefont
  {Ljubotina}}, \bibinfo {author} {\bibfnamefont {M.}~\bibnamefont
  {{\v{Z}}nidari{\v{c}}}},\ and\ \bibinfo {author} {\bibfnamefont
  {T.}~\bibnamefont {Prosen}},\ }\bibfield  {title} {\bibinfo {title} {Spin
  diffusion from an inhomogeneous quench in an integrable system},\ }\href@noop
  {} {\bibfield  {journal} {\bibinfo  {journal} {Nat. Commun.}\ }\textbf
  {\bibinfo {volume} {8}},\ \bibinfo {pages} {16117} (\bibinfo {year}
  {2017})}\BibitemShut {NoStop}%
\bibitem [{\citenamefont {Ljubotina}\ \emph {et~al.}(2019)\citenamefont
  {Ljubotina}, \citenamefont {{\v{Z}}nidari{\v{c}}},\ and\ \citenamefont
  {Prosen}}]{Ljubotina2019}%
  \BibitemOpen
  \bibfield  {author} {\bibinfo {author} {\bibfnamefont {M.}~\bibnamefont
  {Ljubotina}}, \bibinfo {author} {\bibfnamefont {M.}~\bibnamefont
  {{\v{Z}}nidari{\v{c}}}},\ and\ \bibinfo {author} {\bibfnamefont
  {T.}~\bibnamefont {Prosen}},\ }\bibfield  {title} {\bibinfo {title}
  {{Kardar-Parisi-Zhang} physics in the quantum {Heisenberg} magnet},\
  }\href@noop {} {\bibfield  {journal} {\bibinfo  {journal} {Phys. Rev. Lett.}\
  }\textbf {\bibinfo {volume} {122}},\ \bibinfo {pages} {210602} (\bibinfo
  {year} {2019})}\BibitemShut {NoStop}%
\bibitem [{\citenamefont {Takeuchi}\ \emph {et~al.}(2025)\citenamefont
  {Takeuchi}, \citenamefont {Takasan}, \citenamefont {Busani}, \citenamefont
  {Ferrari}, \citenamefont {Vasseur},\ and\ \citenamefont
  {De~Nardis}}]{takeuchi2025partial}%
  \BibitemOpen
  \bibfield  {author} {\bibinfo {author} {\bibfnamefont {K.~A.}\ \bibnamefont
  {Takeuchi}}, \bibinfo {author} {\bibfnamefont {K.}~\bibnamefont {Takasan}},
  \bibinfo {author} {\bibfnamefont {O.}~\bibnamefont {Busani}}, \bibinfo
  {author} {\bibfnamefont {P.~L.}\ \bibnamefont {Ferrari}}, \bibinfo {author}
  {\bibfnamefont {R.}~\bibnamefont {Vasseur}},\ and\ \bibinfo {author}
  {\bibfnamefont {J.}~\bibnamefont {De~Nardis}},\ }\bibfield  {title} {\bibinfo
  {title} {Partial yet definite emergence of the {Kardar-Parisi-Zhang} class in
  isotropic spin chains},\ }\href@noop {} {\bibfield  {journal} {\bibinfo
  {journal} {Phys. Rev. Lett.}\ }\textbf {\bibinfo {volume} {134}},\ \bibinfo
  {pages} {097104} (\bibinfo {year} {2025})}\BibitemShut {NoStop}%
\bibitem [{\citenamefont {Vaillant}\ \emph {et~al.}(2014)\citenamefont
  {Vaillant}, \citenamefont {Jones},\ and\ \citenamefont
  {Potvliege}}]{vaillant2014multichannel}%
  \BibitemOpen
  \bibfield  {author} {\bibinfo {author} {\bibfnamefont {C.}~\bibnamefont
  {Vaillant}}, \bibinfo {author} {\bibfnamefont {M.}~\bibnamefont {Jones}},\
  and\ \bibinfo {author} {\bibfnamefont {R.}~\bibnamefont {Potvliege}},\
  }\bibfield  {title} {\bibinfo {title} {Multichannel quantum defect theory of
  strontium bound {Rydberg} states},\ }\href@noop {} {\bibfield  {journal}
  {\bibinfo  {journal} {J. Phys. B: At. Mol. Opt. Phys.}\ }\textbf {\bibinfo
  {volume} {47}},\ \bibinfo {pages} {155001} (\bibinfo {year}
  {2014})}\BibitemShut {NoStop}%
\bibitem [{\citenamefont {Robertson}\ \emph {et~al.}(2021)\citenamefont
  {Robertson}, \citenamefont {{\v{S}}ibali{\'c}}, \citenamefont {Potvliege},\
  and\ \citenamefont {Jones}}]{robertson2021arc}%
  \BibitemOpen
  \bibfield  {author} {\bibinfo {author} {\bibfnamefont {E.~J.}\ \bibnamefont
  {Robertson}}, \bibinfo {author} {\bibfnamefont {N.}~\bibnamefont
  {{\v{S}}ibali{\'c}}}, \bibinfo {author} {\bibfnamefont {R.~M.}\ \bibnamefont
  {Potvliege}},\ and\ \bibinfo {author} {\bibfnamefont {M.~P.}\ \bibnamefont
  {Jones}},\ }\bibfield  {title} {\bibinfo {title} {Arc 3.0: {An} expanded
  {Python} toolbox for atomic physics calculations},\ }\href@noop {} {\bibfield
   {journal} {\bibinfo  {journal} {Comput. Phys. Commun.}\ }\textbf {\bibinfo
  {volume} {261}},\ \bibinfo {pages} {107814} (\bibinfo {year}
  {2021})}\BibitemShut {NoStop}%
\bibitem [{\citenamefont {Hummel}\ \emph {et~al.}(2024)\citenamefont {Hummel},
  \citenamefont {Weber}, \citenamefont {M{\"o}gerle}, \citenamefont {Menke},
  \citenamefont {King}, \citenamefont {Bloom}, \citenamefont {Hofferberth},\
  and\ \citenamefont {Li}}]{hummel2024engineering}%
  \BibitemOpen
  \bibfield  {author} {\bibinfo {author} {\bibfnamefont {F.}~\bibnamefont
  {Hummel}}, \bibinfo {author} {\bibfnamefont {S.}~\bibnamefont {Weber}},
  \bibinfo {author} {\bibfnamefont {J.}~\bibnamefont {M{\"o}gerle}}, \bibinfo
  {author} {\bibfnamefont {H.}~\bibnamefont {Menke}}, \bibinfo {author}
  {\bibfnamefont {J.}~\bibnamefont {King}}, \bibinfo {author} {\bibfnamefont
  {B.}~\bibnamefont {Bloom}}, \bibinfo {author} {\bibfnamefont
  {S.}~\bibnamefont {Hofferberth}},\ and\ \bibinfo {author} {\bibfnamefont
  {M.}~\bibnamefont {Li}},\ }\bibfield  {title} {\bibinfo {title} {Engineering
  {Rydberg-pair} interactions in divalent atoms with hyperfine-split ionization
  thresholds},\ }\href@noop {} {\bibfield  {journal} {\bibinfo  {journal}
  {Phys. Rev. A}\ }\textbf {\bibinfo {volume} {110}},\ \bibinfo {pages}
  {042821} (\bibinfo {year} {2024})}\BibitemShut {NoStop}%
\bibitem [{\citenamefont {Ravon}\ \emph {et~al.}(2023)\citenamefont {Ravon},
  \citenamefont {M{\'e}haignerie}, \citenamefont {Machu}, \citenamefont
  {Hern{\'a}ndez}, \citenamefont {Favier}, \citenamefont {Raimond},
  \citenamefont {Brune},\ and\ \citenamefont {Sayrin}}]{ravon2023array}%
  \BibitemOpen
  \bibfield  {author} {\bibinfo {author} {\bibfnamefont {B.}~\bibnamefont
  {Ravon}}, \bibinfo {author} {\bibfnamefont {P.}~\bibnamefont
  {M{\'e}haignerie}}, \bibinfo {author} {\bibfnamefont {Y.}~\bibnamefont
  {Machu}}, \bibinfo {author} {\bibfnamefont {A.~D.}\ \bibnamefont
  {Hern{\'a}ndez}}, \bibinfo {author} {\bibfnamefont {M.}~\bibnamefont
  {Favier}}, \bibinfo {author} {\bibfnamefont {J.-M.}\ \bibnamefont {Raimond}},
  \bibinfo {author} {\bibfnamefont {M.}~\bibnamefont {Brune}},\ and\ \bibinfo
  {author} {\bibfnamefont {C.}~\bibnamefont {Sayrin}},\ }\bibfield  {title}
  {\bibinfo {title} {Array of individual circular {Rydberg} atoms trapped in
  optical tweezers},\ }\href@noop {} {\bibfield  {journal} {\bibinfo  {journal}
  {Phys. Rev. Lett.}\ }\textbf {\bibinfo {volume} {131}},\ \bibinfo {pages}
  {093401} (\bibinfo {year} {2023})}\BibitemShut {NoStop}%
\bibitem [{\citenamefont {H{\"o}lzl}\ \emph {et~al.}(2024)\citenamefont
  {H{\"o}lzl}, \citenamefont {G{\"o}tzelmann}, \citenamefont {Pultinevicius},
  \citenamefont {Wirth},\ and\ \citenamefont {Meinert}}]{Holzl2024}%
  \BibitemOpen
  \bibfield  {author} {\bibinfo {author} {\bibfnamefont {C.}~\bibnamefont
  {H{\"o}lzl}}, \bibinfo {author} {\bibfnamefont {A.}~\bibnamefont
  {G{\"o}tzelmann}}, \bibinfo {author} {\bibfnamefont {E.}~\bibnamefont
  {Pultinevicius}}, \bibinfo {author} {\bibfnamefont {M.}~\bibnamefont
  {Wirth}},\ and\ \bibinfo {author} {\bibfnamefont {F.}~\bibnamefont
  {Meinert}},\ }\bibfield  {title} {\bibinfo {title} {Long-lived circular
  {Rydberg} qubits of alkaline-earth atoms in optical tweezers},\ }\href@noop
  {} {\bibfield  {journal} {\bibinfo  {journal} {Phys. Rev. X}\ }\textbf
  {\bibinfo {volume} {14}},\ \bibinfo {pages} {021024} (\bibinfo {year}
  {2024})}\BibitemShut {NoStop}%
\bibitem [{\citenamefont {M\'ehaignerie}\ \emph {et~al.}(2025)\citenamefont
  {M\'ehaignerie}, \citenamefont {Machu}, \citenamefont {Dur\'an~Hern\'andez},
  \citenamefont {Creutzer}, \citenamefont {Papoular}, \citenamefont {Raimond},
  \citenamefont {Sayrin},\ and\ \citenamefont
  {Brune}}]{mehaignerie2025interacting}%
  \BibitemOpen
  \bibfield  {author} {\bibinfo {author} {\bibfnamefont {P.}~\bibnamefont
  {M\'ehaignerie}}, \bibinfo {author} {\bibfnamefont {Y.}~\bibnamefont
  {Machu}}, \bibinfo {author} {\bibfnamefont {A.}~\bibnamefont
  {Dur\'an~Hern\'andez}}, \bibinfo {author} {\bibfnamefont {G.}~\bibnamefont
  {Creutzer}}, \bibinfo {author} {\bibfnamefont {D.}~\bibnamefont {Papoular}},
  \bibinfo {author} {\bibfnamefont {J.}~\bibnamefont {Raimond}}, \bibinfo
  {author} {\bibfnamefont {C.}~\bibnamefont {Sayrin}},\ and\ \bibinfo {author}
  {\bibfnamefont {M.}~\bibnamefont {Brune}},\ }\bibfield  {title} {\bibinfo
  {title} {Interacting circular {Rydberg} atoms trapped in optical tweezers},\
  }\href {https://doi.org/10.1103/PRXQuantum.6.010353} {\bibfield  {journal}
  {\bibinfo  {journal} {PRX Quantum}\ }\textbf {\bibinfo {volume} {6}},\
  \bibinfo {pages} {010353} (\bibinfo {year} {2025})}\BibitemShut {NoStop}%
\bibitem [{\citenamefont {Nguyen}\ \emph {et~al.}(2018)\citenamefont {Nguyen},
  \citenamefont {Raimond}, \citenamefont {Sayrin}, \citenamefont {Cortinas},
  \citenamefont {Cantat-Moltrecht}, \citenamefont {Assemat}, \citenamefont
  {Dotsenko}, \citenamefont {Gleyzes}, \citenamefont {Haroche}, \citenamefont
  {Roux}, \citenamefont {Jolicoeur},\ and\ \citenamefont
  {Brune}}]{nguyen2018towards}%
  \BibitemOpen
  \bibfield  {author} {\bibinfo {author} {\bibfnamefont {T.~L.}\ \bibnamefont
  {Nguyen}}, \bibinfo {author} {\bibfnamefont {J.-M.}\ \bibnamefont {Raimond}},
  \bibinfo {author} {\bibfnamefont {C.}~\bibnamefont {Sayrin}}, \bibinfo
  {author} {\bibfnamefont {R.}~\bibnamefont {Cortinas}}, \bibinfo {author}
  {\bibfnamefont {T.}~\bibnamefont {Cantat-Moltrecht}}, \bibinfo {author}
  {\bibfnamefont {F.}~\bibnamefont {Assemat}}, \bibinfo {author} {\bibfnamefont
  {I.}~\bibnamefont {Dotsenko}}, \bibinfo {author} {\bibfnamefont
  {S.}~\bibnamefont {Gleyzes}}, \bibinfo {author} {\bibfnamefont
  {S.}~\bibnamefont {Haroche}}, \bibinfo {author} {\bibfnamefont
  {G.}~\bibnamefont {Roux}}, \bibinfo {author} {\bibfnamefont {T.}~\bibnamefont
  {Jolicoeur}},\ and\ \bibinfo {author} {\bibfnamefont {M.}~\bibnamefont
  {Brune}},\ }\bibfield  {title} {\bibinfo {title} {Towards quantum simulation
  with circular {Rydberg} atoms},\ }\href@noop {} {\bibfield  {journal}
  {\bibinfo  {journal} {Phys. Rev. X}\ }\textbf {\bibinfo {volume} {8}},\
  \bibinfo {pages} {011032} (\bibinfo {year} {2018})}\BibitemShut {NoStop}%
\bibitem [{\citenamefont {Dobrzyniecki}\ \emph {et~al.}(2025)\citenamefont
  {Dobrzyniecki}, \citenamefont {Heim},\ and\ \citenamefont
  {Tomza}}]{dobrzyniecki2025tunable}%
  \BibitemOpen
  \bibfield  {author} {\bibinfo {author} {\bibfnamefont {J.}~\bibnamefont
  {Dobrzyniecki}}, \bibinfo {author} {\bibfnamefont {P.}~\bibnamefont {Heim}},\
  and\ \bibinfo {author} {\bibfnamefont {M.}~\bibnamefont {Tomza}},\ }\bibfield
   {title} {\bibinfo {title} {Tunable two-species spin models with {Rydberg}
  atoms in circular and elliptical states},\ }\href@noop {} {\bibfield
  {journal} {\bibinfo  {journal} {Phys. Rev. Res.}\ }\textbf {\bibinfo {volume}
  {7}},\ \bibinfo {pages} {013321} (\bibinfo {year} {2025})}\BibitemShut
  {NoStop}%
\bibitem [{\citenamefont {Sheng}\ \emph {et~al.}(2022)\citenamefont {Sheng},
  \citenamefont {Hou}, \citenamefont {He}, \citenamefont {Wang}, \citenamefont
  {Guo}, \citenamefont {Zhuang}, \citenamefont {Mamat}, \citenamefont {Xu},
  \citenamefont {Liu}, \citenamefont {Wang},\ and\ \citenamefont
  {Zhan}}]{sheng2022defect}%
  \BibitemOpen
  \bibfield  {author} {\bibinfo {author} {\bibfnamefont {C.}~\bibnamefont
  {Sheng}}, \bibinfo {author} {\bibfnamefont {J.}~\bibnamefont {Hou}}, \bibinfo
  {author} {\bibfnamefont {X.}~\bibnamefont {He}}, \bibinfo {author}
  {\bibfnamefont {K.}~\bibnamefont {Wang}}, \bibinfo {author} {\bibfnamefont
  {R.}~\bibnamefont {Guo}}, \bibinfo {author} {\bibfnamefont {J.}~\bibnamefont
  {Zhuang}}, \bibinfo {author} {\bibfnamefont {B.}~\bibnamefont {Mamat}},
  \bibinfo {author} {\bibfnamefont {P.}~\bibnamefont {Xu}}, \bibinfo {author}
  {\bibfnamefont {M.}~\bibnamefont {Liu}}, \bibinfo {author} {\bibfnamefont
  {J.}~\bibnamefont {Wang}},\ and\ \bibinfo {author} {\bibfnamefont
  {M.}~\bibnamefont {Zhan}},\ }\bibfield  {title} {\bibinfo {title}
  {Defect-free arbitrary-geometry assembly of mixed-species atom arrays},\
  }\href@noop {} {\bibfield  {journal} {\bibinfo  {journal} {Phys. Rev. Lett.}\
  }\textbf {\bibinfo {volume} {128}},\ \bibinfo {pages} {083202} (\bibinfo
  {year} {2022})}\BibitemShut {NoStop}%
\bibitem [{\citenamefont {Anand}\ \emph {et~al.}(2024)\citenamefont {Anand},
  \citenamefont {Bradley}, \citenamefont {White}, \citenamefont {Ramesh},
  \citenamefont {Singh},\ and\ \citenamefont {Bernien}}]{anand2024dual}%
  \BibitemOpen
  \bibfield  {author} {\bibinfo {author} {\bibfnamefont {S.}~\bibnamefont
  {Anand}}, \bibinfo {author} {\bibfnamefont {C.~E.}\ \bibnamefont {Bradley}},
  \bibinfo {author} {\bibfnamefont {R.}~\bibnamefont {White}}, \bibinfo
  {author} {\bibfnamefont {V.}~\bibnamefont {Ramesh}}, \bibinfo {author}
  {\bibfnamefont {K.}~\bibnamefont {Singh}},\ and\ \bibinfo {author}
  {\bibfnamefont {H.}~\bibnamefont {Bernien}},\ }\bibfield  {title} {\bibinfo
  {title} {A dual-species {Rydberg} array},\ }\href@noop {} {\bibfield
  {journal} {\bibinfo  {journal} {Nat. Phys.}\ }\textbf {\bibinfo {volume}
  {20}},\ \bibinfo {pages} {1744} (\bibinfo {year} {2024})}\BibitemShut
  {NoStop}%
\bibitem [{\citenamefont {Nakamura}\ \emph {et~al.}(2024)\citenamefont
  {Nakamura}, \citenamefont {Kusano}, \citenamefont {Yokoyama}, \citenamefont
  {Saito}, \citenamefont {Higashi}, \citenamefont {Ozawa}, \citenamefont
  {Takano}, \citenamefont {Takasu},\ and\ \citenamefont
  {Takahashi}}]{nakamura2024hybrid}%
  \BibitemOpen
  \bibfield  {author} {\bibinfo {author} {\bibfnamefont {Y.}~\bibnamefont
  {Nakamura}}, \bibinfo {author} {\bibfnamefont {T.}~\bibnamefont {Kusano}},
  \bibinfo {author} {\bibfnamefont {R.}~\bibnamefont {Yokoyama}}, \bibinfo
  {author} {\bibfnamefont {K.}~\bibnamefont {Saito}}, \bibinfo {author}
  {\bibfnamefont {K.}~\bibnamefont {Higashi}}, \bibinfo {author} {\bibfnamefont
  {N.}~\bibnamefont {Ozawa}}, \bibinfo {author} {\bibfnamefont
  {T.}~\bibnamefont {Takano}}, \bibinfo {author} {\bibfnamefont
  {Y.}~\bibnamefont {Takasu}},\ and\ \bibinfo {author} {\bibfnamefont
  {Y.}~\bibnamefont {Takahashi}},\ }\bibfield  {title} {\bibinfo {title}
  {Hybrid atom tweezer array of nuclear spin and optical clock qubits},\
  }\href@noop {} {\bibfield  {journal} {\bibinfo  {journal} {Phys. Rev. X}\
  }\textbf {\bibinfo {volume} {14}},\ \bibinfo {pages} {041062} (\bibinfo
  {year} {2024})}\BibitemShut {NoStop}%
\bibitem [{\citenamefont {Wei}\ \emph {et~al.}(2024)\citenamefont {Wei},
  \citenamefont {Wei}, \citenamefont {Li},\ and\ \citenamefont
  {Yan}}]{wei2024dual}%
  \BibitemOpen
  \bibfield  {author} {\bibinfo {author} {\bibfnamefont {Y.}~\bibnamefont
  {Wei}}, \bibinfo {author} {\bibfnamefont {K.}~\bibnamefont {Wei}}, \bibinfo
  {author} {\bibfnamefont {S.}~\bibnamefont {Li}},\ and\ \bibinfo {author}
  {\bibfnamefont {B.}~\bibnamefont {Yan}},\ }\bibfield  {title} {\bibinfo
  {title} {Dual-species optical tweezer for {Rb} and {K} atoms},\ }\href@noop
  {} {\bibfield  {journal} {\bibinfo  {journal} {Phys. Rev. A}\ }\textbf
  {\bibinfo {volume} {110}},\ \bibinfo {pages} {043118} (\bibinfo {year}
  {2024})}\BibitemShut {NoStop}%
\bibitem [{\citenamefont {Gerhardt}\ \emph {et~al.}(1998)\citenamefont
  {Gerhardt}, \citenamefont {M{\"u}tter},\ and\ \citenamefont
  {Kr{\"o}ger}}]{gerhardt1998metamagnetism}%
  \BibitemOpen
  \bibfield  {author} {\bibinfo {author} {\bibfnamefont {C.}~\bibnamefont
  {Gerhardt}}, \bibinfo {author} {\bibfnamefont {K.-H.}\ \bibnamefont
  {M{\"u}tter}},\ and\ \bibinfo {author} {\bibfnamefont {H.}~\bibnamefont
  {Kr{\"o}ger}},\ }\bibfield  {title} {\bibinfo {title} {Metamagnetism in the
  {XXZ} model with next-to-nearest-neighbor coupling},\ }\href@noop {}
  {\bibfield  {journal} {\bibinfo  {journal} {Phys. Rev. B}\ }\textbf {\bibinfo
  {volume} {57}},\ \bibinfo {pages} {11504} (\bibinfo {year}
  {1998})}\BibitemShut {NoStop}%
\bibitem [{\citenamefont {Batista}(2009)}]{batista2009canted}%
  \BibitemOpen
  \bibfield  {author} {\bibinfo {author} {\bibfnamefont {C.}~\bibnamefont
  {Batista}},\ }\bibfield  {title} {\bibinfo {title} {Canted spiral: {An} exact
  ground state of {XXZ} zigzag spin ladders},\ }\href@noop {} {\bibfield
  {journal} {\bibinfo  {journal} {Phys. Rev. B}\ }\textbf {\bibinfo {volume}
  {80}},\ \bibinfo {pages} {180406} (\bibinfo {year} {2009})}\BibitemShut
  {NoStop}%
\bibitem [{\citenamefont {Sable}\ \emph {et~al.}(2025)\citenamefont {Sable},
  \citenamefont {Myers},\ and\ \citenamefont {Scarola}}]{sable2025toward}%
  \BibitemOpen
  \bibfield  {author} {\bibinfo {author} {\bibfnamefont {H.}~\bibnamefont
  {Sable}}, \bibinfo {author} {\bibfnamefont {N.~M.}\ \bibnamefont {Myers}},\
  and\ \bibinfo {author} {\bibfnamefont {V.~W.}\ \bibnamefont {Scarola}},\
  }\bibfield  {title} {\bibinfo {title} {Toward quantum analog simulation of
  many-body supersymmetry with {Rydberg} atom arrays},\ }\href
  {https://doi.org/10.1103/746s-fv7x} {\bibfield  {journal} {\bibinfo
  {journal} {Phys. Rev. Lett.}\ }\textbf {\bibinfo {volume} {135}},\ \bibinfo
  {pages} {033401} (\bibinfo {year} {2025})}\BibitemShut {NoStop}%
\bibitem [{\citenamefont {Kunimi}\ and\ \citenamefont
  {Tomita}(2025)}]{kunimi_2025_16349138}%
  \BibitemOpen
  \bibfield  {author} {\bibinfo {author} {\bibfnamefont {M.}~\bibnamefont
  {Kunimi}}\ and\ \bibinfo {author} {\bibfnamefont {T.}~\bibnamefont
  {Tomita}},\ }\bibfield  {title} {\bibinfo {title} {Dataset for "{Proposal}
  for realizing {Heisenberg}- type quantum-spin models in {Rydberg} atom
  quantum simulators"},\ }\href {https://doi.org/10.5281/zenodo.16349138}
  {10.5281/zenodo.16349138} (\bibinfo {year} {2025})\BibitemShut {NoStop}%
\bibitem [{\citenamefont {Merkel}\ \emph {et~al.}(2019)\citenamefont {Merkel},
  \citenamefont {Thirumalai}, \citenamefont {Tarlton}, \citenamefont
  {Sch{\"a}fer}, \citenamefont {Ballance}, \citenamefont {Harty},\ and\
  \citenamefont {Lucas}}]{merkel2019magnetic}%
  \BibitemOpen
  \bibfield  {author} {\bibinfo {author} {\bibfnamefont {B.}~\bibnamefont
  {Merkel}}, \bibinfo {author} {\bibfnamefont {K.}~\bibnamefont {Thirumalai}},
  \bibinfo {author} {\bibfnamefont {J.}~\bibnamefont {Tarlton}}, \bibinfo
  {author} {\bibfnamefont {V.}~\bibnamefont {Sch{\"a}fer}}, \bibinfo {author}
  {\bibfnamefont {C.}~\bibnamefont {Ballance}}, \bibinfo {author}
  {\bibfnamefont {T.}~\bibnamefont {Harty}},\ and\ \bibinfo {author}
  {\bibfnamefont {D.}~\bibnamefont {Lucas}},\ }\bibfield  {title} {\bibinfo
  {title} {Magnetic field stabilization system for atomic physics
  experiments},\ }\href@noop {} {\bibfield  {journal} {\bibinfo  {journal}
  {Rev. Sci. Instrum.}\ }\textbf {\bibinfo {volume} {90}},\ \bibinfo {pages}
  {044702} (\bibinfo {year} {2019})}\BibitemShut {NoStop}%
\bibitem [{\citenamefont {Borkowski}\ \emph {et~al.}(2023)\citenamefont
  {Borkowski}, \citenamefont {Reichs{\"o}llner}, \citenamefont {Thekkeppatt},
  \citenamefont {Barb{\'e}}, \citenamefont {van Roon}, \citenamefont {van
  Druten},\ and\ \citenamefont {Schreck}}]{borkowski2023active}%
  \BibitemOpen
  \bibfield  {author} {\bibinfo {author} {\bibfnamefont {M.}~\bibnamefont
  {Borkowski}}, \bibinfo {author} {\bibfnamefont {L.}~\bibnamefont
  {Reichs{\"o}llner}}, \bibinfo {author} {\bibfnamefont {P.}~\bibnamefont
  {Thekkeppatt}}, \bibinfo {author} {\bibfnamefont {V.}~\bibnamefont
  {Barb{\'e}}}, \bibinfo {author} {\bibfnamefont {T.}~\bibnamefont {van Roon}},
  \bibinfo {author} {\bibfnamefont {K.}~\bibnamefont {van Druten}},\ and\
  \bibinfo {author} {\bibfnamefont {F.}~\bibnamefont {Schreck}},\ }\bibfield
  {title} {\bibinfo {title} {Active stabilization of kilogauss magnetic fields
  to the ppm level for magnetoassociation on ultranarrow {Feshbach}
  resonances},\ }\href@noop {} {\bibfield  {journal} {\bibinfo  {journal} {Rev.
  Sci. Instrum.}\ }\textbf {\bibinfo {volume} {94}},\ \bibinfo {pages} {073202}
  (\bibinfo {year} {2023})}\BibitemShut {NoStop}%
\end{thebibliography}%
\newpage
\begin{widetext}
\renewcommand{\thefigure}{S\arabic{figure}}
\renewcommand{\thetable}{S\arabic{table}}
\renewcommand{\theequation}{S\arabic{equation}}
\renewcommand{\thesection}{S\arabic{section}}
\renewcommand{\thesubsection}{S\arabic{section}.\arabic{subsection}}
\setcounter{figure}{0}

\clearpage
\begin{center}
{\bf Supplemental material for ``Proposal for realizing Heisenberg-type quantum-spin models in Rydberg-atom quantum simulators"}
\end{center}

\section{Results for other atomic species}\label{sec:other_atomic_species}

In this section, we show the results for the pairs $\cket{nS_{1/2}, 1/2}\cket{(n+1)S_{1/2}, 1/2}$ and $\cket{nS_{1/2}, -1/2}\cket{(n+1)S_{1/2}, -1/2}$ for ${}^7$Li, ${}^{23}$Na, ${}^{39}$K, ${}^{87}$Rb, and ${}^{133}$Cs atoms. In the following, we focus on the following quantities: the anisotropy parameter $\delta$, $C_6$, $C_6^{\alpha\beta}$, and $R_{\rm c}$. These quantities are calculated in the range $0~{\rm G} \le B \le 200~{\rm G}$ and $20 \le n \le 100$. For the case of the ${}^7$Li atom, we extend the calculation range to $0~{\rm G} \le B \le 400~{\rm G}$, since no F\"orster resonance appears below $200~{\rm G}$.

\subsection{$|nS_{1/2},1/2\rangle|(n+1)S_{1/2},1/2\rangle$ pair}\label{subsec:positive_mJ_pair}
Here, we show the results for $|nS_{1/2},1/2\rangle|(n+1)S_{1/2},1/2\rangle$ pair in Figs.~\ref{fig:supple_delta_positive_mJ}--\ref{fig:supple_rc_positive_mJ}.

\begin{figure}[h]
\centering
\includegraphics[width=17cm,clip]{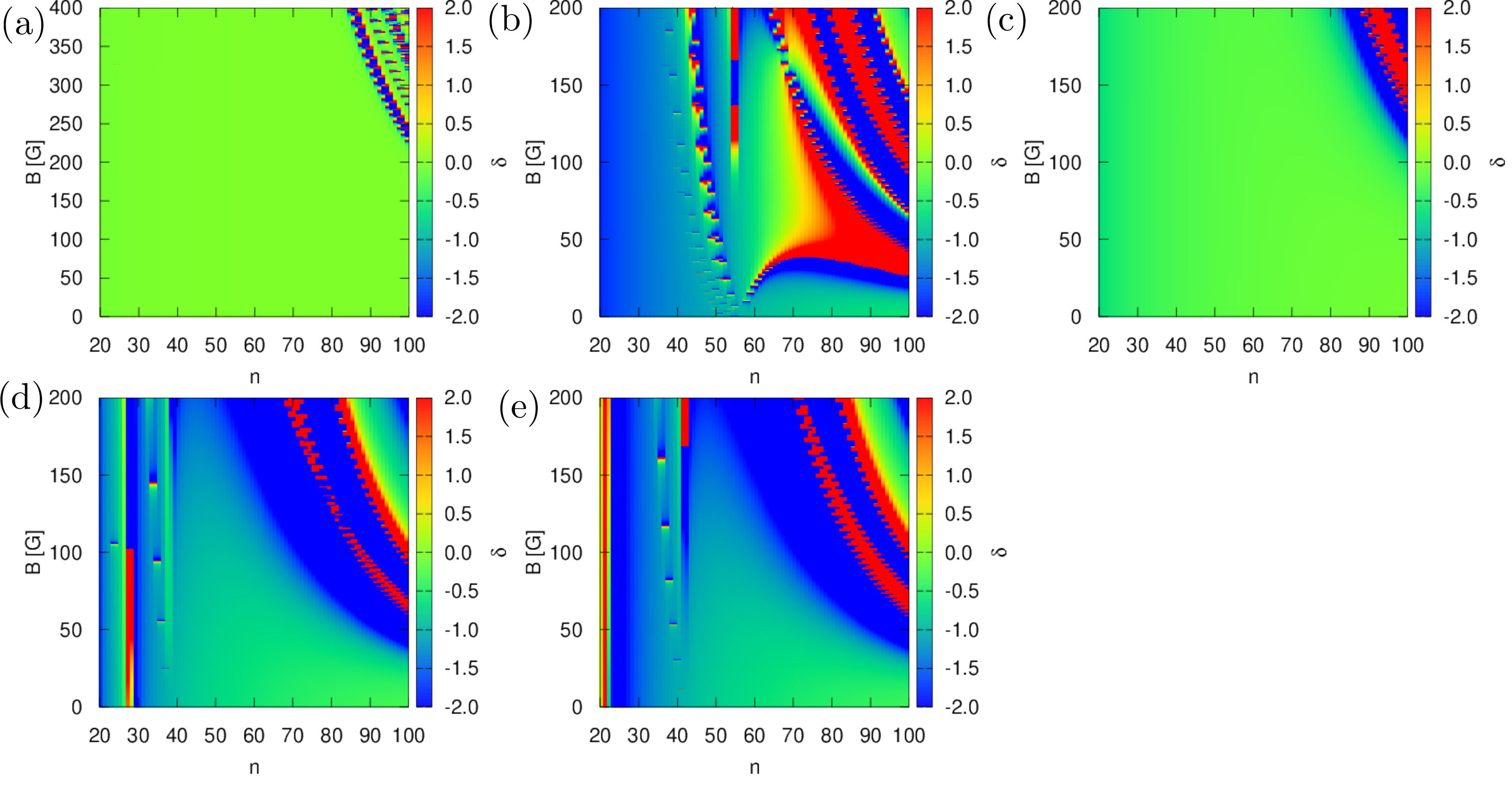}
\caption{Magnetic field and principal quantum number dependence of the anisotropy parameter of the pair $|nS_{1/2},m_J=1/2\rangle$ and $|(n+1)S_{1/2},m_J=1/2\rangle$ for $\theta=\pi/2$. (a) ${}^7$Li atom, (b) ${}^{23}$Na atom, (c) ${}^{39}$K atom, (d) ${}^{87}$Rb atom, (e) ${}^{133}$Cs atom.
}
\label{fig:supple_delta_positive_mJ}
\vspace{-0.75em}
\end{figure}%

\begin{figure}[t]
\centering
\includegraphics[width=17cm,clip]{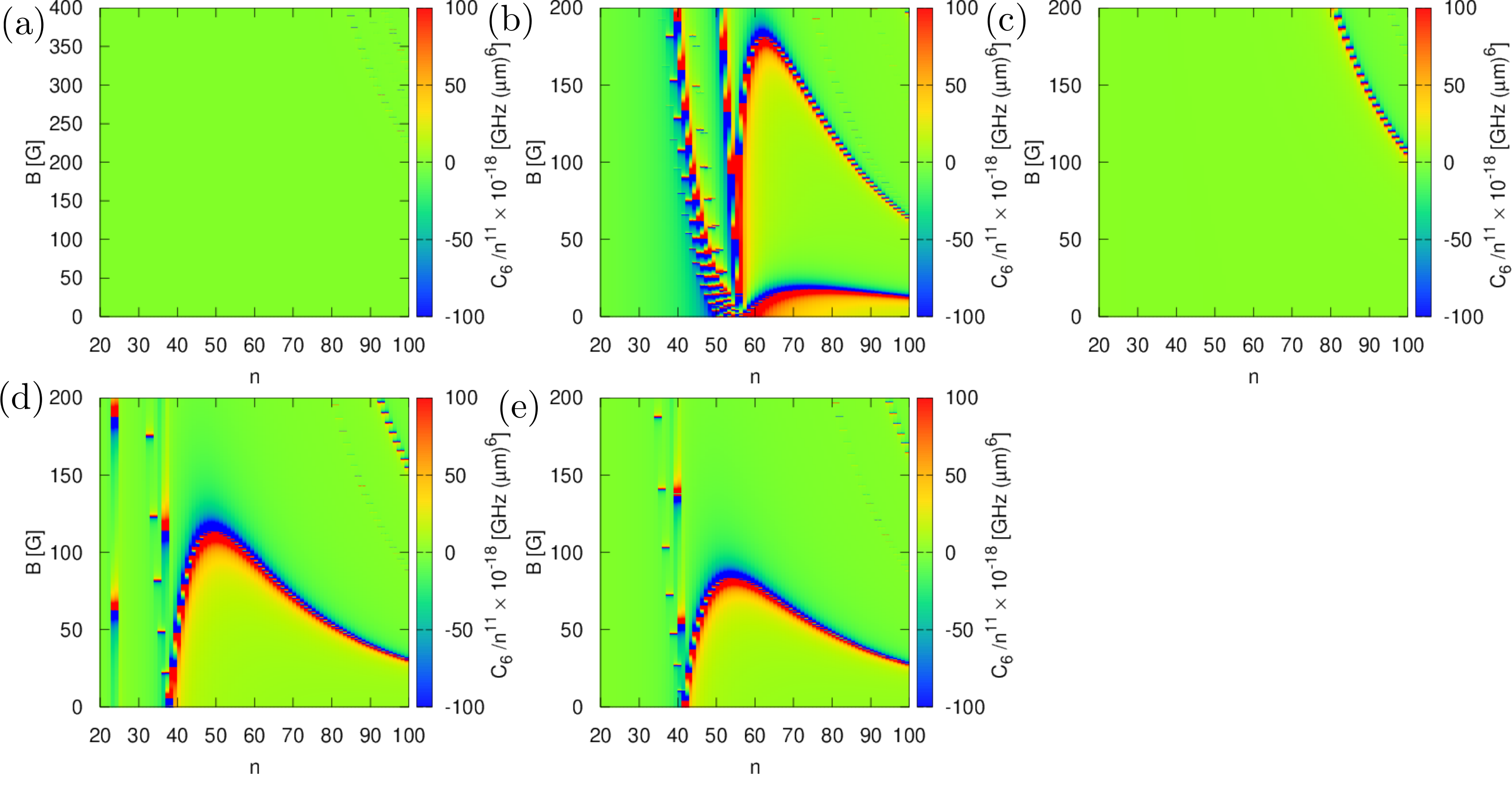}
\caption{Magnetic field and principal quantum number dependence of $C_6$ of the pair $|nS_{1/2},m_J=1/2\rangle$ and $|(n+1)S_{1/2},m_J=1/2\rangle$ for $\theta=\pi/2$. (a) ${}^7$Li atom, (b) ${}^{23}$Na atom, (c) ${}^{39}$K atom, (d) ${}^{87}$Rb atom, (e) ${}^{133}$Cs atom.
}
\label{fig:supple_c6_positive_mJ}
\vspace{-0.75em}
\end{figure}%

\begin{figure}[t]
\centering
\includegraphics[width=17cm,clip]{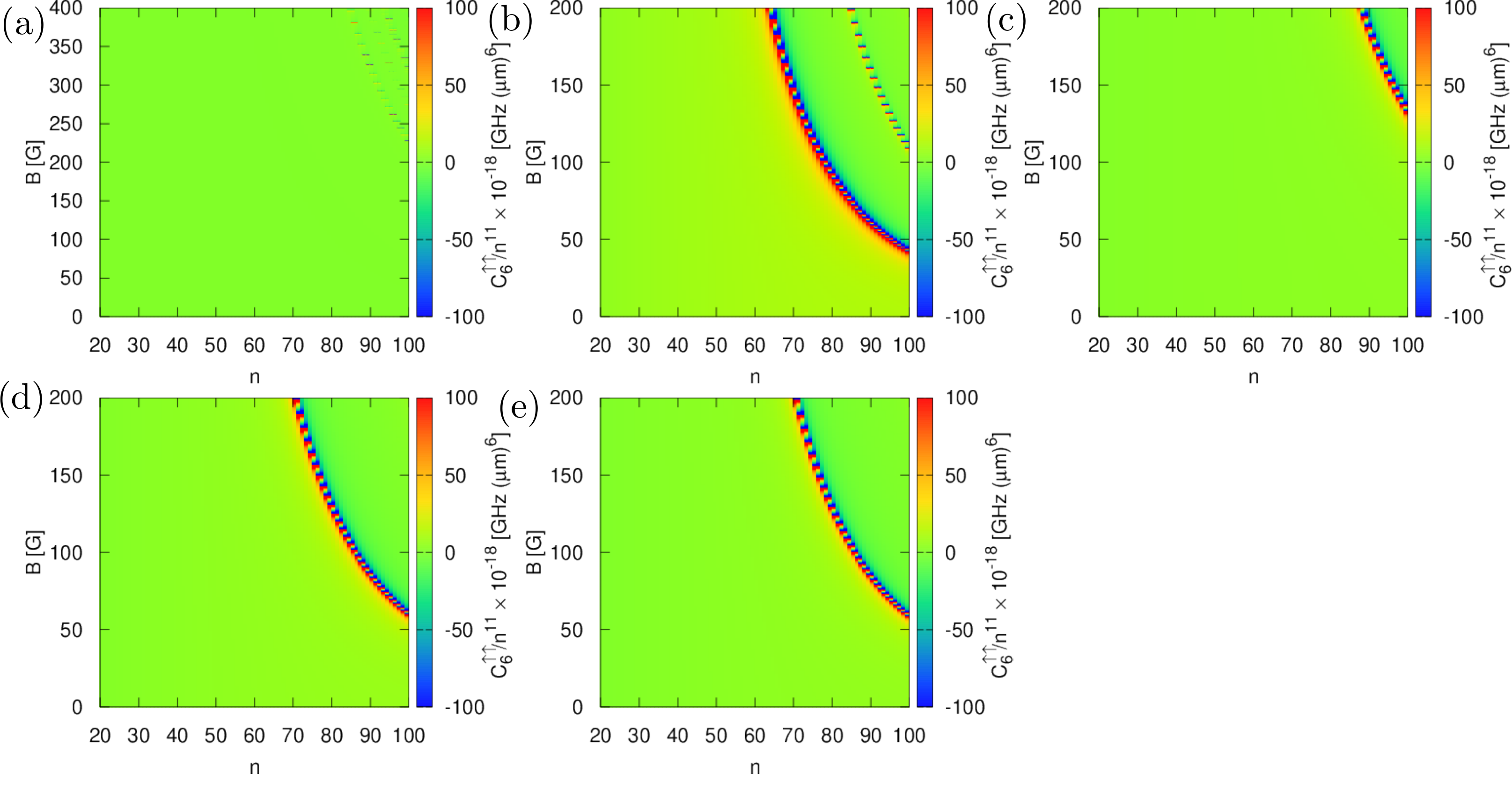}
\caption{Magnetic field and principal quantum number dependence of $C_6^{\uparrow\uparrow}$ of the pair $|nS_{1/2},m_J=1/2\rangle$ and $|(n+1)S_{1/2},m_J=1/2\rangle$ for $\theta=\pi/2$. (a) ${}^7$Li atom, (b) ${}^{23}$Na atom, (c) ${}^{39}$K atom, (d) ${}^{87}$Rb atom, (e) ${}^{133}$Cs atom.
}
\label{fig:supple_c6_uu_positive_mJ}
\vspace{-0.75em}
\end{figure}%

\begin{figure}[t]
\centering
\includegraphics[width=17cm,clip]{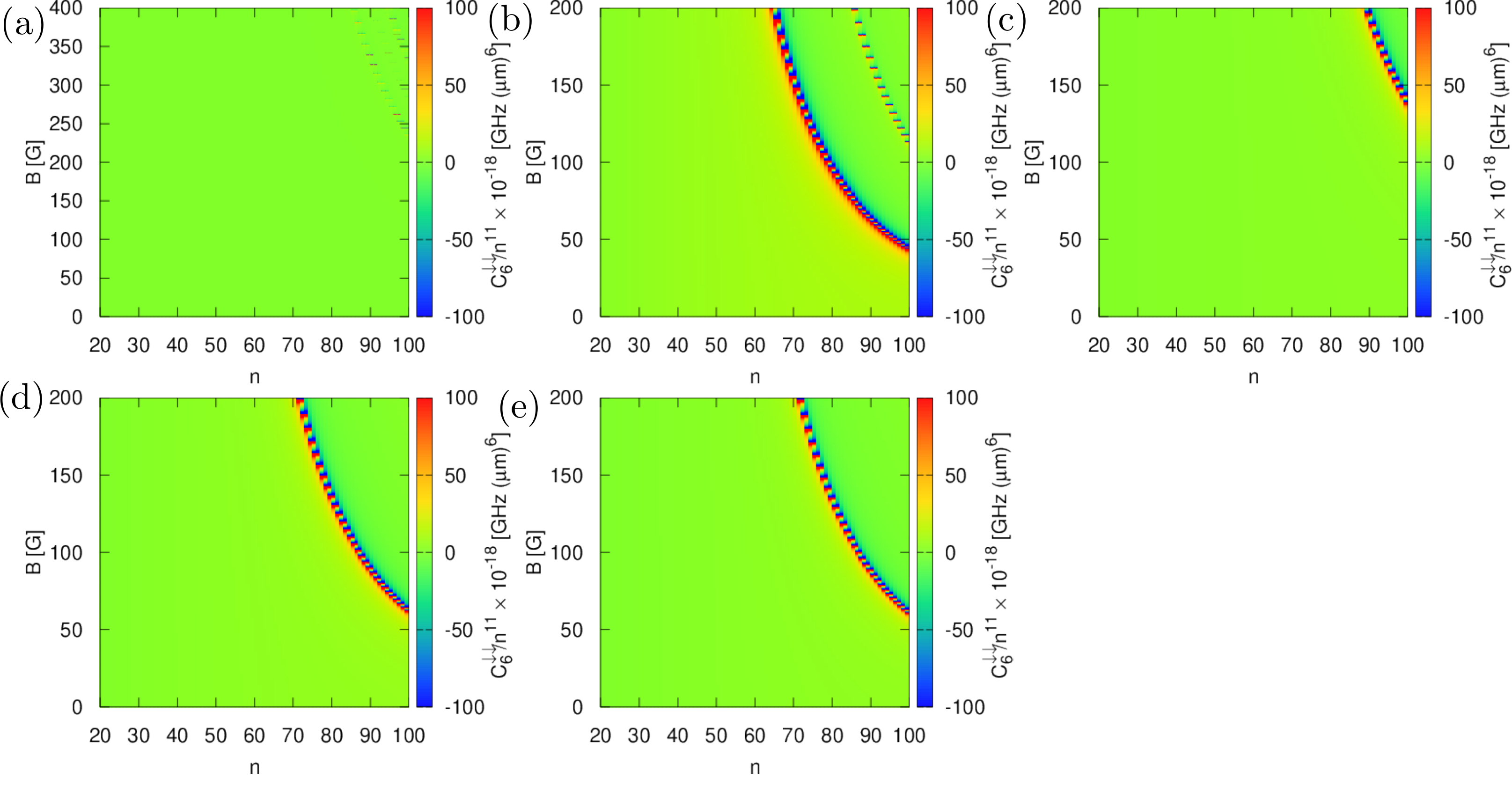}
\caption{Magnetic field and principal quantum number dependence of $C_6^{\downarrow\downarrow}$ of the pair $|nS_{1/2},m_J=1/2\rangle$ and $|(n+1)S_{1/2},m_J=1/2\rangle$ for $\theta=\pi/2$. (a) ${}^7$Li atom, (b) ${}^{23}$Na atom, (c) ${}^{39}$K atom, (d) ${}^{87}$Rb atom, (e) ${}^{133}$Cs atom.
}
\label{fig:supple_c6_dd_positive_mJ}
\vspace{-0.75em}
\end{figure}%

\begin{figure}[t]
\centering
\includegraphics[width=17cm,clip]{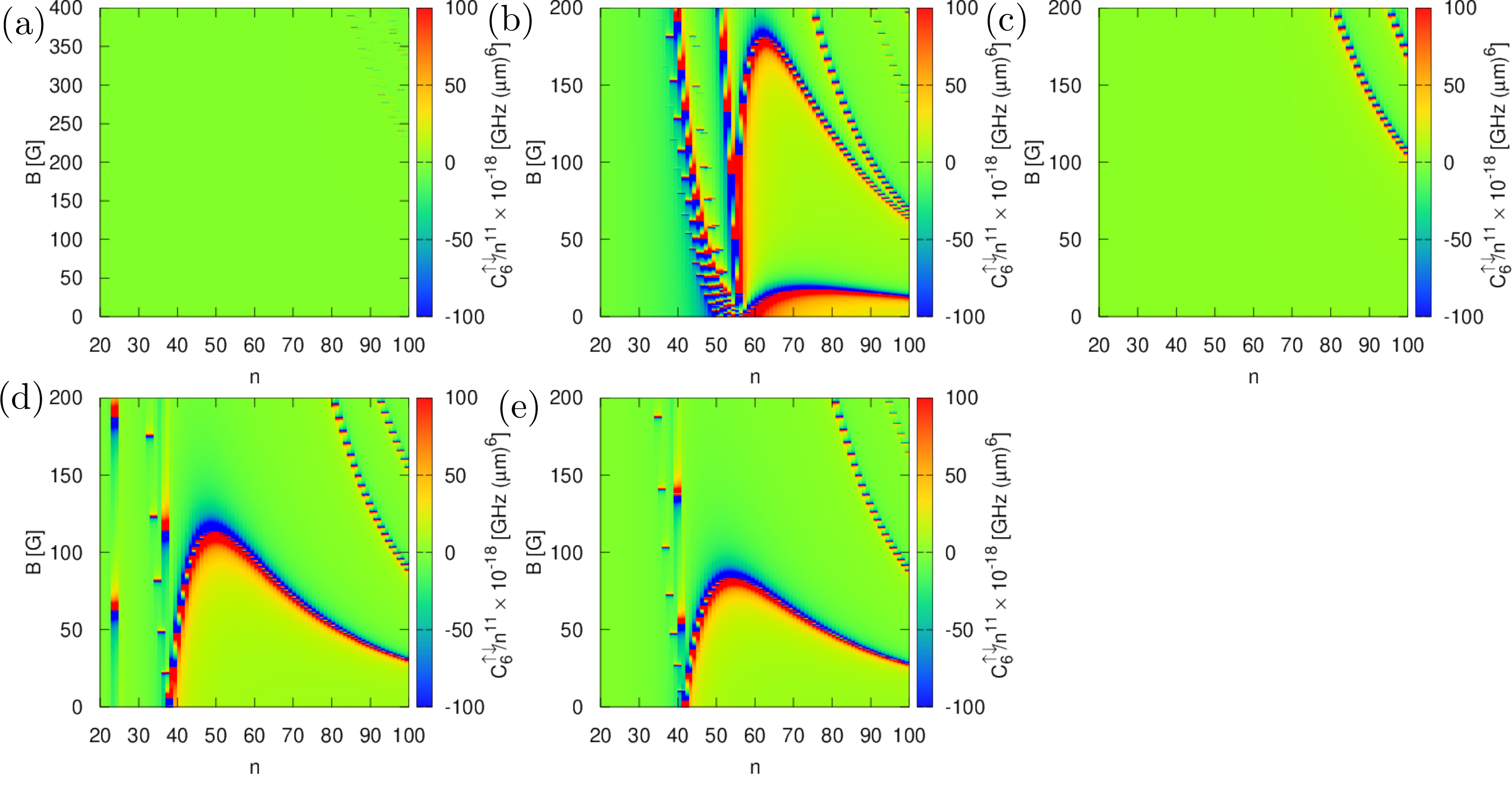}
\caption{Magnetic field and principal quantum number dependence of $C_6^{\uparrow\downarrow}$ of the pair $|nS_{1/2},m_J=1/2\rangle$ and $|(n+1)S_{1/2},m_J=1/2\rangle$ for $\theta=\pi/2$. (a) ${}^7$Li atom, (b) ${}^{23}$Na atom, (c) ${}^{39}$K atom, (d) ${}^{87}$Rb atom, (e) ${}^{133}$Cs atom.
}
\label{fig:supple_c6_ud_positive_mJ}
\vspace{-0.75em}
\end{figure}%

\begin{figure}[t]
\centering
\includegraphics[width=17cm,clip]{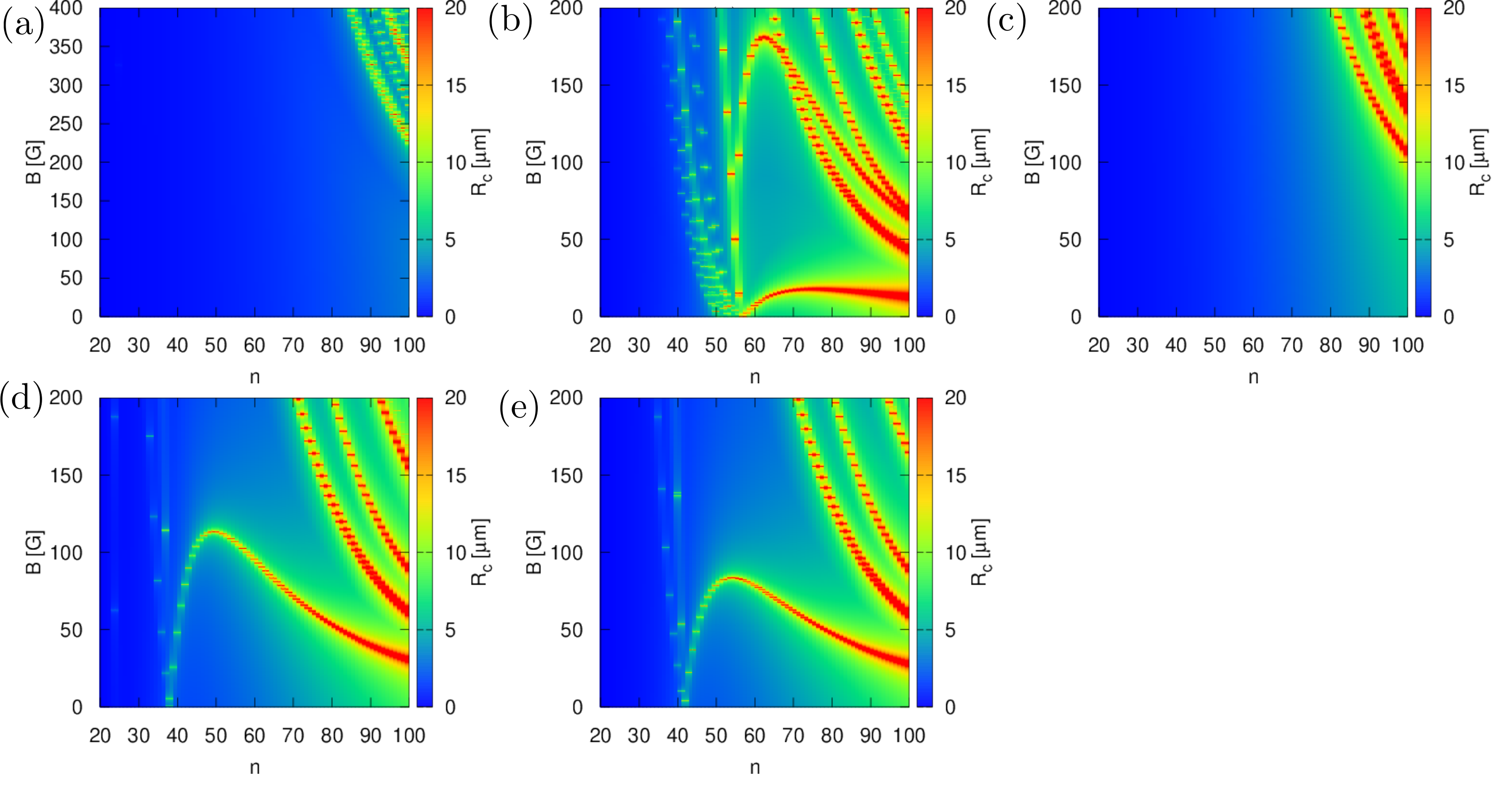}
\caption{Magnetic field and principal quantum number dependence of $R_{\rm c}$ of the pair $|nS_{1/2},m_J=1/2\rangle$ and $|(n+1)S_{1/2},m_J=1/2\rangle$ for $\theta=\pi/2$. (a) ${}^7$Li atom, (b) ${}^{23}$Na atom, (c) ${}^{39}$K atom, (d) ${}^{87}$Rb atom, (e) ${}^{133}$Cs atom.
}
\label{fig:supple_rc_positive_mJ}
\vspace{-0.75em}
\end{figure}%

\clearpage
\subsection{$|nS_{1/2},-1/2\rangle|(n+1)S_{1/2},-1/2\rangle$ pair}\label{subsec:negative_mJ_pair}

Here, we show the results for $|nS_{1/2},-1/2\rangle|(n+1)S_{1/2},-1/2\rangle$ pair in Figs.~\ref{fig:supple_delta_negative_mJ}--\ref{fig:supple_rc_negative_mJ}.

\begin{figure}[t]
\centering
\includegraphics[width=17cm,clip]{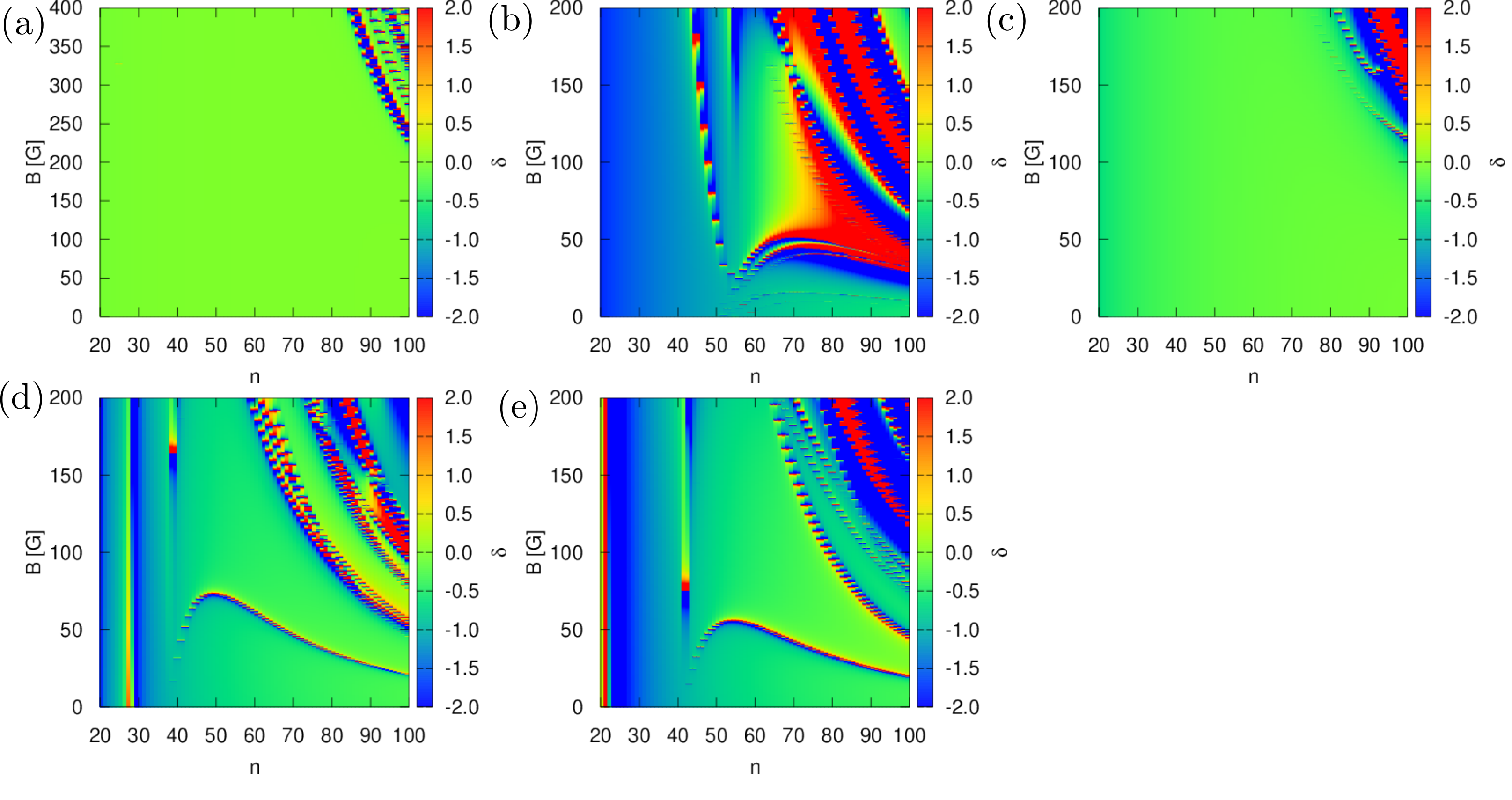}
\caption{Magnetic field and principal quantum number dependence of the anisotropy parameter of the pair $|nS_{1/2},m_J=-1/2\rangle$ and $|(n+1)S_{1/2},m_J=-1/2\rangle$ for $\theta=\pi/2$. (a) ${}^7$Li atom, (b) ${}^{23}$Na atom, (c) ${}^{39}$K atom, (d) ${}^{87}$Rb atom, (e) ${}^{133}$Cs atom.
}
\label{fig:supple_delta_negative_mJ}
\vspace{-0.75em}
\end{figure}%

\begin{figure}[t]
\centering
\includegraphics[width=17cm,clip]{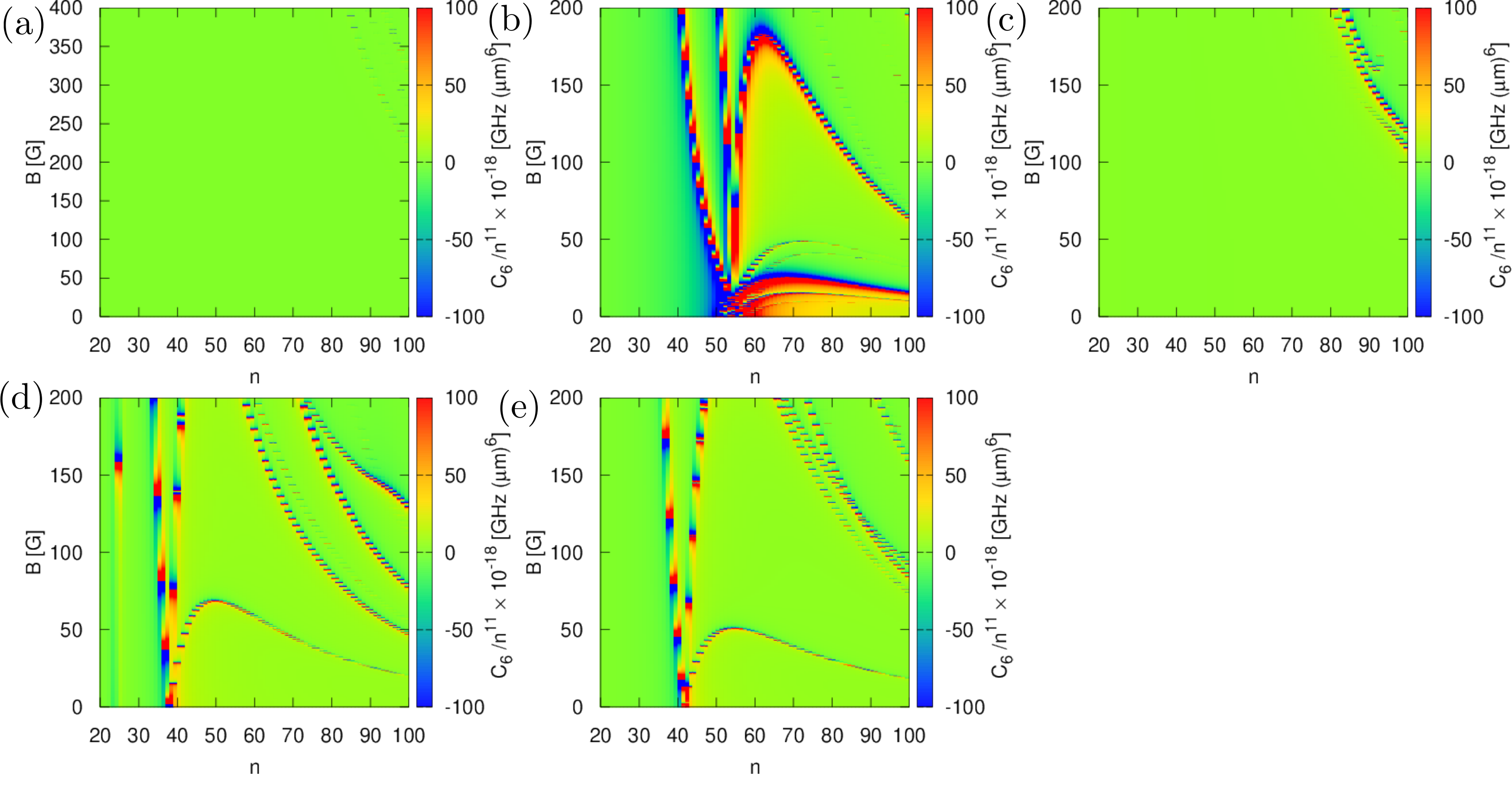}
\caption{Magnetic field and principal quantum number dependence of $C_6$ of the pair $|nS_{1/2},m_J=-1/2\rangle$ and $|(n+1)S_{1/2},m_J=-1/2\rangle$ for $\theta=\pi/2$. (a) ${}^7$Li atom, (b) ${}^{23}$Na atom, (c) ${}^{39}$K atom, (d) ${}^{87}$Rb atom, (e) ${}^{133}$Cs atom.
}
\label{fig:supple_c6_negative_mJ}
\vspace{-0.75em}
\end{figure}%

\begin{figure}[t]
\centering
\includegraphics[width=17cm,clip]{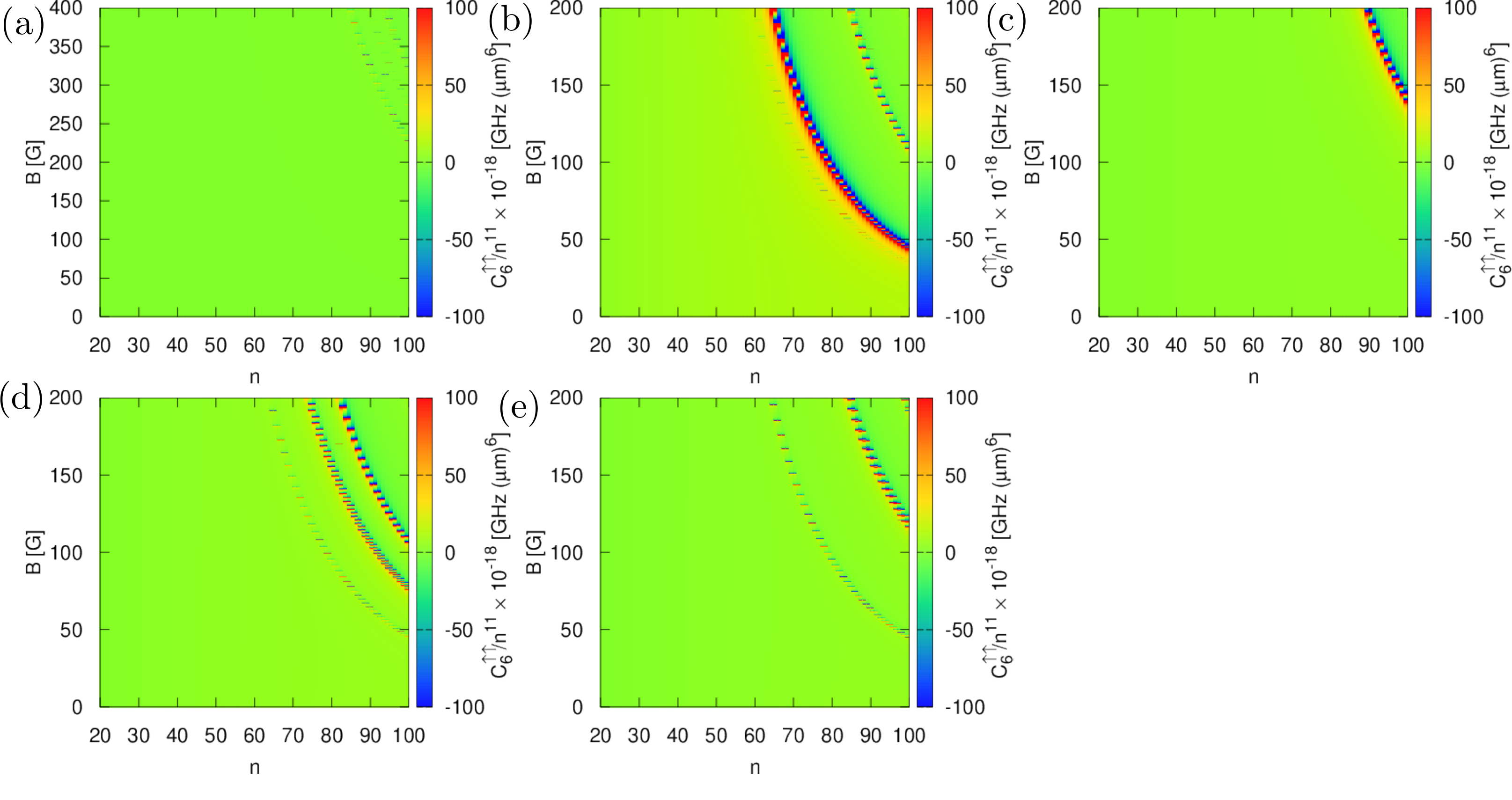}
\caption{Magnetic field and principal quantum number dependence of $C_6^{\uparrow\uparrow}$ of the pair $|nS_{1/2},m_J=-1/2\rangle$ and $|(n+1)S_{1/2},m_J=-1/2\rangle$ for $\theta=\pi/2$. (a) ${}^7$Li atom, (b) ${}^{23}$Na atom, (c) ${}^{39}$K atom, (d) ${}^{87}$Rb atom, (e) ${}^{133}$Cs atom.
}
\label{fig:supple_c6_uu_negative_mJ}
\vspace{-0.75em}
\end{figure}%

\begin{figure}[t]
\centering
\includegraphics[width=17cm,clip]{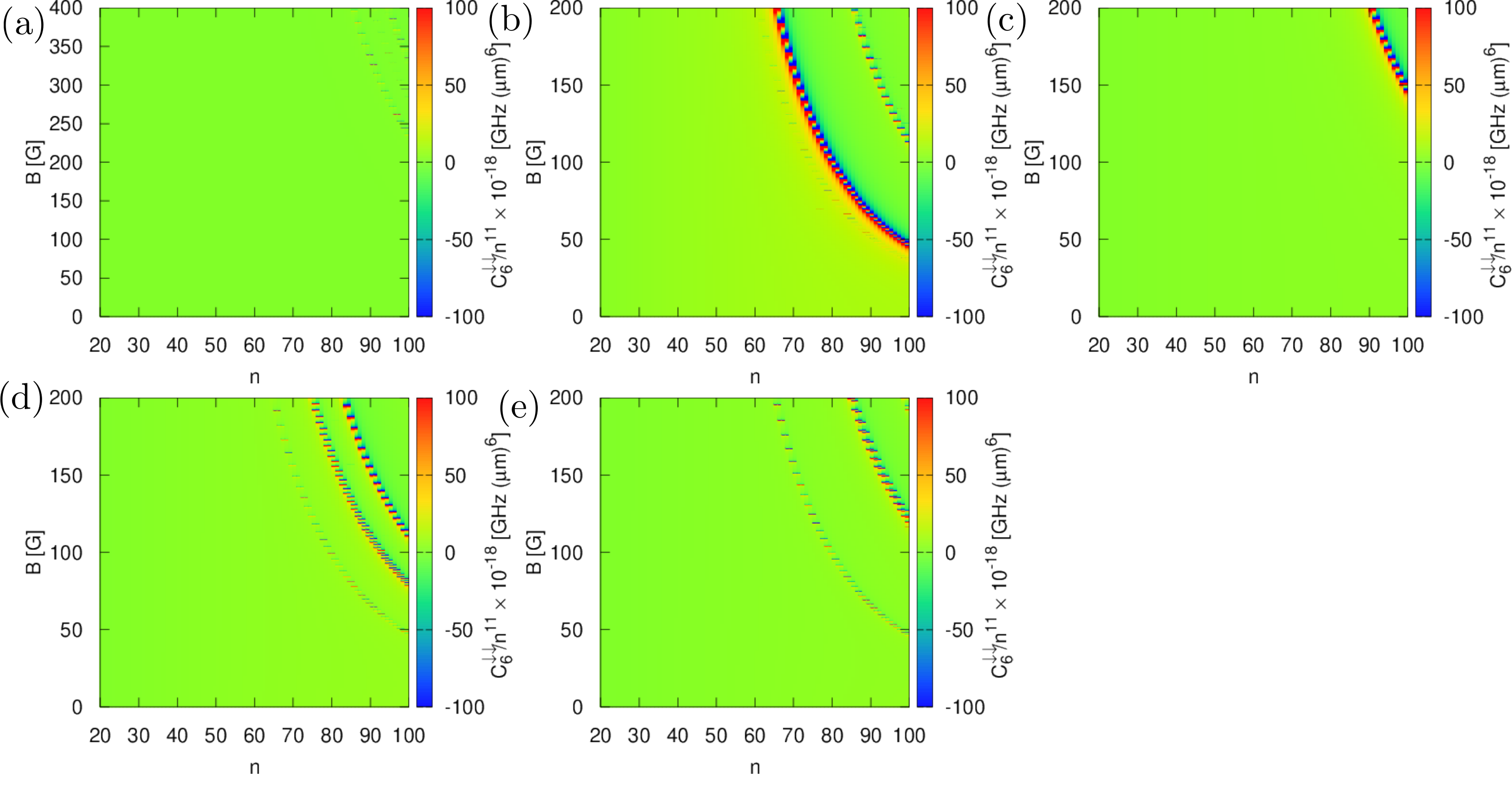}
\caption{Magnetic field and principal quantum number dependence of $C_6^{\downarrow\downarrow}$ of the pair $|nS_{1/2},m_J=-1/2\rangle$ and $|(n+1)S_{1/2},m_J=-1/2\rangle$ for $\theta=\pi/2$. (a) ${}^7$Li atom, (b) ${}^{23}$Na atom, (c) ${}^{39}$K atom, (d) ${}^{87}$Rb atom, (e) ${}^{133}$Cs atom.
}
\label{fig:supple_c6_dd_negative_mJ}
\vspace{-0.75em}
\end{figure}%

\begin{figure}[t]
\centering
\includegraphics[width=17cm,clip]{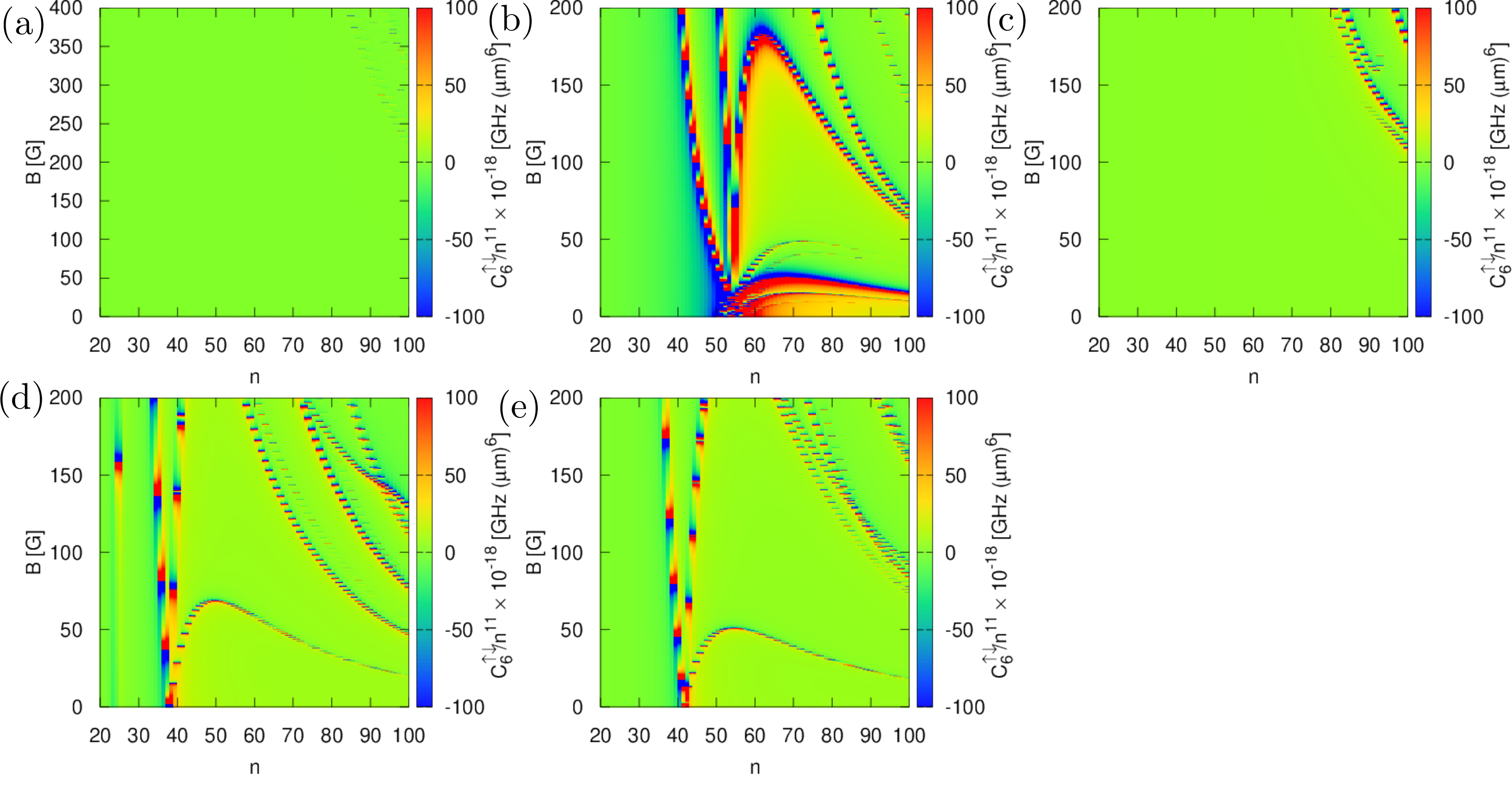}
\caption{Magnetic field and principal quantum number dependence of $C_6^{\uparrow\downarrow}$ of the pair $|nS_{1/2},m_J=-1/2\rangle$ and $|(n+1)S_{1/2},m_J=-1/2\rangle$ for $\theta=\pi/2$. (a) ${}^7$Li atom, (b) ${}^{23}$Na atom, (c) ${}^{39}$K atom, (d) ${}^{87}$Rb atom, (e) ${}^{133}$Cs atom.
}
\label{fig:supple_c6_ud_negative_mJ}
\vspace{-0.75em}
\end{figure}%

\begin{figure}[t]
\centering
\includegraphics[width=17cm,clip]{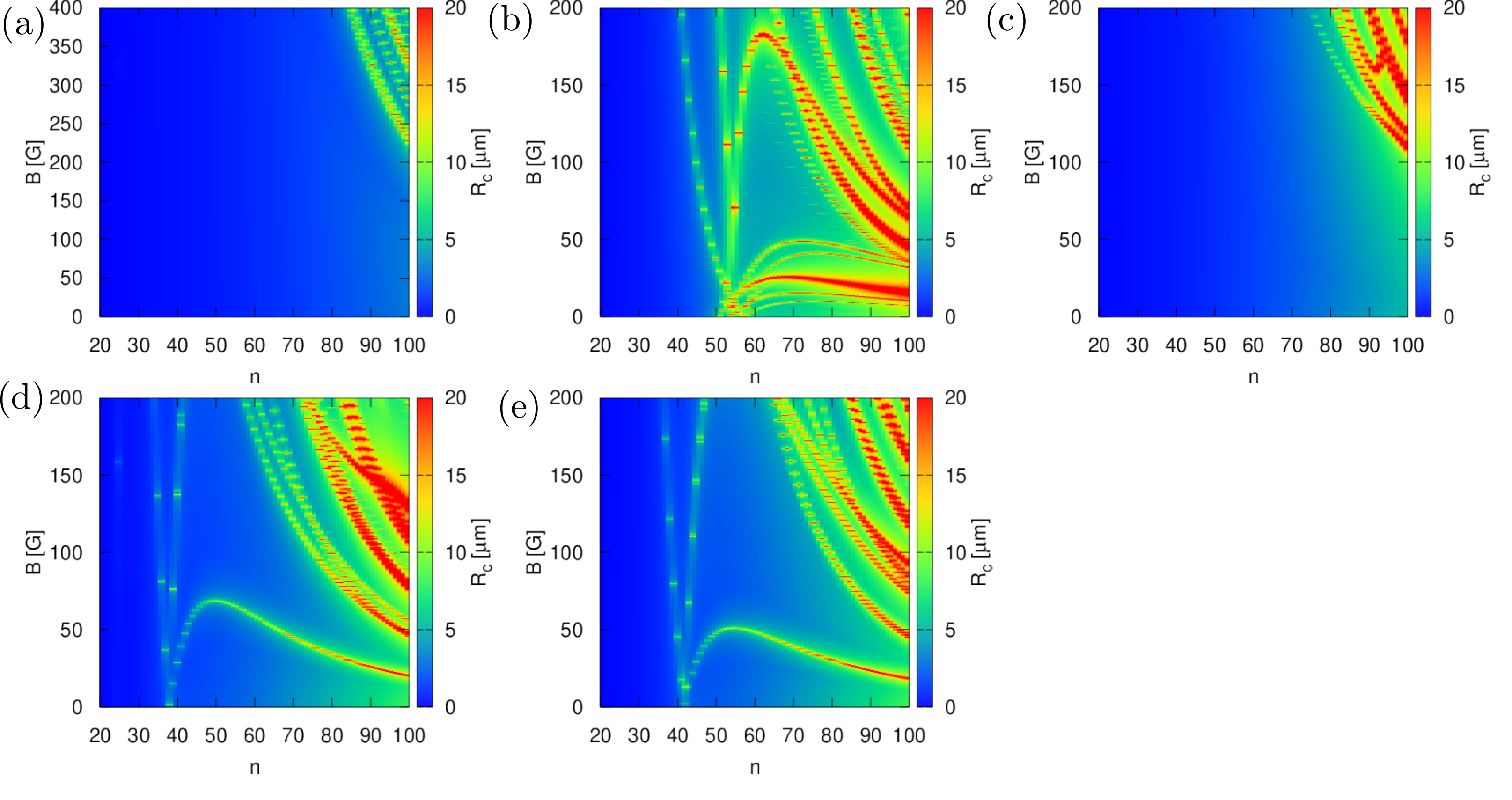}
\caption{Magnetic field and principal quantum number dependence of $R_{\rm c}$ of the pair $|nS_{1/2},m_J=-1/2\rangle$ and $|(n+1)S_{1/2},m_J=-1/2\rangle$ for $\theta=\pi/2$. (a) ${}^7$Li atom, (b) ${}^{23}$Na atom, (c) ${}^{39}$K atom, (d) ${}^{87}$Rb atom, (e) ${}^{133}$Cs atom.
}
\label{fig:supple_rc_negative_mJ}
\vspace{-0.75em}
\end{figure}%

\clearpage
\section{Summary of the Heisenberg points}\label{sec:Heisenberg_point}

In this section, we summarize the representative Heisenberg points based on the results shown in Sec.~\ref{sec:other_atomic_species}. We focus on the following quantities: $n$, $B$, $\delta$, $C_6$, $R_{\rm c}$, $J$, and $d\delta/dB$. The strength of the exchange interaction $J$ is evaluated at $R = 2R_{\rm c}$. The quantity $d\delta/dB$ represents the sensitivity of the anisotropy parameter $\delta$ to fluctuations in the magnetic field. The derivative is numerically calculated using the simple forward difference method.

The results are presented in Tables~\ref{table:supple_list_of_Heisenberg_points_Li_positive}--\ref{table:supple_list_of_Heisenberg_points_Cs_negative}, where we list the data satisfying the condition $|\delta - 1| \le 0.005$. Since the interval for $B$ is $0.05~{\rm G}$ in our calculations, we may miss data points in narrow resonance regions.

\begin{table}[h]
\centering
\caption{List of Heisenberg points for ${}^7$Li atom and $m_J=1/2$ pair. }
\begin{tabular}{cccccccc}\hline\hline
$n$ & $B$ [G] & $\delta$ & $C_6~[{\rm GHz}\cdot(\mu {\rm m})^6]$ & $R_{\rm c}~[\mu{\rm m}]$ & $J/h~[{\rm MHz}]$ & $d\delta/dB~[{\rm mG}^{-1}]$\\ \hline
87 & 398.15 & 1.0036 & 372.63 & 4.68 & 1.1013 & $-2.6 \times 10^{-4}$ \\ 
88 & 359.25 & 1.0007 & -817.04 & 6.01 & -0.5414 & $-6.1 \times 10^{-4}$ \\ 
88 & 382.80 & 0.9991 & 417.07 & 4.81 & 1.0576 & $-2.7 \times 10^{-4}$ \\ 
89 & 368.20 & 0.9962 & 466.02 & 4.93 & 1.0145 & $-2.8 \times 10^{-4}$ \\ 
90 & 354.30 & 0.9964 & 519.64 & 5.06 & 0.9707 & $-2.9 \times 10^{-4}$ \\ 
91 & 341.05 & 1.0027 & 577.79 & 5.19 & 0.9237 & $-3.1 \times 10^{-4}$ \\ 
92 & 328.45 & 1.0033 & 642.55 & 5.32 & 0.8842 & $-3.2 \times 10^{-4}$ \\ 
93 & 316.45 & 1.0018 & 714.05 & 5.45 & 0.8483 & $-3.3 \times 10^{-4}$ \\ 
94 & 286.70 & 0.9992 & -2058.18 & 7.24 & -0.4480 & $-7.4 \times 10^{-4}$ \\ 
94 & 305.00 & 1.0044 & 791.86 & 5.59 & 0.8115 & $-3.5 \times 10^{-4}$ \\ 
95 & 294.10 & 0.9986 & 878.91 & 5.72 & 0.7823 & $-3.7 \times 10^{-4}$ \\ 
96 & 374.50 & 1.0046 & -15266.60 & 10.73 & -0.3120 & $-4.5 \times 10^{-4}$ \\ 
96 & 380.70 & 1.0041 & -4140.80 & 8.88 & -0.2635 & $-1.4 \times 10^{-3}$ \\ 
96 & 383.05 & 1.0010 & -2945.36 & 8.93 & -0.1817 & $2.6 \times 10^{-3}$ \\ 
97 & 362.30 & 0.9953 & -10397.11 & 9.30 & -0.5006 & $-3.2 \times 10^{-4}$ \\ 
98 & 308.55 & 0.9983 & 2280.72 & 7.57 & 0.3781 & $-1.9 \times 10^{-3}$ \\ 
98 & 358.95 & 0.9983 & -3477.62 & 11.70 & -0.0423 & $-1.9 \times 10^{-2}$ \\ 
99 & 255.15 & 1.0009 & 1309.83 & 6.29 & 0.6619 & $-4.4 \times 10^{-4}$ \\   \hline\hline
\end{tabular}
\label{table:supple_list_of_Heisenberg_points_Li_positive}
\end{table}

\begin{table}[h]
\renewcommand{\arraystretch}{0.3}
\centering
\caption{List of Heisenberg points for ${}^{23}$Na atom and $m_J=1/2$ pair. }
\begin{tabular}{cccccccc}\hline\hline
 $n$ & $B$ [G] & $\delta$ & $C_6~[{\rm GHz}\cdot(\mu {\rm m})^6]$ & $R_{\rm c}~[\mu{\rm m}]$ & $J/h~[{\rm MHz}]$ & $d\delta/dB~[{\rm mG}^{-1}]$\\ \hline 
43 & 196.25 & 0.9973 & 2.78 & 1.99 & 1.4144 & $1.5 \times 10^{-4}$ \\ 
44 & 162.30 & 0.9970 & 3.60 & 2.17 & 1.0689 & $2.0 \times 10^{-4}$ \\ 
44 & 191.15 & 1.0004 & 3.65 & 2.29 & 0.7901 & $3.4 \times 10^{-4}$ \\ 
45 & 132.90 & 0.9965 & 4.65 & 2.39 & 0.7892 & $2.9 \times 10^{-4}$ \\ 
45 & 158.80 & 1.0039 & 4.69 & 2.42 & 0.7384 & $3.9 \times 10^{-4}$ \\ 
46 & 130.85 & 1.0020 & 6.02 & 2.56 & 0.6681 & $4.6 \times 10^{-4}$ \\ 
47 & 106.65 & 0.9972 & 7.70 & 2.73 & 0.5834 & $5.7 \times 10^{-4}$ \\ 
48 & 85.65 & 0.9955 & 9.79 & 2.92 & 0.4888 & $7.5 \times 10^{-4}$ \\ 
48 & 99.20 & 1.0045 & 9.81 & 3.75 & 0.1099 & $1.9 \times 10^{-3}$ \\ 
50 & 51.35 & 0.9997 & 15.56 & 3.45 & 0.2902 & $1.6 \times 10^{-3}$ \\ 
52 & 35.10 & 0.9998 & 24.46 & 4.28 & 0.1237 & $2.6 \times 10^{-3}$ \\ 
55 & 103.05 & 1.0014 & 50.22 & 5.90 & 0.0373 & $1.8 \times 10^{-4}$ \\ 
60 & 21.50 & 1.0018 & 128.26 & 5.69 & 0.1182 & $-1.2 \times 10^{-3}$ \\ 
61 & 25.75 & 1.0010 & 155.91 & 5.52 & 0.1731 & $-8.5 \times 10^{-4}$ \\ 
63 & 32.75 & 0.9959 & 229.14 & 5.34 & 0.3083 & $-4.8 \times 10^{-4}$ \\ 
64 & 35.65 & 0.9971 & 276.38 & 5.30 & 0.3882 & $-3.8 \times 10^{-4}$ \\ 
66 & 40.55 & 0.9993 & 399.88 & 5.28 & 0.5772 & $-2.5 \times 10^{-4}$ \\ 
66 & 186.20 & 0.9974 & -3091.39 & 9.91 & -0.1020 & $-1.1 \times 10^{-3}$ \\ 
67 & 42.65 & 0.9996 & 480.06 & 5.28 & 0.6896 & $-2.1 \times 10^{-4}$ \\ 
68 & 44.55 & 1.0043 & 574.33 & 5.30 & 0.8137 & $-1.7 \times 10^{-4}$ \\ 
68 & 159.30 & 1.0022 & 5548.40 & 8.69 & 0.4016 & $1.1 \times 10^{-4}$ \\ 
69 & 46.35 & 1.0032 & 688.54 & 5.31 & 0.9603 & $-1.4 \times 10^{-4}$ \\ 
69 & 142.35 & 1.0008 & 2885.06 & 6.54 & 1.1560 & $4.9 \times 10^{-5}$ \\ 
69 & 183.25 & 1.0005 & -1997.79 & 6.57 & -0.7730 & $4.0 \times 10^{-5}$ \\ 
70 & 48.10 & 0.9990 & 826.56 & 5.32 & 1.1357 & $-1.2 \times 10^{-4}$ \\ 
70 & 128.55 & 1.0000 & 2450.33 & 5.91 & 1.8036 & $4.0 \times 10^{-5}$ \\ 
70 & 183.70 & 1.0011 & -1608.47 & 6.13 & -0.9508 & $5.3 \times 10^{-5}$ \\ 
71 & 49.80 & 0.9993 & 990.17 & 5.34 & 1.3405 & $-9.6 \times 10^{-5}$ \\ 
71 & 116.80 & 0.9993 & 2362.90 & 5.61 & 2.3780 & $3.6 \times 10^{-5}$ \\ 
71 & 181.00 & 1.0014 & -1573.79 & 6.03 & -1.0212 & $6.3 \times 10^{-5}$ \\ 
72 & 51.55 & 1.0009 & 1186.94 & 5.34 & 1.5923 & $-7.7 \times 10^{-5}$ \\ 
72 & 106.35 & 1.0000 & 2397.28 & 5.44 & 2.8945 & $3.2 \times 10^{-5}$ \\ 
72 & 177.00 & 1.0015 & -1651.70 & 6.05 & -1.0488 & $7.3 \times 10^{-5}$ \\ 
73 & 53.55 & 0.9991 & 1430.03 & 5.34 & 1.9320 & $-5.9 \times 10^{-5}$ \\ 
73 & 96.65 & 1.0005 & 2487.46 & 5.34 & 3.3671 & $2.9 \times 10^{-5}$ \\ 
73 & 172.30 & 0.9994 & -1801.71 & 6.14 & -1.0523 & $8.3 \times 10^{-5}$ \\ 
74 & 56.00 & 1.0005 & 1732.25 & 5.31 & 2.4154 & $-4.1 \times 10^{-5}$ \\ 
74 & 87.15 & 0.9999 & 2598.27 & 5.27 & 3.8089 & $2.4 \times 10^{-5}$ \\ 
74 & 167.25 & 0.9982 & -2009.64 & 6.26 & -1.0418 & $9.4 \times 10^{-5}$ \\ 
75 & 60.05 & 0.9997 & 2147.58 & 5.28 & 3.0988 & $-2.0 \times 10^{-5}$ \\ 
75 & 76.75 & 1.0003 & 2679.08 & 5.20 & 4.2374 & $1.5 \times 10^{-5}$ \\ 
75 & 120.05 & 0.9991 & 5412.43 & 13.00 & 0.0351 & $4.6 \times 10^{-3}$ \\ 
75 & 162.05 & 1.0003 & -2272.18 & 6.41 & -1.0228 & $1.1 \times 10^{-4}$ \\ 
76 & 156.75 & 0.9989 & -2603.50 & 6.59 & -0.9960 & $1.2 \times 10^{-4}$ \\ 
77 & 109.30 & 1.0000 & 5924.87 & 13.82 & 0.0266 & $6.5 \times 10^{-3}$ \\ 
77 & 151.50 & 1.0026 & -2999.47 & 6.78 & -0.9660 & $1.3 \times 10^{-4}$ \\ 
78 & 146.25 & 0.9978 & -3493.67 & 7.00 & -0.9306 & $1.5 \times 10^{-4}$ \\ 
79 & 141.15 & 1.0003 & -4073.31 & 7.23 & -0.8947 & $1.7 \times 10^{-4}$ \\ 
80 & 136.15 & 0.9992 & -4776.54 & 7.47 & -0.8566 & $1.9 \times 10^{-4}$ \\ 
81 & 131.30 & 0.9999 & -5613.67 & 7.74 & -0.8180 & $2.1 \times 10^{-4}$ \\ 
82 & 126.60 & 1.0007 & -6614.25 & 8.02 & -0.7789 & $2.3 \times 10^{-4}$ \\ 
83 & 122.05 & 1.0001 & -7814.88 & 8.31 & -0.7395 & $2.6 \times 10^{-4}$ \\ 
84 & 117.65 & 0.9967 & -9260.88 & 8.63 & -0.6999 & $2.8 \times 10^{-4}$ \\ 
87 & 105.45 & 0.9990 & -15473.06 & 9.68 & -0.5862 & $3.9 \times 10^{-4}$ \\ 
88 & 101.70 & 1.0028 & -18391.62 & 10.07 & -0.5502 & $4.3 \times 10^{-4}$ \\ 
91 & 91.30 & 1.0043 & -31270.23 & 11.39 & -0.4468 & $6.0 \times 10^{-4}$ \\ 
92 & 88.10 & 0.9984 & -37522.66 & 11.90 & -0.4139 & $6.7 \times 10^{-4}$ \\ 
92 & 199.30 & 0.9980 & 5045.38 & 7.08 & 1.2550 & $-2.4 \times 10^{-4}$ \\ 
94 & 183.75 & 1.0003 & 6434.51 & 7.46 & 1.1626 & $-2.6 \times 10^{-4}$ \\ 
95 & 176.55 & 1.0023 & 7251.77 & 7.66 & 1.1185 & $-2.7 \times 10^{-4}$ \\ 
97 & 74.00 & 1.0001 & -95719.42 & 15.06 & -0.2561 & $1.2 \times 10^{-3}$ \\ 
97 & 163.20 & 1.0046 & 9177.27 & 8.07 & 1.0385 & $-2.9 \times 10^{-4}$ \\ 
100 & 145.50 & 1.0044 & 12950.55 & 8.70 & 0.9354 & $-3.2 \times 10^{-4}$ \\  \hline\hline
\end{tabular}
\label{table:supple_list_of_Heisenberg_points_Na_positive}
\renewcommand{\arraystretch}{1}
\end{table}

\begin{table}[h]
\centering
\caption{List of Heisenberg points for ${}^{39}$K atom and $m_J=1/2$ pair. }
\begin{tabular}{cccccccc}\hline\hline
$n$ & $B$ [G] & $\delta$ & $C_6~[{\rm GHz}\cdot(\mu {\rm m})^6]$ & $R_{\rm c}~[\mu{\rm m}]$ & $J/h~[{\rm MHz}]$ & $d\delta/dB~[{\rm mG}^{-1}]$\\ \hline
96 & 155.85 & 1.0007 & -15793.31 & 16.59 & -0.0236 & $-8.9 \times 10^{-3}$ \\ \hline\hline
\end{tabular}
\label{table:supple_list_of_Heisenberg_points_K_positive}
\end{table}

\begin{table}[h]
\centering
\caption{List of Heisenberg points for ${}^{87}$Rb atom and $m_J=1/2$ pair. }
\begin{tabular}{cccccccc}\hline\hline
$n$ & $B$ [G] & $\delta$ & $C_6~[{\rm GHz}\cdot(\mu {\rm m})^6]$ & $R_{\rm c}~[\mu{\rm m}]$ & $J/h~[{\rm MHz}]$ & $d\delta/dB~[{\rm mG}^{-1}]$\\ \hline
34 & 143.10 & 0.9951 & 0.06 & 1.33 & 0.3481 & $8.1 \times 10^{-4}$ \\ 
85 & 193.60 & 0.9992 & 1931.34 & 6.15 & 1.1168 & $-1.2 \times 10^{-4}$ \\ 
86 & 185.70 & 0.9970 & 2167.93 & 6.31 & 1.0769 & $-1.3 \times 10^{-4}$ \\ 
87 & 178.15 & 1.0032 & 2427.02 & 6.47 & 1.0324 & $-1.3 \times 10^{-4}$ \\ 
88 & 171.05 & 1.0012 & 2717.69 & 6.63 & 0.9957 & $-1.4 \times 10^{-4}$ \\ 
89 & 164.30 & 1.0011 & 3038.64 & 6.80 & 0.9592 & $-1.4 \times 10^{-4}$ \\ 
90 & 157.90 & 0.9997 & 3394.35 & 6.97 & 0.9250 & $-1.5 \times 10^{-4}$ \\ 
91 & 151.80 & 1.0012 & 3785.34 & 7.14 & 0.8903 & $-1.5 \times 10^{-4}$ \\ 
92 & 146.00 & 1.0030 & 4216.38 & 7.32 & 0.8569 & $-1.6 \times 10^{-4}$ \\ 
93 & 140.50 & 1.0021 & 4693.57 & 7.50 & 0.8265 & $-1.6 \times 10^{-4}$ \\ 
94 & 135.25 & 1.0039 & 5216.48 & 7.68 & 0.7958 & $-1.7 \times 10^{-4}$ \\ 
95 & 130.30 & 0.9972 & 5800.64 & 7.86 & 0.7710 & $-1.7 \times 10^{-4}$ \\ 
96 & 125.55 & 0.9965 & 6436.12 & 8.04 & 0.7439 & $-1.8 \times 10^{-4}$ \\ 
97 & 121.00 & 0.9999 & 7128.08 & 8.23 & 0.7159 & $-1.8 \times 10^{-4}$ \\ 
98 & 116.70 & 0.9959 & 7897.66 & 8.42 & 0.6927 & $-1.9 \times 10^{-4}$ \\ 
99 & 112.55 & 1.0011 & 8725.69 & 8.62 & 0.6660 & $-2.0 \times 10^{-4}$ \\ 
100 & 108.60 & 1.0040 & 9635.08 & 8.82 & 0.6414 & $-2.0 \times 10^{-4}$ \\  \hline\hline
\end{tabular}
\label{table:supple_list_of_Heisenberg_points_Rb_positive}
\end{table}

\begin{table}[h]
\centering
\caption{List of Heisenberg points for ${}^{133}$Cs atom and $m_J=1/2$ pair. }
\begin{tabular}{cccccccc}\hline\hline
$n$ & $B$ [G] & $\delta$ & $C_6~[{\rm GHz}\cdot(\mu {\rm m})^6]$ & $R_{\rm c}~[\mu{\rm m}]$ & $J/h~[{\rm MHz}]$ & $d\delta/dB~[{\rm mG}^{-1}]$\\ \hline
86 & 198.20 & 1.0018 & 1038.50 & 5.37 & 1.3489 & $-9.6 \times 10^{-5}$ \\ 
87 & 190.10 & 0.9993 & 1168.24 & 5.52 & 1.2949 & $-1.0 \times 10^{-4}$ \\ 
88 & 182.40 & 0.9993 & 1311.64 & 5.66 & 1.2412 & $-1.0 \times 10^{-4}$ \\ 
89 & 175.10 & 0.9994 & 1470.63 & 5.81 & 1.1900 & $-1.1 \times 10^{-4}$ \\ 
90 & 168.15 & 1.0023 & 1645.67 & 5.97 & 1.1390 & $-1.1 \times 10^{-4}$ \\ 
91 & 161.60 & 0.9996 & 1841.29 & 6.12 & 1.0949 & $-1.2 \times 10^{-4}$ \\ 
92 & 155.35 & 1.0004 & 2055.97 & 6.28 & 1.0501 & $-1.2 \times 10^{-4}$ \\ 
93 & 149.40 & 1.0025 & 2292.09 & 6.44 & 1.0066 & $-1.3 \times 10^{-4}$ \\ 
94 & 143.80 & 0.9969 & 2556.53 & 6.60 & 0.9705 & $-1.3 \times 10^{-4}$ \\ 
95 & 138.40 & 1.0011 & 2841.78 & 6.76 & 0.9294 & $-1.4 \times 10^{-4}$ \\ 
96 & 133.30 & 0.9995 & 3159.21 & 6.93 & 0.8940 & $-1.4 \times 10^{-4}$ \\ 
97 & 128.45 & 0.9967 & 3509.01 & 7.09 & 0.8610 & $-1.5 \times 10^{-4}$ \\ 
98 & 123.80 & 0.9981 & 3889.25 & 7.26 & 0.8269 & $-1.5 \times 10^{-4}$ \\ 
99 & 119.35 & 1.0020 & 4303.08 & 7.44 & 0.7929 & $-1.6 \times 10^{-4}$ \\ 
100 & 115.15 & 0.9986 & 4764.22 & 7.61 & 0.7646 & $-1.6 \times 10^{-4}$ \\ \hline\hline
\end{tabular}
\label{table:supple_list_of_Heisenberg_points_Cs_positive}
\end{table}

\begin{table}[h]
\centering
\caption{List of Heisenberg points for ${}^7$Li atom and $m_J=-1/2$ pair. }
\begin{tabular}{cccccccc}\hline\hline
$n$ & $B$ [G] & $\delta$ & $C_6~[{\rm GHz}\cdot(\mu {\rm m})^6]$ & $R_{\rm c}~[\mu{\rm m}]$ & $J/h~[{\rm MHz}]$ & $d\delta/dB~[{\rm mG}^{-1}]$\\ \hline
86 & 388.65 & 0.9968 & -586.93 & 5.63 & -0.5753 & $-5.8 \times 10^{-4}$ \\ 
88 & 382.80 & 1.0006 & 416.96 & 4.81 & 1.0560 & $-2.7 \times 10^{-4}$ \\ 
89 & 368.20 & 0.9977 & 465.90 & 4.93 & 1.0130 & $-2.8 \times 10^{-4}$ \\ 
90 & 354.30 & 0.9979 & 519.51 & 5.06 & 0.9693 & $-2.9 \times 10^{-4}$ \\ 
91 & 320.35 & 0.9953 & -1308.61 & 6.60 & -0.4937 & $-6.7 \times 10^{-4}$ \\ 
91 & 341.05 & 1.0043 & 577.64 & 5.19 & 0.9223 & $-3.1 \times 10^{-4}$ \\ 
92 & 328.45 & 1.0048 & 642.38 & 5.32 & 0.8829 & $-3.2 \times 10^{-4}$ \\ 
93 & 316.45 & 1.0034 & 713.86 & 5.45 & 0.8470 & $-3.3 \times 10^{-4}$ \\ 
94 & 286.70 & 1.0035 & -2060.75 & 7.24 & -0.4475 & $-7.3 \times 10^{-4}$ \\ 
95 & 294.10 & 1.0001 & 878.66 & 5.72 & 0.7812 & $-3.7 \times 10^{-4}$ \\ 
95 & 394.50 & 0.9995 & -3889.16 & 8.90 & -0.2441 & $-1.5 \times 10^{-3}$ \\ 
97 & 362.30 & 0.9992 & -10391.64 & 9.31 & -0.4983 & $-3.2 \times 10^{-4}$ \\ 
97 & 367.60 & 0.9991 & -4363.59 & 8.65 & -0.3245 & $-9.8 \times 10^{-4}$ \\ 
98 & 308.55 & 0.9984 & 2280.63 & 7.57 & 0.3781 & $-1.9 \times 10^{-3}$ \\ 
99 & 255.15 & 1.0025 & 1309.42 & 6.29 & 0.6609 & $-4.4 \times 10^{-4}$ \\ \hline\hline
\end{tabular}
\label{table:supple_list_of_Heisenberg_points_Li_negative}
\end{table}

\begin{table}[h]
\centering
\renewcommand{\arraystretch}{0.3}
\caption{List of Heisenberg points for ${}^{23}$Na atom and $m_J=-1/2$ pair. }
\begin{tabular}{cccccccc}\hline\hline
$n$ & $B$ [G] & $\delta$ & $C_6~[{\rm GHz}\cdot(\mu {\rm m})^6]$ & $R_{\rm c}~[\mu{\rm m}]$ & $J/h~[{\rm MHz}]$ & $d\delta/dB~[{\rm mG}^{-1}]$\\ \hline
43 & 196.25 & 0.9973 & 2.78 & 1.99 & 1.4144 & $1.5 \times 10^{-4}$ \\ 
44 & 162.30 & 0.9970 & 3.60 & 2.17 & 1.0689 & $2.0 \times 10^{-4}$ \\ 
44 & 191.15 & 1.0004 & 3.65 & 2.29 & 0.7901 & $3.4 \times 10^{-4}$ \\ 
45 & 132.90 & 0.9965 & 4.65 & 2.39 & 0.7892 & $2.9 \times 10^{-4}$ \\ 
45 & 158.80 & 1.0039 & 4.69 & 2.42 & 0.7384 & $3.9 \times 10^{-4}$ \\ 
46 & 130.85 & 1.0020 & 6.02 & 2.56 & 0.6681 & $4.6 \times 10^{-4}$ \\ 
47 & 106.65 & 0.9972 & 7.70 & 2.73 & 0.5834 & $5.7 \times 10^{-4}$ \\ 
48 & 85.65 & 0.9955 & 9.79 & 2.92 & 0.4888 & $7.5 \times 10^{-4}$ \\ 
48 & 99.20 & 1.0045 & 9.81 & 3.75 & 0.1099 & $1.9 \times 10^{-3}$ \\ 
50 & 51.35 & 0.9997 & 15.56 & 3.45 & 0.2902 & $1.6 \times 10^{-3}$ \\ 
52 & 35.10 & 0.9998 & 24.46 & 4.28 & 0.1237 & $2.6 \times 10^{-3}$ \\ 
55 & 103.05 & 1.0014 & 50.22 & 5.90 & 0.0373 & $1.8 \times 10^{-4}$ \\ 
60 & 21.50 & 1.0018 & 128.26 & 5.69 & 0.1182 & $-1.2 \times 10^{-3}$ \\ 
61 & 25.75 & 1.0010 & 155.91 & 5.52 & 0.1731 & $-8.5 \times 10^{-4}$ \\ 
63 & 32.75 & 0.9959 & 229.14 & 5.34 & 0.3083 & $-4.8 \times 10^{-4}$ \\ 
64 & 35.65 & 0.9971 & 276.38 & 5.30 & 0.3882 & $-3.8 \times 10^{-4}$ \\ 
66 & 40.55 & 0.9993 & 399.88 & 5.28 & 0.5772 & $-2.5 \times 10^{-4}$ \\ 
66 & 186.20 & 0.9974 & -3091.39 & 9.91 & -0.1020 & $-1.1 \times 10^{-3}$ \\ 
67 & 42.65 & 0.9996 & 480.06 & 5.28 & 0.6896 & $-2.1 \times 10^{-4}$ \\ 
68 & 44.55 & 1.0043 & 574.33 & 5.30 & 0.8137 & $-1.7 \times 10^{-4}$ \\ 
68 & 159.30 & 1.0022 & 5548.40 & 8.69 & 0.4016 & $1.1 \times 10^{-4}$ \\ 
69 & 46.35 & 1.0032 & 688.54 & 5.31 & 0.9603 & $-1.4 \times 10^{-4}$ \\ 
69 & 142.35 & 1.0008 & 2885.06 & 6.54 & 1.1560 & $4.9 \times 10^{-5}$ \\ 
69 & 183.25 & 1.0005 & -1997.79 & 6.57 & -0.7730 & $4.0 \times 10^{-5}$ \\ 
70 & 48.10 & 0.9990 & 826.56 & 5.32 & 1.1357 & $-1.2 \times 10^{-4}$ \\ 
70 & 128.55 & 1.0000 & 2450.33 & 5.91 & 1.8036 & $4.0 \times 10^{-5}$ \\ 
70 & 183.70 & 1.0011 & -1608.47 & 6.13 & -0.9508 & $5.3 \times 10^{-5}$ \\ 
71 & 49.80 & 0.9993 & 990.17 & 5.34 & 1.3405 & $-9.6 \times 10^{-5}$ \\ 
71 & 116.80 & 0.9993 & 2362.90 & 5.61 & 2.3780 & $3.6 \times 10^{-5}$ \\ 
71 & 181.00 & 1.0014 & -1573.79 & 6.03 & -1.0212 & $6.3 \times 10^{-5}$ \\ 
72 & 51.55 & 1.0009 & 1186.94 & 5.34 & 1.5923 & $-7.7 \times 10^{-5}$ \\ 
72 & 106.35 & 1.0000 & 2397.28 & 5.44 & 2.8945 & $3.2 \times 10^{-5}$ \\ 
72 & 177.00 & 1.0015 & -1651.70 & 6.05 & -1.0488 & $7.3 \times 10^{-5}$ \\ 
73 & 53.55 & 0.9991 & 1430.03 & 5.34 & 1.9320 & $-5.9 \times 10^{-5}$ \\ 
73 & 96.65 & 1.0005 & 2487.46 & 5.34 & 3.3671 & $2.9 \times 10^{-5}$ \\ 
73 & 172.30 & 0.9994 & -1801.71 & 6.14 & -1.0523 & $8.3 \times 10^{-5}$ \\ 
74 & 56.00 & 1.0005 & 1732.25 & 5.31 & 2.4154 & $-4.1 \times 10^{-5}$ \\ 
74 & 87.15 & 0.9999 & 2598.27 & 5.27 & 3.8089 & $2.4 \times 10^{-5}$ \\ 
74 & 167.25 & 0.9982 & -2009.64 & 6.26 & -1.0418 & $9.4 \times 10^{-5}$ \\ 
75 & 60.05 & 0.9997 & 2147.58 & 5.28 & 3.0988 & $-2.0 \times 10^{-5}$ \\ 
75 & 76.75 & 1.0003 & 2679.08 & 5.20 & 4.2374 & $1.5 \times 10^{-5}$ \\ 
75 & 120.05 & 0.9991 & 5412.43 & 13.00 & 0.0351 & $4.6 \times 10^{-3}$ \\ 
75 & 162.05 & 1.0003 & -2272.18 & 6.41 & -1.0228 & $1.1 \times 10^{-4}$ \\ 
76 & 156.75 & 0.9989 & -2603.50 & 6.59 & -0.9960 & $1.2 \times 10^{-4}$ \\ 
77 & 109.30 & 1.0000 & 5924.87 & 13.82 & 0.0266 & $6.5 \times 10^{-3}$ \\ 
77 & 151.50 & 1.0026 & -2999.47 & 6.78 & -0.9660 & $1.3 \times 10^{-4}$ \\ 
78 & 146.25 & 0.9978 & -3493.67 & 7.00 & -0.9306 & $1.5 \times 10^{-4}$ \\ 
79 & 141.15 & 1.0003 & -4073.31 & 7.23 & -0.8947 & $1.7 \times 10^{-4}$ \\ 
80 & 136.15 & 0.9992 & -4776.54 & 7.47 & -0.8566 & $1.9 \times 10^{-4}$ \\ 
81 & 131.30 & 0.9999 & -5613.67 & 7.74 & -0.8180 & $2.1 \times 10^{-4}$ \\ 
82 & 126.60 & 1.0007 & -6614.25 & 8.02 & -0.7789 & $2.3 \times 10^{-4}$ \\ 
83 & 122.05 & 1.0001 & -7814.88 & 8.31 & -0.7395 & $2.6 \times 10^{-4}$ \\ 
84 & 117.65 & 0.9967 & -9260.88 & 8.63 & -0.6999 & $2.8 \times 10^{-4}$ \\ 
87 & 105.45 & 0.9990 & -15473.06 & 9.68 & -0.5862 & $3.9 \times 10^{-4}$ \\ 
88 & 101.70 & 1.0028 & -18391.62 & 10.07 & -0.5502 & $4.3 \times 10^{-4}$ \\ 
91 & 91.30 & 1.0043 & -31270.23 & 11.39 & -0.4468 & $6.0 \times 10^{-4}$ \\ 
92 & 88.10 & 0.9984 & -37522.66 & 11.90 & -0.4139 & $6.7 \times 10^{-4}$ \\ 
92 & 199.30 & 0.9980 & 5045.38 & 7.08 & 1.2550 & $-2.4 \times 10^{-4}$ \\ 
94 & 183.75 & 1.0003 & 6434.51 & 7.46 & 1.1626 & $-2.6 \times 10^{-4}$ \\ 
95 & 176.55 & 1.0023 & 7251.77 & 7.66 & 1.1185 & $-2.7 \times 10^{-4}$ \\ 
97 & 74.00 & 1.0001 & -95719.42 & 15.06 & -0.2561 & $1.2 \times 10^{-3}$ \\ 
97 & 163.20 & 1.0046 & 9177.27 & 8.07 & 1.0385 & $-2.9 \times 10^{-4}$ \\ 
100 & 145.50 & 1.0044 & 12950.55 & 8.70 & 0.9354 & $-3.2 \times 10^{-4}$ \\  \hline\hline
\end{tabular}
\label{table:supple_list_of_Heisenberg_points_Na_negative}
\renewcommand{\arraystretch}{1}
\end{table}

\begin{table}[h]
\centering
\caption{List of Heisenberg points for ${}^{39}$K atom and $m_J=-1/2$ pair. }
\begin{tabular}{cccccccc}\hline\hline
$n$ & $B$ [G] & $\delta$ & $C_6~[{\rm GHz}\cdot(\mu {\rm m})^6]$ & $R_{\rm c}~[\mu{\rm m}]$ & $J/h~[{\rm MHz}]$ & $d\delta/dB~[{\rm mG}^{-1}]$\\ \hline
84 & 155.95 & 1.0019 & 2450.87 & 15.09 & 0.0065 & $-1.3 \times 10^{-2}$ \\ 
92 & 188.20 & 1.0016 & -17831.58 & 17.93 & -0.0168 & $-6.2 \times 10^{-3}$ \\  \hline\hline
\end{tabular}
\label{table:supple_list_of_Heisenberg_points_K_negative}
\end{table}

\begin{table}[h]
\centering
\caption{List of Heisenberg points for ${}^{87}$Rb atom and $m_J=-1/2$ pair. }
\begin{tabular}{cccccccc}\hline\hline
$n$ & $B$ [G] & $\delta$ & $C_6~[{\rm GHz}\cdot(\mu {\rm m})^6]$ & $R_{\rm c}~[\mu{\rm m}]$ & $J/h~[{\rm MHz}]$ & $d\delta/dB~[{\rm mG}^{-1}]$\\ \hline
39 & 178.20 & 1.0003 & 0.32 & 1.58 & 0.6370 & $-1.2 \times 10^{-4}$ \\ 
50 & 74.35 & 0.9955 & 6.36 & 3.26 & 0.1656 & $-1.7 \times 10^{-3}$ \\ 
51 & 74.00 & 1.0041 & 8.00 & 3.37 & 0.1704 & $-1.7 \times 10^{-3}$ \\ 
54 & 71.40 & 1.0002 & 15.76 & 3.73 & 0.1818 & $-1.6 \times 10^{-3}$ \\ 
56 & 68.85 & 1.0048 & 24.13 & 4.00 & 0.1841 & $-1.6 \times 10^{-3}$ \\ 
63 & 195.35 & 0.9974 & 121.63 & 3.68 & 1.5377 & $-1.8 \times 10^{-4}$ \\ 
64 & 56.75 & 1.0021 & 115.62 & 5.23 & 0.1772 & $-1.6 \times 10^{-3}$ \\ 
64 & 168.80 & 0.9955 & 133.48 & 4.47 & 0.5196 & $-5.6 \times 10^{-4}$ \\ 
64 & 188.55 & 0.9989 & 151.78 & 3.77 & 1.6421 & $-1.6 \times 10^{-4}$ \\ 
65 & 182.30 & 0.9969 & 193.07 & 3.83 & 1.9117 & $-1.3 \times 10^{-4}$ \\ 
65 & 189.45 & 1.0033 & 281.65 & 5.22 & 0.4364 & $2.5 \times 10^{-4}$ \\ 
67 & 52.30 & 1.0029 & 197.22 & 5.75 & 0.1705 & $-1.7 \times 10^{-3}$ \\ 
70 & 137.25 & 1.0049 & 423.54 & 5.10 & 0.7556 & $-3.9 \times 10^{-4}$ \\ 
71 & 132.85 & 0.9958 & 520.31 & 5.26 & 0.7642 & $-3.6 \times 10^{-4}$ \\ 
72 & 138.15 & 1.0000 & 328.00 & 6.47 & 0.1402 & $1.7 \times 10^{-4}$ \\ 
73 & 124.90 & 0.9965 & 820.48 & 5.75 & 0.7082 & $-2.7 \times 10^{-4}$ \\ 
76 & 199.55 & 0.9993 & -3021.02 & 8.49 & -0.2513 & $-4.0 \times 10^{-4}$ \\ 
78 & 38.50 & 0.9961 & 1156.18 & 7.95 & 0.1435 & $-1.9 \times 10^{-3}$ \\ 
78 & 112.10 & 0.9962 & 757.91 & 7.51 & 0.1320 & $-1.1 \times 10^{-3}$ \\ 
79 & 37.45 & 1.0015 & 1336.15 & 8.17 & 0.1403 & $-2.0 \times 10^{-3}$ \\ 
79 & 102.75 & 0.9985 & 2035.55 & 8.57 & 0.1611 & $5.9 \times 10^{-4}$ \\ 
80 & 97.65 & 0.9991 & 1521.49 & 7.98 & 0.1844 & $6.2 \times 10^{-4}$ \\ 
81 & 101.90 & 0.9978 & 1593.89 & 7.60 & 0.2595 & $-7.5 \times 10^{-4}$ \\ 
82 & 34.50 & 1.0028 & 2054.35 & 8.86 & 0.1328 & $-2.1 \times 10^{-3}$ \\ 
82 & 98.70 & 0.9957 & 1914.58 & 7.69 & 0.2887 & $-7.1 \times 10^{-4}$ \\ 
83 & 95.60 & 1.0049 & 2261.02 & 7.82 & 0.3094 & $-6.9 \times 10^{-4}$ \\ 
86 & 139.30 & 0.9974 & -2062.91 & 7.53 & -0.3538 & $2.7 \times 10^{-4}$ \\ 
87 & 136.55 & 1.0015 & -1304.27 & 7.72 & -0.1924 & $4.0 \times 10^{-4}$ \\ 
88 & 81.85 & 0.9999 & 4915.74 & 8.36 & 0.4492 & $-5.6 \times 10^{-4}$ \\ 
89 & 129.15 & 0.9985 & -1405.31 & 11.21 & -0.0221 & $-3.9 \times 10^{-3}$ \\ 
90 & 27.90 & 1.0050 & 6002.87 & 10.85 & 0.1148 & $-2.3 \times 10^{-3}$ \\ 
91 & 127.70 & 1.0020 & 3661.07 & 8.72 & 0.2594 & $-1.3 \times 10^{-4}$ \\ 
91 & 139.85 & 0.9982 & 24412.88 & 12.46 & 0.2036 & $1.3 \times 10^{-4}$ \\ 
93 & 66.05 & 1.0039 & 8130.13 & 8.70 & 0.5878 & $-6.5 \times 10^{-4}$ \\ 
93 & 102.75 & 0.9989 & -36585.23 & 18.69 & -0.0268 & $8.9 \times 10^{-4}$ \\ 
93 & 139.10 & 1.0005 & 56569.66 & 19.78 & 0.0295 & $3.4 \times 10^{-3}$ \\ 
93 & 148.20 & 0.9980 & -110530.80 & 21.34 & -0.0365 & $3.4 \times 10^{-3}$ \\ 
94 & 68.60 & 0.9992 & 11327.22 & 8.81 & 0.7575 & $-4.4 \times 10^{-4}$ \\ 
94 & 100.10 & 1.0007 & -35674.01 & 14.38 & -0.1261 & $-4.1 \times 10^{-4}$ \\ 
96 & 64.85 & 1.0016 & 14731.56 & 8.88 & 0.9363 & $-4.0 \times 10^{-4}$ \\ 
97 & 63.10 & 0.9978 & 16815.37 & 8.89 & 1.0677 & $-3.8 \times 10^{-4}$ \\ 
98 & 61.40 & 1.0024 & 19075.91 & 8.91 & 1.1930 & $-3.7 \times 10^{-4}$ \\ 
99 & 59.80 & 0.9955 & 21765.97 & 8.87 & 1.3936 & $-3.4 \times 10^{-4}$ \\  \hline\hline
\end{tabular}
\label{table:supple_list_of_Heisenberg_points_Rb_negative}
\end{table}

\begin{table}[h]
\centering
\caption{List of Heisenberg points for ${}^{133}$Cs atom and $m_J=-1/2$ pair. }
\begin{tabular}{cccccccc}\hline\hline
$n$ & $B$ [G] & $\delta$ & $C_6~[{\rm GHz}\cdot(\mu {\rm m})^6]$ & $R_{\rm c}~[\mu{\rm m}]$ & $J/h~[{\rm MHz}]$ & $d\delta/dB~[{\rm mG}^{-1}]$\\ \hline
42 & 95.95 & 0.9985 & 0.54 & 1.94 & 0.3195 & $-6.2 \times 10^{-5}$ \\ 
47 & 44.25 & 0.9999 & 2.17 & 3.20 & 0.0634 & $-4.0 \times 10^{-3}$ \\ 
55 & 57.05 & 0.9966 & 14.38 & 3.67 & 0.1825 & $-1.5 \times 10^{-3}$ \\ 
58 & 55.85 & 1.0021 & 27.01 & 3.99 & 0.2098 & $-1.3 \times 10^{-3}$ \\ 
59 & 55.15 & 1.0048 & 33.05 & 4.10 & 0.2168 & $-1.3 \times 10^{-3}$ \\ 
60 & 54.35 & 1.0039 & 40.36 & 4.22 & 0.2235 & $-1.2 \times 10^{-3}$ \\ 
65 & 49.45 & 1.0046 & 104.00 & 4.87 & 0.2451 & $-1.1 \times 10^{-3}$ \\ 
74 & 39.95 & 1.0046 & 477.55 & 6.24 & 0.2529 & $-1.1 \times 10^{-3}$ \\ 
74 & 162.95 & 1.0042 & 319.33 & 8.46 & 0.0271 & $-4.1 \times 10^{-3}$ \\ 
75 & 158.15 & 1.0021 & 381.23 & 8.64 & 0.0287 & $-3.9 \times 10^{-3}$ \\ 
77 & 37.05 & 0.9963 & 763.38 & 6.75 & 0.2524 & $-1.1 \times 10^{-3}$ \\ 
79 & 98.30 & 1.0014 & 3199.84 & 8.96 & 0.1930 & $7.1 \times 10^{-4}$ \\ 
80 & 94.10 & 0.9972 & 3622.35 & 9.11 & 0.1983 & $6.9 \times 10^{-4}$ \\ 
81 & 90.15 & 1.0041 & 4100.08 & 9.29 & 0.1996 & $6.9 \times 10^{-4}$ \\ 
81 & 131.75 & 1.0026 & 984.66 & 10.06 & 0.0297 & $-3.8 \times 10^{-3}$ \\ 
82 & 86.40 & 0.9991 & 4638.30 & 9.44 & 0.2044 & $6.8 \times 10^{-4}$ \\ 
84 & 83.85 & 1.0038 & 6111.68 & 12.70 & 0.0455 & $3.2 \times 10^{-3}$ \\ 
89 & 65.20 & 1.0028 & 10786.28 & 10.76 & 0.2170 & $6.6 \times 10^{-4}$ \\ 
90 & 62.75 & 0.9993 & 12125.60 & 10.95 & 0.2194 & $6.6 \times 10^{-4}$ \\ 
94 & 54.10 & 0.9991 & 19182.76 & 11.80 & 0.2220 & $6.7 \times 10^{-4}$ \\ 
96 & 23.10 & 0.9995 & 9911.24 & 10.63 & 0.2146 & $-1.2 \times 10^{-3}$ \\ 
96 & 180.65 & 0.9986 & -55854.05 & 16.30 & -0.0929 & $1.7 \times 10^{-3}$ \\   \hline\hline
\end{tabular}
\label{table:supple_list_of_Heisenberg_points_Cs_negative}
\end{table}

\clearpage
\section{Details of the atom positions}\label{sec:details_of_atom_position}

In this section, we present the details of the atom positions for spin-1/2 models. Figure~\ref{fig:supple_position_of_Rydberg_atoms_ladder} illustrates the atom positions. We place the leftmost atom at the origin $\bm{r}_1\equiv (x_1,y_1)=(0,0)$. The coordinates of the remaining atoms are given by
\begin{align}
\bm{r}_{2n+1}&\equiv (x_{2n+1},0)=\left(2nR\sin\left(\frac{\theta_0}{2}\right),0\right),\quad n=1,2,\cdots,\label{eq:position_of_odd_atoms}\\
\bm{r}_{2n}&\equiv (x_{2n},y_{2n})=\left((2n-1)R\sin\left(\frac{\theta_0}{2}\right),\; R\cos\left(\frac{\theta_0}{2}\right)\right),\quad n=1,2,\ldots.\label{eq:position_of_even_atoms}
\end{align}
From these expressions, the distance $R_n$ between the $n$-th neighbor atoms is calculated as
\begin{align}
R_{2n+1}&=\sqrt{\left(n+\frac{1}{2}\right)^24\sin^2\left(\frac{\theta_0}{2}\right)+\cos^2\left(\frac{\theta_0}{2}\right)}R,\quad n=0,1,2,\ldots,\label{eq:distance_2n+1_th_neighbors_J1J2}\\
R_{2n}&=2nR\sin\left(\frac{\theta_0}{2}\right),\quad n=1,2,\ldots.\label{eq:distance_2n_th_neighbors_J1J2}
\end{align}
When we set $\sin(\theta_0/2)=2^{-5/6}\;(\theta_0\simeq 68.3^{\circ})$, the distance between the second-neighbor atoms is given by $R_2=2^{1/6}R\simeq 1.12R$.

\begin{figure}[t]
\centering
\includegraphics[width=8.6cm,clip]{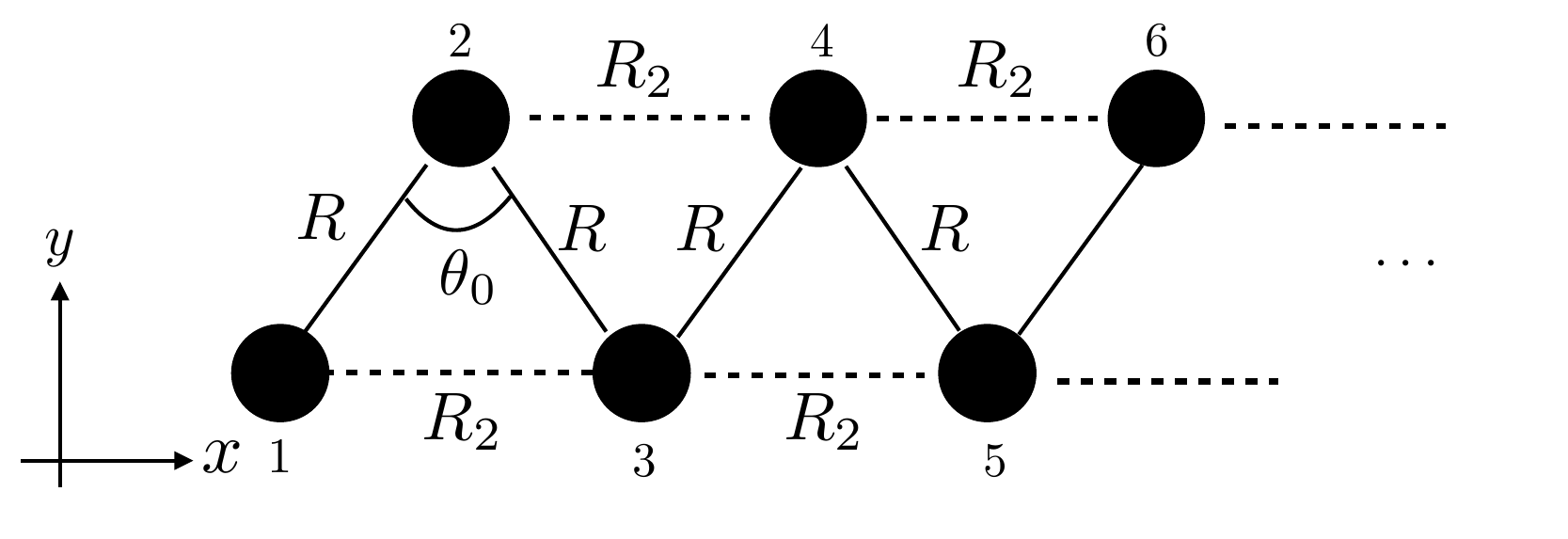}
\caption{Atom configuration for the spin-1/2 $J_1$-$J_2$ model. Black circles represent the positions of Rydberg atoms. Here, $\theta_0$ denotes the angle between the vertices of the same length.
}
\label{fig:supple_position_of_Rydberg_atoms_ladder}
\vspace{-0.75em}
\end{figure}%

\section{Perturbation theory for spin-1 systems}\label{sec:perturbation_theory}

Here, we derive the effective Hamiltonian for spin-1 models~\cite{hida1992crossover,hung2005numerical}. For simplicity, we consider four atoms ($M=2$ case) arranged in the Gelfand ladder configuration [see Fig.~4(b) in the main text]. Our starting point is the spin-1/2 {\it XXZ} Hamiltonian:
\begin{align}
\hat{H}&=J_1(\hat{S}_1^x\hat{S}_2^x+\hat{S}_1^y\hat{S}_2^y+\delta\hat{S}_1^z\hat{S}_2^z)+J_1(\hat{S}_3^x\hat{S}_4^x+\hat{S}_3^y\hat{S}_4^y+\delta\hat{S}_3^z\hat{S}_4^z)\notag \\
&+J_2(\hat{S}_1^x\hat{S}_3^x+\hat{S}_1^y\hat{S}_3^y+\delta\hat{S}_1^z\hat{S}_3^z)+J_2(\hat{S}_2^x\hat{S}_4^x+\hat{S}_2^y\hat{S}_4^y+\delta\hat{S}_2^z\hat{S}_4^z)\notag \\
&+J_3(\hat{S}_1^x\hat{S}_4^x+\hat{S}_1^y\hat{S}_4^y+\delta\hat{S}_1^z\hat{S}_4^z)+J_3(\hat{S}_2^x\hat{S}_3^x+\hat{S}_2^y\hat{S}_3^y+\delta\hat{S}_2^z\hat{S}_3^z).\label{eq:supple_definition_of_Gelfand_Ladder_Hamiltonian}
\end{align}
Here, we define the nonperturbative Hamiltonian $\hat{H}_0$ as
\begin{align}
\hat{H}_0&\equiv J_1(\hat{\bm{S}}_1\cdot\hat{\bm{S}}_2+\hat{\bm{S}}_3\cdot\hat{\bm{S}}_4).\label{eq:supple_definition_of_nonperturbative_Hamiltonian}
\end{align}
The perturbation Hamiltonian is then defined by $\hat{V}\equiv \hat{H}-\hat{H}_0$.

First, we consider the nonperturbative part. Because $\hat{H}_0$ is decoupled into pairs of sites $1, 2$ and $3, 4$, we can easily diagonalize the nonperturbative Hamiltonian. The eigenstates corresponding to sites $j$ and $j+1$ are given by:
\begin{align}
J_1 \hat{\bm{S}}_j \cdot \hat{\bm{S}}_{j+1} \cket{J_{1,m}}_{j,j+1} &= \frac{J_1}{4} \cket{J_{1,m}}_{j,j+1}, \quad m = -1, 0, 1, \label{eq:supple_triplet_eigenstate} \\
J_1 \hat{\bm{S}}_j \cdot \hat{\bm{S}}_{j+1} \cket{J_{0,0}}_{j,j+1} &= -\frac{3J_1}{4} \cket{J_{0,0}}_{j,j+1}. \label{eq:supple_singlet_eigenstate}
\end{align}
Here, we define the triplet and singlet states as
\begin{align}
\cket{J_{1,1}}_{j,j+1} &\equiv \cket{\uparrow_j \uparrow_{j+1}}, \label{eq:supple_definition_of_J11} \\
\cket{J_{1,0}}_{j,j+1} &\equiv \frac{1}{\sqrt{2}} (\cket{\uparrow_j \downarrow_{j+1}} + \cket{\downarrow_j \uparrow_{j+1}}), \label{eq:supple_definition_of_J10} \\
\cket{J_{1,-1}}_{j,j+1} &\equiv \cket{\downarrow_j \downarrow_{j+1}}, \label{eq:supple_definition_of_J1-1} \\
\cket{J_{0,0}}_{j,j+1} &\equiv \frac{1}{\sqrt{2}} (\cket{\uparrow_j \downarrow_{j+1}} - \cket{\downarrow_j \uparrow_{j+1}}). \label{eq:supple_definition_of_J00}
\end{align}
The eigenstates of $\hat{H}_0$ are given by product states of triplet and/or singlet states on each bond. If $|J_1| \gg |(1 - \delta) J_1|$, $|J_2|$, and $|J_3|$ hold, the energy difference between the triplet and singlet sectors is large. Therefore, we can apply the standard perturbation theory to the system.

To construct the effective Hamiltonian, we define the target subspace $\mathcal{H}_P$, which is spanned by the triplet states:
\begin{align}
\mathcal{H}_P \equiv \mathrm{Span}\left\{\cket{J_{1,m}}_{1,2} \otimes \cket{J_{1,m'}}_{3,4} \mid m,m' = 1,0,-1 \right\}.
\label{eq:supple_definition_of_target_subspace_for_perturbation}
\end{align}
The total Hilbert space $\mathcal{H}$ can be decomposed into $\mathcal{H}=\mathcal{H}_P\oplus\mathcal{H}_Q$. We define the projection operator onto $\mathcal{H}_P$ as $\hat{P}$, and its complement as $\hat{Q} \equiv \hat{1} - \hat{P}$, where $\hat{1}$ is the identity operator. The zeroth-order effective Hamiltonian is given by
\begin{align}
\hat{H}_{\rm eff}^{(0)} &= \hat{P} \hat{H}_0 \hat{P} = \frac{J_1}{2}.
\label{eq:supple_zeroth_order_effective_Hamiltonian}
\end{align}
The first-order effective Hamiltonian is defined as
\begin{align}
\hat{H}_{\rm eff}^{(1)} \equiv \hat{P} \hat{V} \hat{P}.
\label{eq:supple_definition_of_first_order_perturbation_theory}
\end{align}
A direct calculation shows that the first-order effective Hamiltonian is given by
\begin{align}
\hat{H}_{\rm eff}^{(1)} &= \frac{J_1}{2}(\delta - 1)\left[-1 + (\hat{\tau}_1^z)^2 + (\hat{\tau}_2^z)^2 \right] + \frac{(J_2 + J_3)\delta}{2} \hat{\tau}_1^z \hat{\tau}_2^z + \frac{J_2 + J_3}{2} \left( \hat{\tau}_1^x \hat{\tau}_2^x + \hat{\tau}_1^y \hat{\tau}_2^y \right),
\label{eq:supple_first_order_effective_Hamiltonian_Gelfand}
\end{align}
where $\hat{\tau}_j^{\mu}~(\mu = x,y,z)$ are spin-1 operators acting on site $j$. Here, the pair of sites 1 and 2 (respectively, 3 and 4) corresponds to site 1 (respectively, site 2) in the effective spin-1 system [see Fig.~4(b) in the main text]. When we set $\delta = 1$, the effective Hamiltonian reduces to the spin-1 Heisenberg model:
\begin{align}
\hat{H}_{\rm eff}^{(1)} &= \frac{J_2 + J_3}{2} \hat{\bm{\tau}}_1 \cdot \hat{\bm{\tau}}_2=J_1^{S=1}\hat{\bm{\tau}}_1 \cdot \hat{\bm{\tau}}_2.
\label{eq:supple_Heisenberg_point}
\end{align}
The second-order effective Hamiltonian is defined by
\begin{align}
\hat{H}_{\rm eff}^{(2)} &\equiv -\sum_n \frac{\hat{P} \hat{V} \hat{Q}_n \hat{V} \hat{P}}{E_n - J_1/2},
\label{eq:supple_definition_of_second-order_effective_Hamiltonian}
\end{align}
where $\hat{Q}_n \equiv \cket{\varphi_n} \bra{\varphi_n}$, and $\cket{\varphi_n}$ is an eigenstate of $\hat{H}_0$ with eigenvalue $E_n$ in the subspace $\mathcal{H}_Q$. A direct calculation shows that the second-order effective Hamiltonian is given by
\begin{align}
\hat{H}_{\rm eff}^{(2)} &= \frac{(J_2 - J_3)^2}{8J_1} \left[
\cket{+_1 -_2} \bra{+_1 -_2}
- \delta \cket{+_1 -_2} \bra{0_1 0_2}
+ \cket{+_1 -_2} \bra{-_1 +_2}
- \delta \cket{0_1 0_2} \bra{+_1 -_2} \right. \notag \\
&\left.\hspace{5.0em}
+ \delta^2 \cket{0_1 0_2} \bra{0_1 0_2}
- \delta \cket{0_1 0_2} \bra{-_1 +_2}
+ \cket{-_1 +_2} \bra{+_1 -_2}
- \delta \cket{-_1 +_2} \bra{0_1 0_2}
+ \cket{-_1 +_2} \bra{-_1 +_2}
\right],
\label{eq:supple_second_order_effective_Hamiltonian_bracket_form}
\end{align}
where $\cket{+_j}$, $\cket{0_j}$, and $\cket{-_j}$ are eigenstates of $\hat{\tau}_j^z$ with eigenvalues $+1$, $0$, and $-1$, respectively. The matrix representation of this term becomes
\begin{align}
\hat{H}_{\rm eff}^{(2)}&\to\frac{(J_2-J_3)^2}{8J_1}
\begin{bmatrix}
0 &  &  &  &  &  &  &  &  \\
 & 0 & 0 &  &  &  &  &  &  \\
 & 0 & 0 &  &  &  &  &  &  \\
 &  &  & 1 & -\delta & 1 &  &  &  \\
 &  &  & -\delta & \delta^2 & -\delta &  &  &  \\
 &  &  & 1 & -\delta & 1 &  &  &  \\
 &  &  &  &  &  & 0 & 0 &  \\
 &  &  &  &  &  & 0 & 0 &  \\
 &  &  &  &  &  &  &  & 0
\end{bmatrix}
\begin{pmatrix}
\cket{++} \\
\cket{+0} \\
\cket{0+} \\
\cket{+-} \\
\cket{00} \\
\cket{-+} \\
\cket{0-}\\
\cket{-0}\\
\cket{--}
\end{pmatrix}
.\label{eq:supple_matrix_representation_second_order_effective_Hamiltonian}
\end{align}
When $\delta = 1$, the second-order term reduces to a simpler form:
\begin{align}
\hat{H}_{\rm eff}^{(2)} &= \frac{(J_2 - J_3)^2}{8J_1} \left[ \left( \hat{\bm{\tau}}_1 \cdot \hat{\bm{\tau}}_2 \right)^2 - 1 \right].
\label{eq:supple_second_order_term_Heisenberg_point}
\end{align}
Therefore, at the Heisenberg point ($\delta=1$), the effective Hamiltonian becomes the bilinear-biquadratic model:
\begin{align}
\hat{H}_{\rm eff}&=\frac{J_1}{2}+\frac{J_2+J_3}{2}\hat{\bm{\tau}}_1\cdot\hat{\bm{\tau}}_2+\frac{(J_2 - J_3)^2}{8J_1} \left[ \left( \hat{\bm{\tau}}_1 \cdot \hat{\bm{\tau}}_2 \right)^2 - 1 \right].\label{eq:supple_total_effective_Hamiltonian_for_M=4_case}
\end{align}

Then, we extend the above results to the many-spin case. A straightforward calculation shows that the effective Hamiltonian becomes
\begin{align}
\hat{H}_{\rm eff} &= \sum_{j=1}^{M-1} \left[J_1^{S=1}\hat{\bm{\tau}}_j \cdot \hat{\bm{\tau}}_{j+1}
+ J'^{S=1}_1\left( \hat{\bm{\tau}}_j \cdot \hat{\bm{\tau}}_{j+1} \right)^2\right]
+J_2^{S=1}\sum_{j=1}^{M-2} \hat{\bm{\tau}}_j \cdot \hat{\bm{\tau}}_{j+2},
\label{eq:supple_effective_Hamiltonian_many-body_and_NNN_case}\\
J_2^{S=1}&\equiv \frac{J_4+J_5}{2},\quad J'^{S=1}_1\equiv \frac{(J_2 - J_3)^2}{8J_1},\label{eq:supple_definition_of_J2_and_J1'}
\end{align}
where constant terms are omitted, the next-nearest-neighbor interaction is included as a first-order perturbation, and $J_4\equiv 2h C_6/(2R_2)^6$ and $J_5\equiv 2h C_6/(R_1^2+4R_2^2)^3$ are the fourth and fifth neighbor interaction strength of the spin-1/2 systems. The reason for including the next-nearest-neighbor terms will be discussed below.

Here, we evaluate the magnitude of the interaction strengths. We rewrite $J_1^{S=1}$, $J_1'^{S=1}$, and $J_2^{S=1}$ as functions of $J_2 / J_1$ as follows:
\begin{align}
J_{1}^{S=1} &= J_1 \left\{ \frac{1}{2} \frac{J_2}{J_1} + \frac{1}{2} \frac{J_2 / J_1}{\left[1 + (J_2 / J_1)^{1/3} \right]^3} \right\}, \label{eq:supple_spin1_NN_interaction_as_J2/J1} \\
J_1'^{S=1} &= \frac{J_1}{8} \left\{ \frac{J_2}{J_1} - \frac{J_2 / J_1}{\left[1 + (J_2 / J_1)^{1/3} \right]^3} \right\}^2, \label{eq:supple_spin1-NN_biquadratic_term_J2/J1} \\
J_{2}^{S=1} &= J_1 \left\{ \frac{1}{128} \frac{J_2}{J_1} + \frac{1}{2} \frac{J_2 / J_1}{\left[4 + (J_2 / J_1)^{1/3} \right]^3} \right\}. \label{eq:supple_spin1-NNN_interaction_as_J2/J1}
\end{align}
From these expressions, the ratios $J_1'^{S=1} / J_1^{S=1}$ and $J_2^{S=1} / J_1^{S=1}$ are approximately $0.03$ and $0.02$, respectively, when $J_2 / J_1 = 0.2$. This implies that the second-order perturbative term is comparable to the first-order next-nearest-neighbor term. Therefore, both terms must be considered simultaneously to maintain consistency.

\end{widetext}
\end{document}